\documentclass[a4paper]{article}

\RequirePackage{amsthm,amsmath,amsfonts,amssymb}
\RequirePackage[authoryear]{natbib}
\RequirePackage[colorlinks,citecolor=blue,urlcolor=blue]{hyperref}
\RequirePackage{graphicx}
\RequirePackage{geometry}

\theoremstyle{plain}

\theoremstyle{remark}

\newcommand{\bmu}{\mbox{\boldmath $\mu$}}

\newcommand{\balpha}{\mbox{\boldmath $\alpha$}}
\newcommand{\bbeta}{\mbox{\boldmath $\beta$}}
\newcommand{\bgamma}{\mbox{\boldmath $\gamma$}}

\newcommand{\bsigma}{{\bf \Sigma}}

\newcommand{\T}{\ensuremath{{\mathsf{T}}}}

\newcommand{\E}{\textrm{E}}

\newcommand{\logit}{\textrm{logit}}
\newcommand{\diag}{\textrm{diag}}

\newcommand{\Corr}{\textrm{Corr}}

\newcommand{\median}{\textrm{median}}
\newcommand{\med}{\textrm{med}}

\newcommand{\aRI}{\textrm{aRI}}

\newcommand{\A}{\textrm{A}}
\newcommand{\B}{\textrm{B}}
\newcommand{\Acat}{\textrm{A-cat}}
\newcommand{\Bcat}{\textrm{B-cat}}
\newcommand{\Y}{{\bf Y}}
\newcommand{\X}{{\bf X}}
\newcommand{\K}{{\bf K}}

\newcommand{\naiveleer}{na\"ive }
\newcommand{\naive}{na\"ive}

\newcommand{\SIMEX}{\textrm{\footnotesize SIMEX}}
\newcommand{\NCI}{\textrm{\footnotesize NCI}}
\newcommand{\MI}{\textrm{\footnotesize MI}}
\newcommand{\rel}{\mathrm{\footnotesize rel}}
\newcommand{\Hdrei}{\textrm{\footnotesize H3}}
\newcommand{\Hvier}{\textrm{\footnotesize H4}}
\newcommand{\y}{{\bf y}}

\title{Correcting for bias due to categorisation based on cluster analysis using multiple continuous error-prone exposures}

\author{Timm Intemann$^{1}$ \and Iris Pigeot$^{1, 2}$}
\date{%
	$^1$Biometry and Data Management, Leibniz Institute for Prevention Research and Epidemiology - BIPS, Bremen, Germany\\%
	$^2$Faculty of Mathematics and Computer Science, University Bremen, Bremen, Germany\\[2ex]%
	\small{intemann@leibniz-bips.de, pigeot@leibniz-bips.de}
}

\begin{document}
\maketitle

\begin{abstract}
The association between multidimensional exposure patterns and outcomes is commonly investigated by first applying cluster analysis algorithms to derive patterns and then estimating the associations. However, errors in the underlying continuous, possibly skewed, exposure variables lead to misclassified exposure patterns and therefore to biased effect estimates. This is often the case for lifestyle exposures in epidemiology, e.g. for dietary variables measured on daily basis. 
We introduce three new algorithms for correcting the biased effect estimates, which are based on regression calibration (RC), simulation extrapolation (SIMEX) and multiple imputation (MI). In addition, the naïve method ignoring the measurement error structure is considered for comparison. These methods are combined with the $k$-means cluster algorithm and the Gaussian mixture model to derive exposure patterns. The performance of the correction methods is compared in a simulation study regarding absolute, maximum and relative bias. The simulated data mimic a typical situation in nutritional epidemiology when diet is assessed using repeated 24-hour dietary recalls. Continuous and binary outcomes are considered.
Simulation results show, that the correction method based on RC and MI perform better than the naïve and the SIMEX-based method. Furthermore, the MI-based approach, which can use outcome information in the error model, is superior to the RC-based approach in most scenarios. Therefore, we recommend using the MI-based approach.
\end{abstract}



\section{Introduction}
Measurement errors (ME) occur in many variables in epidemiological research and can lead to wrong conclusions when associations between exposures and health outcomes are investigated \citep{Carroll06}. Reasons for ME are manifold, e.g. biological and random variability or shortcomings of the assessment method. In addition, there are estimation and recall problems when data are based on study participants’ reports \citep{Subar03, Bornhorst13a}. Daily variation due to assessment on a daily basis can also be considered as ME when the variable of interest is the usual or habitual behaviour \citep{Kipnis09}. This is for example the case when accelerometers or 24-hour dietary recalls (24HDR) are used for the assessment of physical activity or dietary intake on multiple days. An additional complication is that the distribution of these daily measurements is often skewed.

 Accelerometers or 24-hour dietary recalls do not only capture a single exposure but a high number of exposures, which may be correlated with each other \citep{Boeing14}. Furthermore, usually not single but  a combination of many of such exposures is associated with chronic diseases, such as cancer \citep{WCRF07}. To handle the multidimensional character for multifactorial health outcomes besides principal component analysis or appropriate indices researchers often use cluster analysis to derive exposure patterns, e.g. dietary patterns \citep{Boeing14} or lifestyle patterns \citep{Bel-Serrat13a}. Additionally, this approach might be helpful when the single exposure effects are too small and can only be proven if the exposures are bundled \citep{Moeller07}. In contrast to considering all continuous variables as single exposures, using exposure patterns may lead to simpler models and may allow for better model interpretability. Moreover, including a high number of exposure variables in a regression model leads to problems regarding multicollinearity, model extrapolation and multiple testing. First performing a cluster analysis may avoid these problems. Two popular cluster analysis methods are the $k$-means cluster algorithm \citep{Hartigan79, Ocke13} and the Gaussian mixture model (GMM) \citep{Fahey07}, which are based on Euclidean distances and on the maximum likelihood method, respectively. However, the underlying measurement error of the exposures is usually not considered and leads to misclassification in the exposure patterns, which may bias the association with the health outcome. 

Counterintuitively, categorising an error-prone continuous exposure leads to differential misclassification even if the error of the original continuous exposure is non-differential \citep{Flegal91}. Nevertheless, this approach can sometimes reduce bias compared to using the original continuous exposure \citep{Gustafson02}. \citet{Keogh12} developed and systematically compared different correction methods for categorised error-prone exposures in a simulation study but only for univariate exposures. Among others, they considered correction methods based on regression calibration (RC) \citep{Carroll06}, simulation extrapolation (SIMEX) \citep{Cook94, Stefanski95} and multiple imputation (MI) \citep{Carroll95}. RC, more precisely the NCI-method \citep{Kipnis09}, was also used in a study \citep{Hebestreit17}, in which exposure patterns based on multiple error-prone repeated measurements of several dietary components were derived. In this study, the NCI-method was used to estimate the individual usual intake of food groups. Subsequently, the individual estimates were used in a cluster analysis to derive dietary patterns. Finally, based on these patterns, the association between parental and children’s dietary patterns was investigated. Such a three stage approach combines the advantages of clustering and error correction. Since a systematic comparison of such correction methods for multiple error-prone exposures is lacking it is not clear how effective such a non-linear method is and if approaches based on SIMEX or MI may reduce the bias even more. In addition, the STRATOS (STRengthening Analytical Thinking for Observational Studies) group “Measurement error and Misclassification” \citep{Freedman18} recommends extending measurement correction methods and especially MI for non-linear exposure-outcome associations \citep{Shaw20}. 

Therefore, we will first introduce correction methods for misclassified exposure patterns due to clustering of multiple error-prone exposures, second apply them on real data and third compare them in a simulation study. The paper is structured as follows. Section~\ref{s:measurement} describes the underlying measurement error as well as the linear and logistic health models. In Section~\ref{s:correction}, the correction methods based on RC, SIMEX and MI are introduced. These correction methods are used in Section~\ref{s:app} to estimate the association between dietary patterns derived from 24HDR data and long-term blood sugar $\textrm{HbA}_{1c}$ based on real data from the I.Family study \citep{Ahrens17}. This example is mimicked by a simulation study, whose design and results are described in Section~\ref{s:simulation}. Furthermore, measures are introduced in this section to assess the different methods based on the simulation results. A final discussion and a recommendation for the application of the correction methods are given in Section~\ref{s:discussion}.

\section{Measurement error and health model}\label{s:measurement}
\citet{Dodd06}, \citet{Tooze06} and \citet{Kipnis09} developed a measurement error model for single dietary intake components. This model can be used for error-prone exposures measured on a daily basis. To account for skewness of the dietary intake components, the Box-Cox transformation is applied. Instead of only considering one single component we consider $M$ true usual components $Y_{mi}$, $m=1,...,M, i=1,...,I$, for $I$ individuals. The corresponding true component on day $t$ is denoted by $Y_{mit}$, $t = 1,..., T_i$. We assume that $Y_{mit}$ is unobservable and can only be assessed with errors using a 24-hour dietary recall (24HDR) on day $t$. The resulting error-prone version of $Y_{mit}$ is denoted by $Y_{mit}^{*}$. Following conventions in nutritional epidemiology, $Y_{mit}^{*}$ is assumed to be unbiased for $Y_{mi}$ \citep{Dodd06}. Furthermore, we assume that a Box-Cox transformation $g_{\lambda_m}$, $g_{\lambda_m} = (\nu^{\lambda_m} -1) / \lambda_m$, $\lambda_m > 0$, exists such that the non-linear mixed effects error model

\begin{align}
	g_{\lambda_m}(Y_{mit}^{*}) &= \E (g_{\lambda_m}(Y_{mit}^{*}) ) + \varepsilon_{mit} \nonumber \\
	&= \beta_{0m} + \X_i \bbeta_m + u_{mi} + \varepsilon_{mit}\label{EM}
\end{align}
holds with mutually independent random variables $u_{mi} \sim \mathcal{N}(0, \sigma_{u m}^{2})$ and $\varepsilon_{mit} \sim\mathcal{N}(0, \sigma_{\varepsilon m}^{2})$, covariate vector $\X_i$, parameter vector $\bbeta_m$ and intercept $\beta_{0m}$. Implicitly, we assume that the expectation of the transformed error-prone exposure follows the linear regression model $\E (g_{\lambda_m}(Y_{mit}^{*}) ) = \beta_{0m} + \X_i  \bbeta_m+ u_{mi}$. Furthermore, the assumption of unbiasedness means $Y_{mi} = E(Y_{mit}^{*}| \X_i, u_{mi})$. For details please see \citet{Kipnis09}.

The health model including the exposure pattern depending on the true exposures and the health outcome $H_i$ is assumed to be either a linear or a logistic regression model. The exposure pattern is assumed to be a categorical variable with $C$ separate categories, such that each individual can be assigned to exactly one group depending on a classification function $K$. Typically, the classification function $K$ and the true exposures $\Y_i = (Y_{1i},...,Y_{Mi})$ are unknown and must be estimated from the data. Assuming both is known, the linear health model has the form
\begin{equation}\label{LinHM}
	H_i = \alpha_0 + \K(\Y_i) \balpha_K + \X_i \balpha_X + e_i
\end{equation}
if $H_i$ is continuous and the logistic health model has the form 
\begin{equation}\label{LogHM}
	\logit(H_i|\K(\Y_i), \X) = \gamma_0 + \K(\Y_i) \bgamma_K + \X_i \bgamma_X
\end{equation}
if $H_i$ is dichotomous where $\alpha_0, \balpha_K, \balpha_X, \gamma_0, \bgamma_K$ and $\bgamma_X$ are parameters,  $e_i \sim \mathcal{N}(0, \sigma_{e}^{2})$ and the covariate vector $\K(\Y_i)$ summarises the individual dummy-coded categories of $K$. The corresponding $(C-1)$ coefficient parameters $\alpha_{K12},..., \alpha_{K1 C}$ and $\gamma_{K12},..., \gamma_{K1 C}$ contrasting the first with each of the other clusters are summarised in $\balpha_K$ and $\bgamma_K$, respectively. For sake of simplicity, we use the same covariate vector $\X_i$ as in the measurement error model. 

Furthermore, we assume that if the true exposures $\Y_i$, $i=1,...,I$, are known, $K$ can be obtained correctly using a chosen cluster algorithm. For this purpose, the $k$-means cluster algorithm \citep{Hartigan79} and the Gaussian mixture model (GMM) \citep{Banfield93} are considered. For both methods input values are standardised beforehand, which is regarded as one step of the cluster methods in this paper. The classification function of the $k$-means algorithm is based on smallest distances between the individual and derived cluster means, whereas the classification function of the GMM is based on the maximum of the estimated weighted density functions for each individual. Using an estimate $\hat{K}$ for $K$ based on an error-prone version $\Y_i^{*}$ of $\Y_i$ instead of the true variable $K(\Y_i)$ leads to misclassification, which results in biased estimates of the parameters in the health model. Therefore, measurement error correction methods have to be applied to reduce bias.

\section{Correction Methods}\label{s:correction}
Usually, correction methods consider either errors in continuous covariates or misclassification in categorical covariates and ignore misclassification due to categorisation of an error-prone continuous covariate. Examples of correction methods for this latter situation can  for example be found in \citet{Keogh12}. \citet{Gustafson03} and \cite{Buonaccorsi10}, who theoretically investigated the case of dichotomising an error-prone continuous covariate, showed that even if the error in the continuous variables is non-differential the error in the dichotomised variables actually is differential. The same can be shown for the non-differential error model and the classification of the health model from Section \ref{s:measurement} (see Appendix \ref{appA}).
Nevertheless, categorising a single error-prone continuous variable can sometimes reduce bias \citep{Gustafson02}. To further reduce the bias of estimates for $\balpha_K$ and $\bgamma_K$ from health models~(\ref{LinHM}) and (\ref{LogHM}) we suggest in the following three stage methods which consist of: (i) estimation of the usual exposure, (ii) clustering and classification of individuals and (iii) fit of the health model. They are therefore referred to as three stage approach (3-SA).

\subsection{Na\"ive Method}\label{ss:naive}
The simplest way to estimate $\balpha_K$ and $\bgamma_K$ is the \naiveleer method. Its three stages are the following. 
First, the individual mean exposures $\bar{\Y}_i^{*} = (\bar{Y}_{1i\bullet}^*,..., \bar{Y}_{Mi\bullet}^*)$,  $\bar{Y}_{mi\bullet}^* = 1/ T_i \sum_{t=1}^{T_i}Y_{mit}^*$, $m=1,...,M$, are calculated ignoring the measurement error structure. Second, $\bar{\Y}_i^{*}$ is used instead of the unknown true exposures $\Y_i$  in a cluster analysis for classification of all individuals. The obtained classification $\hat{K}$ is then used to derive the covariate vector $\hat{\K}(\bar{\Y}_i^{*})$ summarising the dummy-coded categories. The classification function is derived using either the $k$-means algorithm or GMM. Third, $\hat{\K}(\bar{\Y}_i^{*})$ is used instead of $\K(\Y_i)$ in the health models~(\ref{LinHM}) and (\ref{LogHM}) to obtain the corresponding effect estimates. 

Since $\bar{Y}_{mi\bullet}^*$ approaches the true individual intake for $T_i\rightarrow\infty$ the same is true for the corresponding classification and effect estimators. Nevertheless, in epidemiological studies the number of measurements $T_i$ cannot be increased such that $\bar{Y}_{mi\bullet}^*$ is negligible due to cost restrictions and the increasing burden for participants. Therefore $\hat{\K}(\bar{\Y}_i^{*})$ is subject to misclassification and the resulting effect estimates are usually biased. 

\subsection{Regression Calibration-based Correction Algorithm}\label{ss:regressioncalibration}
The idea of regression calibration is to replace the unknown true exposure in the health model by the estimated conditional mean exposure given reported exposures and covariates $\X_i$. This approach is transferred to the situation of underlying measurement error combined with the health models introduced in Section~\ref{s:measurement}. The general algorithm is similar to the \naiveleer method, but the assumption of the measurement error model~(\ref{EM}) is used to estimate the usual exposures in the first stage. The approach is denoted by RC-3-SA and its three stages are defined as follows.

At the first stage the measurement error model is taken into account to estimate the usual exposures $Y_{mi}$ following the regression calibration approach and by using the NCI-method \citep{Kipnis09}. The NCI-method can be briefly described as follows. Initially, the non-linear mixed effects error model~(\ref{EM}) is fitted using maximum likelihood estimation. Following the error model (\ref{EM}) the true exposure $Y_{mi}$ is given by
\begin{align}
	Y_{mi} &= \E(Y_{mit}^{*} | \X_i, u_i) \nonumber\\
	&= \E(g^{-1}_{\lambda_m}(\beta_{0m} + \X_i \bbeta_m + u_{mi} + \varepsilon_{mit}) | \X_i, u_{mi}), \label{ForTaylor}
\end{align}
where $g_{\lambda_m}^{-1}$ denotes the inverse Box-Cox transformation $g_{\lambda_m}^{-1}(\nu) = (\lambda_m \nu + 1 )^{1/\lambda_m}$. Formula (\ref{ForTaylor}) can be approximated using Taylor series expansion, in which the estimated parameters of error model~(\ref{EM}) are substituted to estimate $Y_{mi}$ for given $Y_{mi1}^{*},..., Y_{miT}^{*}$ (for details see \citet{Kipnis09}). The corresponding estimator is denoted by $\hat{Y}_{mi}^{\NCI}$.

As with the \naiveleer approach, at the second stage, the exposures $\hat{\Y}_{i}^{\NCI} = (\hat{Y}_{1i}^{\NCI},..., \hat{Y}_{Mi}^{\NCI})$ are used instead of the true exposures to obtain the classification $\hat{K}$. At the third stage, $\hat{K}$ is in turn used to derive the corresponding dummy-coded covariate vector $\hat{\K}(\hat{\Y}_{i}^{\NCI})$, which is used in the health models~(\ref{LinHM}) and (\ref{LogHM}) instead of the true exposure to estimate the parameters $\balpha_K$ and $\bgamma_K$.

As many other correction method RC-3-SA can also be seen as plug-in estimator \citep{Shaw20} and it can be justified as follows. Compared to the \naiveleer approach RC-3-SA is beneficial in several aspects. First, it considers the skewed exposure distribution, different numbers of reported days $T_i$, the inter- and intraindividual variation as well as covariates to estimate the individual exposure. Second, even without any covariates $\X_i$ the variance of $\hat{Y}_{mi}^{\NCI}$ is smaller than $\bar{Y}_{mi\bullet}^*$. Third, with additional covariates $\X_i$ the variance is further reduced (see Supplementary Material of \citet{Kipnis09}). Overall, it can therefore be argued that exploiting the model assumptions and the individual additional information of the covariates lead to a considerably better individual exposure estimate that reduces misclassification. Albeit the reduction is not complete, it might be sufficient for practical application.

\subsection{Simulation Extrapolation (SIMEX)-based Correction Algorithm}\label{ss:simex}
The idea of SIMEX is to generate exposures with varying additional measurement error to obtain a function describing the association between measurement error and effect estimate. This function is extrapolated to the hypothetical case of zero measurement error to provide the corrected effect estimate. The fundamental requirement for application of SIMEX is that the error model can be simulated using a Monte-Carlo method, which is fulfilled by the assumed error model~(\ref{EM}). Furthermore, assuming this error model for $M=1$, \citet{Intemann19b} showed that for Box-Cox transformed data the key property of SIMEX according to \citet{Carroll06} holds, i.e. that the mean squared error of the generated error-prone exposure converges to zero if its variance converges to zero. Therefore, the approaches of \citet{Intemann19b} and \citet{Keogh12} are combined to a new SIMEX-based algorithm. The algorithm is denoted by SIMEX-3-SA and its steps are defined as follows:

\begin{enumerate}
	\item \textbf{Initial model:} Since $\lambda_m$ and $\sigma_{\varepsilon m}^2$ are unknown, the non-linear mixed effects error model~(\ref{EM}) is fitted as in the RC-based approach. Furthermore, as in the \naiveleer method the individual mean exposures $\bar{\Y}_i^{*} = (\bar{Y}_{1i}^{*},..., \bar{Y}_{Mi}^{*})$ are used in a cluster analysis to derive a classification function $\hat{K}$.
	\item \textbf{Simulation step:} Analogously to \citet{Intemann19b}, the SIMEX data $Y^{(l)}_{mit}(\zeta)$ are generated as:
	\[
	Y^{(l)}_{mit}(\zeta) = g_{\lambda_m}^{-1}\left(g_{\lambda_m}(Y_{mit}^*) + Z^{(l)}_{mit} (\zeta) \right), 
	\]
	where $l, l=1,...,L$, refers to the $l$-th data set for each multiple of the error variance $\zeta, \zeta \in \mathcal{Z}$, and pseudo-random variable $Z^{(l)}_{mit} (\zeta)$ with $Z^{(l)}_{mit} (\zeta) \sim \mathcal{N}(\mu_{mi}(\zeta), \zeta \hat{\sigma}^{2}_{\varepsilon m})$. The so-called corrective expected value $\mu_{mi}(\zeta)$ ensures that $Y^{(l)}_{mit}(\zeta)$ is unbiased (for details see Appendix \ref{appB} or \citet{Intemann19b}). We used $\mathcal{Z} = \{1/4, 2/4, ..., 8/4\}$ as set of multiples of the error variance.
	
	\item \textbf{Mean exposure:} For each of the $\# \mathcal{Z} \times L$ generated data sets the individual mean exposure $\bar{Y}_{mi\bullet}^{(l)}(\zeta),$ $m=1,...,M$, is calculated.
	
	\item \textbf{Classification:} According to the classification function $\hat{K}$ (from step (1)),  the individuals are classified based on the individual mean exposures $\bar{\Y}_{i\bullet}^{(l)}(\zeta) = (\bar{Y}_{1i\bullet}^{(l)}(\zeta),..., \bar{Y}_{Mi\bullet}^{(l)}(\zeta))$ for each $\zeta \in \mathcal{Z}$. Based on $\hat{K} \left( \bar{\Y}_{i\bullet}^{(l)}(\zeta) \right)$ the corresponding dummy-coded covariate $\hat{\K}_i^{(l)}(\zeta)$ is derived. For $\hat{\K}_i(0)$ the classification based on the original data $\bar{\Y}_{i\bullet}^*$ as in step (1) is used.
	
	\item \textbf{Health model:} The health models (\ref{LinHM}) and (\ref{LogHM}) are fitted, in which $\K(\Y_i)$ is replaced by $\hat{\K}_i^{(l)}(\zeta)$ for $\zeta \in \mathcal{Z}\bigcup\{0\}$ and $l=1,...,L$. The resulting parameter estimates for $\balpha_K$ and $\bgamma_K$ are $\hat{\balpha}_K^{(l)}(\zeta) = (\hat{\alpha}_{K12}^{(l)}(\zeta),..., \hat{\alpha}_{K1 C}^{(l)}(\zeta))^\T$ in case of the linear health model and $\hat{\bgamma}_K^{(l)}(\zeta) = (\hat{\gamma}_{K12}^{(l)}(\zeta),..., \hat{\gamma}_{K1 C}^{(l)}(\zeta))^\T$ for the logistic health model.
	
	\item \textbf{Mean parameter estimates:} The last two steps are the same as in the original SIMEX algorithm. For each $\zeta \in \mathcal{Z}$ the means of the parameter estimates are calculated:
	\begin{align*}
		\bar{\hat{\balpha}}^{(\bullet)}_K(\zeta)=& (\bar{\hat{\alpha}}_{K12}^{(\bullet)}(\zeta),..., \bar{\hat{\alpha}}_{K1 C}^{(\bullet)}(\zeta))^\T \\
		= &(\hat{\alpha}_{K12}(\zeta),..., \hat{\alpha}_{K1 C}(\zeta))^\T = \hat{\balpha}_K(\zeta).
	\end{align*}
	Then the univariate estimates $\hat{\alpha}_{K1c}(\zeta)$, $c = 2,...,C$, are plotted against $\zeta \in \mathcal{Z}\bigcup\{0\}$ resulting in $C-1$ SIMEX plots. The corresponding associations between $\zeta$ and $\hat{\alpha}_{K1c}(\zeta)$, $c = 2,...,C$, are estimated using polynomials (analogously for $\hat{\bgamma}_K(\zeta)$). 
	
	\item \textbf{Extrapolation:} Each of these polynomials is extrapolated to $\zeta = -1$. The resulting estimates  $\hat{\alpha}_{K1c}(-1)$ and $\hat{\gamma}_{K1c}(-1)$ are the SIMEX estimates $\hat{\alpha}_{K1c}^{\SIMEX}$ and $\hat{\gamma}_{K1c}^{\SIMEX}$, $c=2,...,C$, summarised as vectors $\hat{\balpha}_{K}^{\SIMEX}$ and $\hat{\bgamma}_{K}^{\SIMEX}$. Analogously, the corrected estimates for $\alpha_0, \balpha_X, \gamma_0$ and $\bgamma_X$ of health models~(\ref{LinHM}) and (\ref{LogHM}) can be derived.
\end{enumerate}

The rationale behind this approach is the following. Since $Y^{(l)}_{mit}(\zeta)$ is error-free for $\zeta \rightarrow -1$, the corresponding classification for a given classification function is then also error-free, which means in turn that also the corresponding parameter estimate is error-free. Since we are not able to calculate $Y^{(l)}_{mit}(\zeta)$ or the corresponding classification for $\zeta \rightarrow -1$ we assume an extrapolation function which provides a parameter estimate for the hypothetical case of $\zeta \rightarrow -1$.

\subsection{Multiple Imputation-based Correction Algorithm}\label{ss:multipleimputation}
MI is a procedure usually applied for missing data problems \citep{Sterne09}. In this case, the missing values of one variable are imputed multiple times by values drawn from a distribution of this variable given the observed data. The actual analysis is then conducted multiple times based on the imputed data sets. The means of the corresponding estimates are the so-called MI estimates. Since every measurement error problem can be understood as a missing data problem, where the unobservable variable is missing for all observations \citep{Carroll95}, MI becomes also an attractive method for ME problems \citep{Cole06}. \citet{Keogh12} developed an MI approach for dichotomised error-prone continuous exposures. This approach needs to be extended to situations described in Section \ref{s:measurement}. For this purpose, we slightly modify the error model~(\ref{EM}) and add the health outcome as a covariate in the error model to make use of all relevant information. Therefore, we assume the following error model
\begin{equation}\label{EM_MI}
	g_{\lambda'_m}(Y_{mit}^{*}) = \beta'_{0m} + \X'_i \bbeta'_m + u'_{mi} + \varepsilon'_{mit},
\end{equation}
where the covariate vector $\X'_i$ includes the discrete or continuous health outcome $H_i$. The function $g_{\lambda'_m}$, coefficients and random variables, $u'_{mi} \sim \mathcal{N}(0, \sigma_{u' m}^{2})$ and $\varepsilon'_{mit} \sim \mathcal{N}(0, \sigma_{u' m}^{2})$ are defined analogously to error model~(\ref{EM}). As before, we assume  health model~(\ref{LinHM}) or (\ref{LogHM}). For estimating its model parameters we propose the following MI-based algorithm (MI-3-SA) similar to the SIMEX-based approach:

\begin{enumerate}
	\item \textbf{Initial model:} Model~(\ref{EM_MI}) is fitted to the data to obtain estimates for $\sigma_{u' m}^{2}$ and $\sigma_{\varepsilon' m}^{2}$, $m=1,...,M$. Furthermore, following the NCI-method we get estimates for the usual exposures on the original scale. Based on these estimates we apply a cluster algorithm, which yields a classification function $\hat{K}^{\MI}$.
	
	\item \textbf{Usual exposure on the Box-Cox-transformed scale:} Additional to the estimator for the usual exposure on the original scale, we derive the estimator for the usual exposure on the Box-Cox-transformed scale as the empirical best linear unbiased predictor (BLUP)
	\begin{align*}
		\widehat{g_{\lambda'_m}(Y_{mi})} & =  \hat{\E}(g_{\lambda'_m}(Y_{mi}) | \Y_{mi}^*, \X_i, H_i) \\
		& =   \frac{\hat{\sigma}^2_{u'm}}{\hat{\sigma}^2_{u'm} + \hat{\sigma}^2_{\varepsilon'm}}
		(\hat{\beta}'_{0m}  + \X_i \hat{\bbeta}'_{m} + \hat{\beta}'_{Hm} H_i) +           	(1 - \frac{\hat{\sigma}^2_{u'm}}{\hat{\sigma}^2_{u'm} + \hat{\sigma}^2_{\varepsilon'm}})
		\overline{g_{\lambda'_m}(Y_{mi\bullet}^*)},
	\end{align*}
	using  the individual mean $\overline {g_{\lambda'_m}(Y_{mi\bullet}^*)} = 1/ T_i \sum_{t = 1}^{T_i}g_{\lambda'_m}(Y_{mit}^*)$ and the estimated model parameters of the initial model from step (1).
	
	\item \textbf{Imputed individual mean:} From error model~(\ref{EM_MI}) it follows
	\[
	Y_{mit}^{*} = g_{\lambda'_m}^{-1}(\beta'_{0m} + \X'_i \bbeta'_m + u'_{mi} + \varepsilon'_{mit}).
	\]
	In this equation, we replace $\beta'_{0m} + \X'_i \bbeta'_m + u'_{mi}$ by prediction $\widehat{g_{\lambda'_m}(Y_{mi})}$ and $ \varepsilon'_{mit}$ by pseudo random variable $Z_{mi}^{(l)}$, which follows the normal distribution $\mathcal{N}(0,\hat{\sigma}_{\varepsilon' m}^2)$,  to obtain the imputed individual reported exposure for the $l$-th data set as:
	\begin{equation}\label{impute_MI}
	Y_{mi}^{\MI(l)} = g_{\lambda'_m}^{-1}(\widehat{g_{\lambda'_m}(Y_{mi})} + Z_{mi}^{(l)}).
	\end{equation}
	\item \textbf{Classification and imputed data sets:} Using the classification function $\hat{K}^{\MI}$ of the initial model from step (1) and the imputed individual exposure $Y_{mi}^{\MI(l)}$ the individuals are classified according to
	\begin{equation}\label{classi_MI}
	\hat{K}^{\MI}((Y_{1i}^{\MI(l)},..., Y_{Mi}^{\MI(l)})) = \hat{K}^{\MI(l)}_i.
	\end{equation}
	The resulting $\hat{K}^{\MI(l)}_i$, $l=1,..., L$, are saved in $L$ imputed data sets.
	
	\item \textbf{Health model and mean parameter estimates:} In the last step, the health models
	\[
	H_i = \alpha_0^{(l)} + \K_i^{\MI(l)} \balpha_K^{(l)} + \X_i \balpha_X^{(l)} + e_i^{(l)}
	\]
	are fitted to the imputed data sets. The mean of the resulting parameter estimates 
	\[
	\frac{1}{L}\sum_{l=1}^{L}\hat{\balpha}_K^{(l)} = \bar{\hat{\balpha}}_K^{(\bullet)} = \balpha_K^{\MI}
	\]
	is the estimate of the parameter vector ${\balpha}_K$ of health model~(\ref{LinHM}) (analogously for parameters
	$\alpha_0$  and $\balpha_X$  and for parameters $\gamma_0$, $\bgamma_X$ and $\bgamma_K$ of the logistic health model~(\ref{LogHM})).
\end{enumerate}

The idea of this approach, was already successfully applied in \citet{Keogh12}. Furthermore, MI for error correction using repeated measures was also already introduced \citep{Keogh14}. MI-3-SA is based on the assumption that drawing from the distribution of $K(\Y_i) | \Y^*_i, H_i, \X_i$ can be achieved by the imputation and classification models (\ref{impute_MI}) and (\ref{classi_MI}) (which is derived from the error model (\ref{EM_MI})). It should be noted that the parameters of the models are estimated from the data and that the resulting estimates are plug-in estimates, as it is often the case for MI in practice., i.e. the approach does not follow a strict Bayesian paradigm.

\section{Example: Association between chrononutrition and insulin level in European children and adolescents}\label{s:app}
Chrononutrition is an emerging field in nutritional epidemiology, which focuses on the timing of meals and the inner clock. In this example, we estimated  the association between different patterns of chrononutrition and the long-term insulin marker $\textrm{HbA}_{1c}$ based on data from European children and adolescents 6 to 15 years of the I.Family study \citep{Ahrens17}. The study is a follow-up of the IDEFICS cohort \citep{Ahrens11}, in which children from eight European countries (Belgium, Cyprus, Estonia, Germany, Hungary, Italy, Spain and Sweden) were examined to investigate the causes of lifestyle-induced health effects. SACANA \citep{Hebestreit19}, a web-based 24HDR, was used to assess the dietary behaviour of the study participants. Blood samples were drawn to measure $\textrm{HbA}_{1c}$. In addition, the body mass index (BMI) was measured and the highest parental educational level according to the International Standard Classification of Education (ISCED) was assessed. All institutional and governmental regulations concerning the ethical use of human volunteers were followed. Each survey centre obtained ethical approval from the local responsible authority in accordance with the ethical standards of the 1964 Declaration of Helsinki and its later amendments.

Since the analysis was conducted mainly for illustrative purposes, we only included in total $I=1.498$ female children and adolescents with complete information on 24HDR, $\textrm{HbA}_{1c}$, age, BMI and ISCED. Age-specific z-scores were calculated for $\textrm{HbA}_{1c}$ and BMI since both are strongly age dependent. The \naiveleer method, RC-3-SA, SIMEX-3-SA and MI-3-SA were applied to estimate the association between the patterns of chrononutrition and $\textrm{HbA}_{1c}$. The patterns of chrononutrition included the following $M=5$ dietary components derived from the 24HDR data: energy intake from 5:00 to 11:00 in kcal ($m =1$), energy intake from 17:00 to 24:00 in kcal ($m=2$), time from first to last energy intake in hours ($m = 3$), time between last energy intake and bedtime ($m=4$) and the number of meals ($m=5$). In a second analysis, we additionally included further 4 dietary components (i.e. $M=9$) to illustrate that the number of components can be increased if needed:  total energy intake in kcal ($m=6$), intake of unhealthy food in kcal according to \citet{Hebestreit17} based on fat, sugar and fibre content ($m=7$), carbohydrate intake from 5:00 to 11:00 in gram ($m=8$) and carbohydrate intake from 17:00 to 24:00 in gram ($m=9$).  

The corresponding error-prone components $Y_{mit}^*$  for each individual ($i=1,...,1498$) on multiple days ($t=1,...,T_i$) were assumed to follow the error model~(\ref{EM}) or in case of MI-3-SA the error model~(\ref{EM_MI}), where $\X_i$ consists of the covariates age, BMI z-score and ISCED and $\X_i'$ additionally contains the health outcome of the respective health models as a covariate.

Both health models~(\ref{LinHM}) and (\ref{LogHM}) were assumed to estimate the association between the dietary patterns and the continuous $\textrm{HbA}_{1c}$ z-score, $H_i$, as well as the dichotomised version of $H_i$ using the 90\%-quantile as cut-off, respectively. In both health models, the covariate vector $\X_i$ contains the same covariates as in model~(\ref{EM}). In this illustrative example, for sake of simplicity, we considered the numbers of cluster to be $C=3$. All four correction methods were applied using the $k$-means algorithm and GMM to estimate the classification function $K$. For GMM, a model with identical $M\times M$-covariance matrices $\Sigma_c = \diag(\sigma,...,\sigma)$, $c = 1,2,3$, was assumed.  For the \naiveleer and the RC-based approach no further specifications were necessary. For MI-3-SA and SIMEX-3-SA, $L = 300$ data sets were generated. Furthermore, for SIMEX-3-SA a quadratic function was applied as extrapolation function. 

In epidemiological studies, besides the interpretability of cluster solutions, the total within sums of squares, silhouette coefficients \citep{Rousseeuw15}, reproducibility of cluster solutions \citep{Lo11} and, for the GMM, likelihood-based goodness-of-fit criteria can be used  to decide on the appropriate number of clusters. Graphical tools can be used to facilitate the interpretation and presentation of cluster solutions, e.g. radar charts as  depicted in Figure SM2 (see Supplementary Material).

The results, including the fitted error models, the estimated cluster means and the resulting estimates for $\balpha_K$ and $\bgamma_K$ can be found in Tables SM3 - SM7 (see Supplementary Material). As an example, the results for $M=5$ dietary components based on the GMM cluster solution are briefly described here. The resulting cluster means of RC-3-SA and MI-3-SA were basically the same. The cluster means of SIMEX-3-SA and the \naiveleer approach were the same by definition and similar to the means of the other correction methods, which allows a comparison of  parameter estimates between all methods. The resulting corrected estimates were (first cluster as reference):  $(-0.03, -0.06)^\T$, $(-0.01, 0.00)^\T$, $(-0.05, -0.10)^\T$ and $(0.02, 0.04)^\T$ for $\balpha_K$ and $(0.17, 0.10)^\T$, $(0.13, 0.18)^\T$, $(0.36, 0.22)^\T$ and $(0.16, 0.18)^\T$ for $\bgamma_K$ using the \naiveleer approach, RC-3-SA, SIMEX-3-SA and MI-3-SA, respectively. That means, RC-3-SA and MI-3-SA led to similar estimates. The \naiveleer approach led to different results, which was as expected because it ignores the measurement error structure. The deviation of SIMEX-3-SA to RC-3-SA and MI-3-SA has two reasons. First, SIMEX-3-SA assumed slightly different underlying dietary clusters. Second, the complex relationship between the level of additional measurement error and resulting effect estimates cannot be captured by the extrapolation function (see Section \ref{ss:results} for more details).

\section{Simulation study}\label{s:simulation}
For comparison of the correction methods RC-3-SA, SIMEX-3-SA and MI-3-SA a simulation study was conducted using the example from Section \ref{s:app} as a template. The same error and health models were assumed and the same cluster methods were used.

\subsection{Design}
We simulated $S = 1000$ data sets with the information of $I=1500$ individuals. The following distribution was assumed for the number of reported measurements per individual, $T_i =1$: 35\%, $T_i = 2$: 20\%, $T_i= 3$: 40\% and $T_i=4$: 5\%, roughly representing the distribution in our study. This resulted in $3225$ observations in each data set. 
First, three covariates, age, BMI z-score and ISCED, were simulated using the estimated multivariate normal distribution from Appendix \ref{appC}. The third covariate was subsequently dichotomised. The error term $\varepsilon_{mit}$ was generated according to the corresponding estimated error models. The term for the interindividual variation $u_{mi}$ was generated in two ways, following the  estimated multivariate normal distribution and alternatively assuming $\Corr(u_{ki}, u_{li}) = 0$ for $k\neq l$ to investigate the effect of uncorrelated components. The covariates and the simulated values for $u_{mi}$ and $\varepsilon_{mit}$ were used in error model~(\ref{EM}) to simulate the daily reported intake on the Box-Cox transformed scale $g_{\lambda_m}(Y_{mit}^*)$, $m=1,...,M, i=1,..,1500, t=1,...,T_i$ for $M=5$ or $M =9$ components (see Table SM3 and SM8 for parameters used). Applying $g_{\lambda_m}^{-1}$ yielded the intake on the original scale $Y_{mit}^*$. In addition to the simulated up to four daily measurements, for each individual the true usual intake was calculated by the mean of $1000$ simulated daily measurements (the same simulation approach was used in \citet{Kipnis09}). Accordingly, 7 and 28 daily measurements per individual were simulated, which may be considered as gold standard in this context (similar to \citet{Carroll06}).

Simulating reported intake on the Box-Cox transformed scale based on the error model (\ref{EM}) can lead to $g_{\lambda_m}(y_{mit}^{*}) < -1/\lambda$ for some simulated values $g_{\lambda_m}(y_{mit}^{*})$. Since $g_{\lambda_m}^{-1}(\nu)$ is only defined for $\nu \geq-1/\lambda$, applying $g_{\lambda_m}^{-1}$ on these values leads to missing values for reported intake on the original scale. In these rare cases, the missing values are imputed with random numbers of the uniform distribution on $[y_{m, \min}^*, y_{m, \textrm{\footnotesize 5\%}}^*]$, where $y_{m, \min}^*$ and $y_{m, \textrm{\footnotesize 5\%}}^*$ denote the minimum and the 5\%-quantile of all successfully calculated values of $g_{\lambda_m}^{-1}(g_{\lambda_m}(y_{mit}^{*}))$.

Subsequently, the continuous health outcome was simulated using a linear regression models with $\X_i$ and the simulated true intake as covariates (variant A) and additionally considering the reported intake (variant B) (similar to health models from Appendix \ref{appC} but replacing the estimated intake with the true intake). Table SM9 summarises the parameters chosen for the simulations study. In order not to prefer a cluster solution and method in advance and to ensure comparability between the correction methods, it was not possible to specify a favoured cluster solution, which is why the continuous intakes had to be used for simulating the health outcome (see section \ref{ss:measures} for details). The resulting health outcomes are denoted by $H_i^{\A}$ and $H_i^{\B}$, respectively. As in our example, dichotomised health outcomes $H_i^{\Acat}$ and $H_i^{\Bcat}$ are derived using the 90th percentiles of $H_i^{\A}$ and $H_i^{\B}$ as cut-offs. The described procedure was repeated $1000$ times to create all simulation data sets. 

All correction methods described in Section \ref{s:correction} were applied to the simulated data. In addition, the \naiveleer method was applied to the average of the simulated 7 and 28 measurements, which is considered as the gold standard approach. For the correction methods the same specifications as in Section \ref{s:app} were used. For the cluster solutions of the $k$-means algorithm and the GMM we considered $C = 3, 4, 5$ clusters, which are numbers of clusters often used in dietary patterns analysis. In summary, a total of 96 scenarios were investigated. The scenarios differed by:
\begin{itemize}
	\item number of cluster ($C=3,4,5$), 
	\item cluster method ($k$-means or GMM), 
	\item number of exposures ($M=5$ or $M=9$),
	\item correlation between interindividual random effect terms (no correlation or as in the data example, see Table SM8), 
	\item simulation model for the outcome (see Appendix \ref{appC}) and
	\item used health model (linear or logistic).  
\end{itemize}

Additional to these complex scenarios based on the example from the previous section, we conducted a simulation study in a simple setting ($M=1$ and $C=2$ fix clusters in a linear health model). This gives us the opportunity to compare the methods visually to potentially gain further insights into the performance of the different methods (see Appendix \ref{appD} for details).

\subsubsection{Continuous versus categorised exposures}\label{sss:contvscat}
In order to assess the influence of categorising the intake variables, an additional analysis was conducted, where three linear health models with continuous exposures without categorising them were fitted to the simulated data: 
\begin{enumerate}
	\item using the estimated usual intake $\hat{Y}_{mi}^{\NCI}$  (see health model~(\ref{HM_HA}) in Appendix \ref{appC}), 
	\item same as in 1. but replacing $\hat{Y}_{mi}^{\NCI}$ by the simulated true exposures and 
	\item same as in 1. but replacing $\hat{Y}_{mi}^{\NCI}$ by the \naiveleer exposures.
\end{enumerate}
The linear health models were fitted using $H_i^{\A}$ and $H_i^{\B}$ as outcome. Analogously, the logistic health models with the outcomes $H_i^{\Acat}$ and $H_i^{\Bcat}$ were fitted. Thus, for each outcome error-corrected (1), true (2) and \naiveleer (3) parameter estimates of models assuming continuous exposures instead of clusters were calculated.

\subsection{Measures for assessing the correction methods}\label{ss:measures}
The resulting parameter estimates for $\balpha_K$ and $\bgamma_{K}$ depend on the classification function $K$, which is not the same for every simulated data set. Therefore, a direct comparison of estimates based on different simulated data sets is not possible. To summarise the resulting bias for different data sets in a meaningful way, we propose the following measures.
In health model~(\ref{LinHM}), the elements of $\balpha_{K}$ can be interpreted as difference of $H_i$ between the reference cluster and all other clusters. Without loss of generality, we assume the first cluster to be the reference cluster. Since this is arbitrary and all other cluster comparisons are also of interest, not only the elements of $\balpha_{K}$ must be considered, but also the other cluster comparisons. For example, if $C=5$, the parameter vector is $\balpha_{K} = (\alpha_{K12}, \alpha_{K13}, \alpha_{K14}, \alpha_{K15})^\T$ and the $10$ cluster comparisons of interest are $\alpha_{Kcc'}$ with $c < c', c' = 1,...,5$.

To ensure that the derived clusters correspond to the true clusters and a meaningful comparison is possible, the same classification function must be used. Therefore, we assumed $\hat{K} = K$. This implies the assumption that $K$ can be derived adequately using the data or that in any case $\hat{K}$ is at least a meaningful classification function, which can be considered as the true classification function. 

For each repeated data set, $s=1,...,S$, we define the absolute bias of each parameter estimate $\hat{\alpha}_{Kcc'}^{(s)}$, $c, c' = 1,...,C, c < c'$, as $\Delta_{cc'}^{(s)} = |\alpha_{Kcc'}^{(s)} - \hat{\alpha}_{Kcc'}^{(s)}|$. The corresponding mean absolute bias of all cluster comparisons may be calculated as 
\begin{align*}
	\bar{\Delta}^{(s)} = & \frac{1}{\#\{ (c, c') |\: c, c' = 1,...,C,\: c < c'\}} \sum\limits_{c, c' = 1,...,C,\: c < c'} \Delta_{c c'}^{(s)}\\
	= & \frac{2}{C(C-1)}\sum\limits_{c, c' = 1,...,C,\: c < c'}\Delta_{c c'}^{(s)}.
\end{align*}
Analogously, the maximal absolute bias  is calculated as 
\[
\Delta_{\max}^{(s)} = \max\{\Delta_{cc'}^{(s)} | c, c' = 1,..., C, c < c'\}.
\]
And similarly, using the relative absolute bias ${\Delta_{\rel}^{(s)}}_{cc'}= |\alpha_{Kcc'}^{(s)} - \hat{\alpha}_{Kcc'}^{(s)}| / |\alpha_{Kcc'}^{(s)}|$, the mean relative absolute bias is calculated as
\[
\bar{\Delta}_{\rel}^{(s)} = \frac{2}{C(C-1)} \sum\limits_{c, c' = 1,...,C,\: c < c'}{\Delta_{\rel}^{(s)}}_{cc'}.
\]
Taking all the simulated data sets together, we get:
\begin{align*}
	\bar{\Delta} & = \frac{1}{S} \sum\limits_{s = 1}^{S}\bar{\Delta}^{(s)}, \\
	\bar{\Delta}_{\max} & =  \frac{1}{S} \sum\limits_{s = 1}^{S}\Delta_{\max}^{(s)}, \textrm{ and} \\
	\med(\bar{\Delta}_{\rel}) & =  \median\{\bar{\Delta}_{\rel}^{(s)} | s =1,..., S \}.
\end{align*}
Analogously, the measures can be defined for the elements of $\bgamma_{K}$ from the logistic health model~(\ref{LogHM}).
Furthermore, for the evaluation of the resulting cluster solutions the misclassification rate (MR) and the adjusted Rand index (aRI) \citep{Hubert85} were calculated.

\subsection{Results}\label{ss:results}

\begin{table}	
	\centering
	\caption{Simulation results of correction methods (CM) and gold standard methods based on 7 and 28 days (GS7 and GS28) for scenarios with linear health model and outcome $H_i^{\A}$ (simulated using the true intake) and cluster method with $C$ clusters regarding empirical adjusted Rand index, mean absolute bias, maximal absolute bias and median mean relative absolute bias}
	\label{t:HA}
	\begin{tabular}{llrrrr}
		\hline
		$C$& CM & $\bar{\aRI}$ & $\bar{\Delta}$   &  $\bar{\Delta}_{\max}$  & $\med(\bar{\Delta}_{\rel})$ \\  
		\hline
		\multicolumn{6}{c}{Cluster method: $k$-means} \\
		\hline
		3 &\naive&0.16&\textbf{0.07}&\textbf{0.11}&0.85\\
		&RC&0.21&0.08&\textbf{0.11}&0.77\\
		&SIMEX-Q&0.16&0.10 &0.13&1.11\\
		&MI&0.21&0.08&\textbf{0.11}&\textbf{0.66}\\
		&GS7&0.35&0.06&0.08&0.57\\
		&GS28&0.61&0.04&0.05&0.40\\
		\hline
		4 &\naive&0.13&\textbf{0.09}&0.16&1.04\\
		&RC&0.17&\textbf{0.09}&0.14&0.88\\
		&SIMEX-Q&0.13&0.12&0.18&1.27\\
		&MI&0.17&\textbf{0.09}&\textbf{0.13}&\textbf{0.73}\\
		&GS7&0.31&0.07&0.10&0.69\\
		&GS28&0.57&0.05&0.07&0.46\\
		\hline
		5&\naive&0.11&0.10 &0.20 &1.20\\
		&RC&0.14&0.10 &0.17&1.01\\
		&SIMEX-Q&0.11&0.13&0.23&1.51\\
		&MI&0.14&\textbf{0.09}&\textbf{0.16}&\textbf{0.78}\\
		&GS7&0.28&0.07&0.13&0.76\\
		&GS28&0.54&0.05&0.09&0.62\\
		\hline 
		\multicolumn{6}{c}{Cluster method: GMM}\\
		\hline
		3&\naive&0.20 &0.20 &0.30 &1.16\\
		&RC&0.25&0.09&0.13&0.87\\
		&SIMEX-Q&0.20 &0.27&0.39&1.64\\
		&MI&0.25&\textbf{0.08}&\textbf{0.11}&\textbf{0.68}\\
		&GS7&0.44&0.12&0.16&0.75\\
		&GS28&0.68&0.07&0.10 &0.54\\
		\hline
		4&\naive&0.19&0.22&0.40 &1.26\\
		&RC&0.21&\textbf{0.10}&0.17&0.97\\
		&SIMEX-Q&0.19&0.30 &0.48&1.63\\
		&MI&0.21&\textbf{0.10}&\textbf{0.16}&\textbf{0.73}\\
		&GS7&0.41&0.12&0.19&0.81\\
		&GS28&0.65&0.08&0.13&0.63\\
		\hline
		5&\naive&0.20 &0.20 &0.42&1.53\\
		&RC&0.19&0.12&0.22&1.03\\
		&SIMEX-Q&0.20 &0.29&0.50 &2.02\\
		&MI&0.19&\textbf{0.11}&\textbf{0.20}&\textbf{0.79}\\
		&GS7&0.38&0.11&0.20 &0.94\\
		&GS28&0.61&0.07&0.13&0.62\\
		\hline
	\end{tabular}
\end{table}

\begin{table}
	\centering
	\caption{Simulation results of correction methods (CM) and gold standard methods based on 7 and 28 days  (GS7 and GS28) for scenarios with logistic health model and outcome $H_i^{\Acat}$ (simulated using the true intake) and cluster method with $C$ clusters regarding empirical adjusted Rand index, mean absolute bias, maximal absolute bias and median mean relative absolute bias}
	\label{t:HAcat}
	\begin{center}
		\begin{tabular}{llrrrr}
			\hline
			$C$& CM & $\bar{\aRI}$ & $\bar{\Delta}$   &  $\bar{\Delta}_{\max}$  & $\med(\bar{\Delta}_{\rel})$ \\  
			\hline
			\multicolumn{6}{c}{Cluster method: $k$-means} \\
			\hline
			3 &\naive&0.16&0.23&0.34&1.12\\
			&RC&0.21&0.24&0.33&1.09\\
			&SIMEX-Q&0.16&0.34&0.46&1.51\\
			&MI&0.21&\textbf{0.21}&\textbf{0.28}&\textbf{0.83}\\
			&GS7&0.35&0.18&0.25&0.81\\
			&GS28&0.61&0.13&0.18&0.62\\
			\hline
			4 &\naive&0.13&0.27&0.48&1.37\\
			&RC&0.17&0.27&0.42&1.21\\
			&SIMEX-Q&0.13&0.40 &0.63&1.84\\
			&MI&0.16&\textbf{0.23}&\textbf{0.37}&\textbf{0.89}\\
			&GS7&0.31&0.22&0.34&1.02\\
			&GS28&0.57&0.16&0.25&0.7\\
			\hline
			5&\naive&0.11&0.30 &0.62&1.64\\
			&RC&0.14&0.30 &0.52&1.32\\
			&SIMEX-Q&0.11&0.46&0.79&2.09\\
			&MI&0.14&\textbf{0.26}&\textbf{0.46}&\textbf{0.93}\\
			&GS7&0.28&0.24&0.42&1.11\\
			&GS28&0.54&0.18&0.31&0.87\\
			\hline 
			\multicolumn{6}{c}{Cluster method: GMM}\\
			\hline
			3&\naive&0.20 &2.12&3.19&1.37\\
			&RC&0.25&0.35&0.47&1.12\\
			&SIMEX-Q&0.20 &2.91&4.42&2.13\\
			&MI&0.25&\textbf{0.25}&\textbf{0.35}&\textbf{0.85}\\
			&GS7&0.44&0.89&1.31&1.02\\
			&GS28&0.68&0.46&0.70 &0.83\\
			\hline
			4&\naive&0.19&1.91&3.68&1.72\\
			&RC&0.21&0.39&0.63&1.24\\
			&SIMEX-Q&0.19&2.74&4.96&2.31\\
			&MI&0.21&\textbf{0.35}&\textbf{0.57}&\textbf{0.90}\\
			&GS7&0.41&0.85&1.44&1.14\\
			&GS28&0.65&0.52&0.88&0.93\\
			\hline
			5&\naive&0.20 &1.47&3.38&1.93\\
			&RC&0.19&0.49&0.95&1.32\\
			&SIMEX-Q&0.20 &2.14&4.49&2.49\\
			&MI&0.19&\textbf{0.41}&\textbf{0.80}&\textbf{0.95}\\
			&GS7&0.38&0.56&1.03&1.2\\
			&GS28&0.61&0.32&0.63&0.92\\
			\hline
		\end{tabular}
	\end{center}
\end{table}

\begin{table}
	\centering
	\caption{Simulation results of correction methods (CM) and gold standard methods based on 7 and 28 days  (GS7 and GS28) for scenarios with linear health model and outcome $H_i^{\B}$ (simulated using true and reported intake) and cluster method with $C$ clusters regarding empirical adjusted Rand index, mean absolute bias, maximal absolute bias and median mean relative absolute bias}
	\label{t:HB}
	\begin{tabular}{llrrrr}
		\hline
		$C$& CM & $\bar{\aRI}$ & $\bar{\Delta}$   &  $\bar{\Delta}_{\max}$  & $\med(\bar{\Delta}_{\rel})$ \\  
		\hline
		\multicolumn{6}{c}{Cluster method: $k$-means} \\
		\hline
		3 &\naive&0.16&\textbf{0.23}&0.35&0.91\\
		&RC&0.21&0.29&0.41&\textbf{0.83}\\
		&SIMEX-Q&0.16&\textbf{0.23}&\textbf{0.32}&0.92\\
		&MI&0.21&0.35&0.48&0.94\\
		&GS7&0.35&0.08&0.11&0.28\\
		&GS28&0.61&0.04&0.06&0.16\\
		\hline
		4 &\naive&0.13&\textbf{0.23}&0.41&0.99\\
		&RC&0.17&0.29&0.45&\textbf{0.87}\\
		&SIMEX-Q&0.13&\textbf{0.23}&\textbf{0.36}&1.03\\
		&MI&0.17&0.33&0.51&0.94\\
		&GS7&0.31&0.09&0.14&0.36\\
		&GS28&0.57&0.05&0.08&0.21\\
		\hline
		5&\naive&0.11&\textbf{0.25}&0.51&1.05\\
		&RC&0.14&0.30 &0.50 &\textbf{0.91}\\
		&SIMEX-Q&0.11&0.26&\textbf{0.45}&1.12\\
		&MI&0.14&0.34&0.56&0.95\\
		&GS7&0.28&0.09&0.16&0.39\\
		&GS28&0.54&0.06&0.10 &0.27\\
		\hline 
		\multicolumn{6}{c}{Cluster method: GMM}\\
		\hline
		3&\naive&0.20 &0.39&0.59&1.07\\
		&RC&0.25&\textbf{0.31}&\textbf{0.43}&\textbf{0.83}\\
		&SIMEX-Q&0.20 &0.44&0.61&1.28\\
		&MI&0.25&0.36&0.49&0.94\\
		&GS7&0.44&0.14&0.20 &0.38\\
		&GS28&0.68&0.08&0.11&0.21\\
		\hline
		4&\naive&0.19&0.42&0.77&1.16\\
		&RC&0.21&\textbf{0.34}&\textbf{0.52}&\textbf{0.84}\\
		&SIMEX-Q&0.19&0.47&0.75&1.42\\
		&MI&0.21&0.40 &0.61&0.94\\
		&GS7&0.41&0.15&0.24&0.47\\
		&GS28&0.65&0.09&0.15&0.28\\
		\hline
		5&\naive&0.20 &0.44&0.94&1.19\\
		&RC&0.19&\textbf{0.34}&\textbf{0.57}&\textbf{0.89}\\
		&SIMEX-Q&0.20&0.48&0.85&1.39\\
		&MI&0.19&0.40&0.66&0.94\\
		&GS7&0.38&0.13&0.23&0.47\\
		&GS28&0.61&0.08&0.14&0.31\\
		\hline
	\end{tabular}
\end{table}

\begin{table}
	\centering
	\caption{Simulation results of  correction methods (CM) and gold standard methods based on 7 and 28 days  (GS7 and GS28) for scenarios with logistic health model and outcome $H_i^{\Bcat}$ (simulated using true and reported intake) and  cluster method with $C$ clusters regarding empirical adjusted Rand index, mean absolute bias, maximal absolute bias and median mean relative absolute bias}
	\label{t:HBcat}
	\begin{tabular}{llrrrr}
		\hline
		$C$&CM &  $\bar{\aRI}$ & $\bar{\Delta}$   &  $\bar{\Delta}^{\max}$  & $\med(\bar{\Delta}_{\rel})$ \\  
		\hline
		\multicolumn{6}{c}{Cluster method: $k$-means} \\
		\hline
		3 &\naive&0.16&\textbf{0.47}&\textbf{0.71}&1.03\\
		&RC&0.21&0.59&0.81&0.88\\
		&SIMEX-Q&0.16&0.52&0.72&1.15\\
		&MI&0.21&0.60 &0.83&\textbf{0.83}\\
		&GS7&0.35&0.21&0.28&0.41\\
		&GS28&0.61&0.14&0.19&0.26\\
		\hline
		4 &\naive&0.13&\textbf{0.47}&\textbf{0.87}&1.21\\
		&RC&0.17&0.60 &0.94&0.93\\
		&SIMEX-Q&0.13&0.55&\textbf{0.87}&1.37\\
		&MI&0.17&0.60 &0.94&\textbf{0.87}\\
		&GS7&0.31&0.24&0.38&0.56\\
		&GS28&0.57&0.16&0.26&0.34\\
		\hline
		5&\naive&0.11&\textbf{0.53}&1.08&1.33\\
		&RC&0.14&0.62&1.05&1.02\\
		&SIMEX-Q&0.11&0.62&1.07&1.61\\
		&MI&0.14&0.61&\textbf{1.02}&\textbf{0.88}\\
		&GS7&0.28&0.27&0.47&0.61\\
		&GS28&0.54&0.19&0.33&0.47\\
		\hline 
		\multicolumn{6}{c}{Cluster method: GMM}\\
		\hline
		3 &\naive&0.20 &2.77&4.16&1.15\\
		&RC&0.25&0.74&1.02&0.90\\
		&SIMEX-Q&0.20&3.50&5.29&1.55\\
		&MI&0.25&\textbf{0.70}&\textbf{1.00}&\textbf{0.85}\\
		&GS7&0.44&1.37&2.07&0.65\\
		&GS28&0.68&0.85&1.33&0.41\\
		\hline
		4 &\naive&0.19&2.61&5.01&1.50\\
		&RC&0.21&0.87&1.42&0.91\\
		&SIMEX-Q&0.19&3.39&6.21&1.88\\
		&MI&0.22&\textbf{0.82}&\textbf{1.37}&\textbf{0.86}\\
		&GS7&0.41&1.14&2.09&0.82\\
		&GS28&0.65&0.80 &1.53&0.56\\
		\hline
		5&\naive&0.20 &2.08&4.77&1.51\\
		&RC&0.19&0.91&1.75&0.97\\
		&SIMEX-Q&0.20 &2.55&5.61&1.84\\
		&MI&0.19&\textbf{0.89}&\textbf{1.73}&\textbf{0.89}\\
		&GS7&0.38&0.64&1.31&0.81\\
		&GS28&0.61&0.49&1.00&0.58\\
		\hline
	\end{tabular}
\end{table} 

\begin{table}
	\centering
	\caption{Comparison of the approaches considering the continuous exposures only (\naiveleer and regression calibration) with MI-3-SA considering categorised exposures regarding the median mean relative absolute bias for the four different health outcomes ($H_i$) in all scenarios}
	\label{t:countVScat}
	\begin{tabular}{lllcccc}
		\hline
		$M$ & $\Corr(u_{ki}, u_{li})$ for $k\neq l$ &$H_i$ & \multicolumn{2}{c}{Cont. exp.} &  \multicolumn{2}{c}{MI-3-SA ($C=3,4,5$)}  \\  
		\cline{4-5} \cline{6-7}
		& & &\naiveleer &  RC &   $k$-means &GMM \\
		\hline
		5 & $=0$ &$H_i^{\A}$	&0.84	&1.14&  \textbf{0.66}, \textbf{0.73}, \textbf{0.78} &\textbf{0.68}, \textbf{0.73}, \textbf{0.79} \\
		& &$H_i^{\Acat}$ &	 0.93 &	1.64& \textbf{0.83, 0.89, 0.93}	& \textbf{0.85}, \textbf{0.90}, 0.95\\
		& &	$H_i^{\B}$ 		& 0.88	& \textbf{0.73}& 0.94, 0.94, 0.95& 0.94, 0.94, 0.94\\
		& &$H_i^{\Bcat}$ & 0.92 &	0.95&  \textbf{0.83}, \textbf{0.87}, \textbf{0.88} & \textbf{0.85}, \textbf{0.86}, \textbf{0.89} \\
		\hline
		5 & $\neq0$ &$H_i^{\A}$	&	0.92 &	1.24 & \textbf{0.73,         0.80,         0.86}&\textbf{0.86,         0.84,         0.90} \\
		& &$H_i^{\Acat}$ &	 0.95&	1.68 & \textbf{0.85},         0.97,        0.98& \textbf{0.97},         1.03,         1.13 \\
		& &$H_i^{\B}$ 		& 1.00	&1.04& \textbf{0.99,         0.98,        1.00 }& \textbf{0.96,        0.97,        0.99}\\
		& &$H_i^{\Bcat}$ & 1.03&	1.31& 1.03,         1.07,        1.12&\textbf{0.91},      \textbf{1.01},         1.13\\
		\hline
		9 & $=0$ &$H_i^{\A}$	&\textbf{0.64}&	0.83		&0.71,         0.79,        0.82 &0.74         0.82,         0.85 \\
		& &$H_i^{\Acat}$ &	 \textbf{0.66}&	1.01		& 0.85,         0.92,         0.94 	& 0.94         1.02,         1.07\\
		& &$H_i^{\B}$ 		& \textbf{0.85}&	1.06		& 0.97,        0.97,         0.97& 0.97         0.97,         0.97\\
		& &$H_i^{\Bcat}$ & \textbf{0.81	}&0.95		&  0.91,        0.94,         0.97 & 0.91         0.94,         0.99 \\
		\hline
		9 & $\neq0$ &	$H_i^{\A}$	&1.49&	1.93	&   \textbf{0.54,      0.68,         0.81 }&\textbf{0.60,    0.71,         0.80} \\
		& &$H_i^{\Acat}$ &	 1.17&	1.65		& \textbf{0.75,         0.88,        1.02}	&\textbf{0.83,       0.97,        1.41}\\
		& &$H_i^{\B}$ 		& 1.06&	1.78		& \textbf{0.97,         0.98,         1.02}& \textbf{1.00,         0.98,        1.00} \\
		& &$H_i^{\Bcat}$ &\textbf{ 1.00}&	1.66		&\textbf{1.00},        1.11,         1.35  &1.11,         1.19,         1.55\\
		\hline
	\end{tabular}
\end{table} 

\begin{figure}[htb]
	\begin{center}
		\includegraphics[width=14cm]{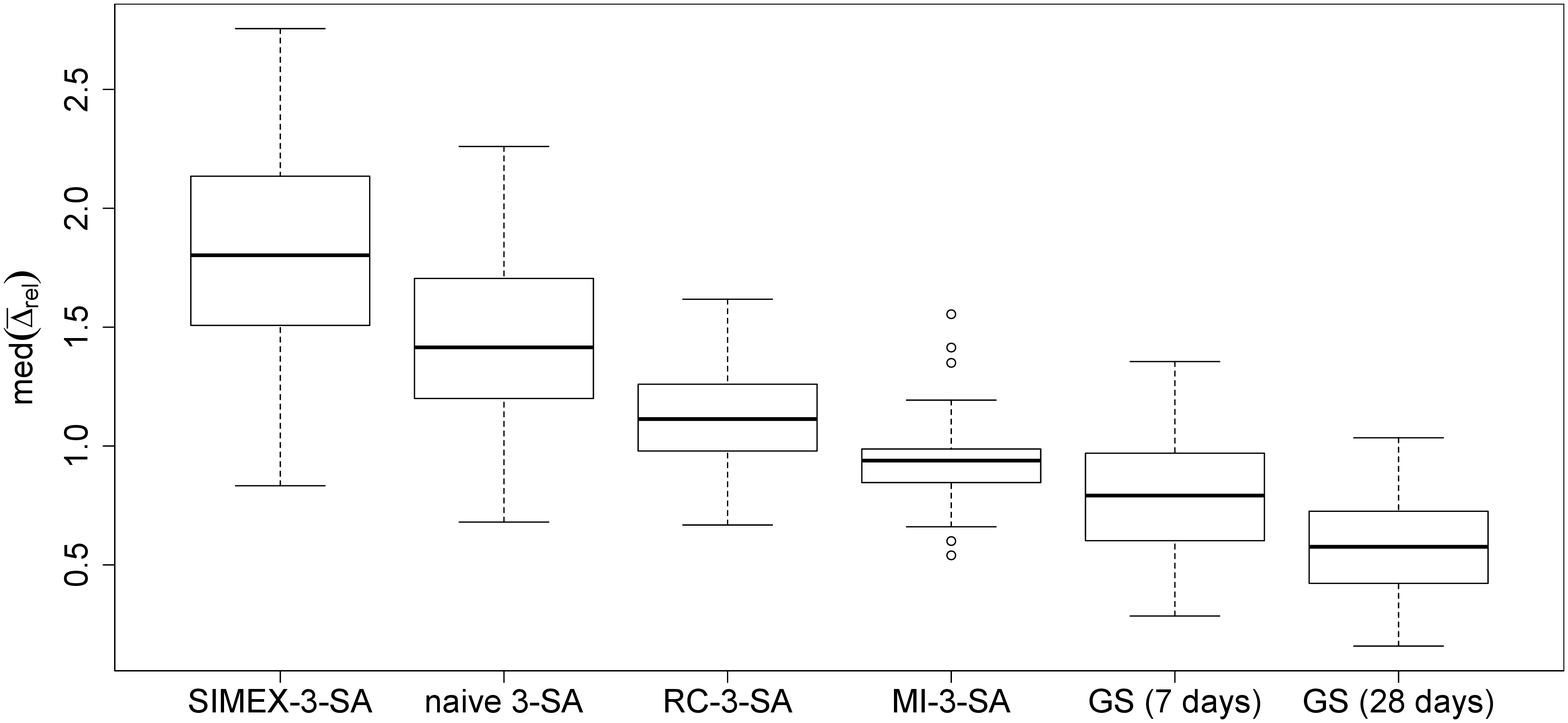}
		\caption{Box plot of the median relative absolute bias of the different correction approaches across all 96 scenarios} \label{Fig1_summ_boxplot}
	\end{center}
\end{figure}

\begin{figure}[htb]
	\begin{center}
		\includegraphics[width=14cm]{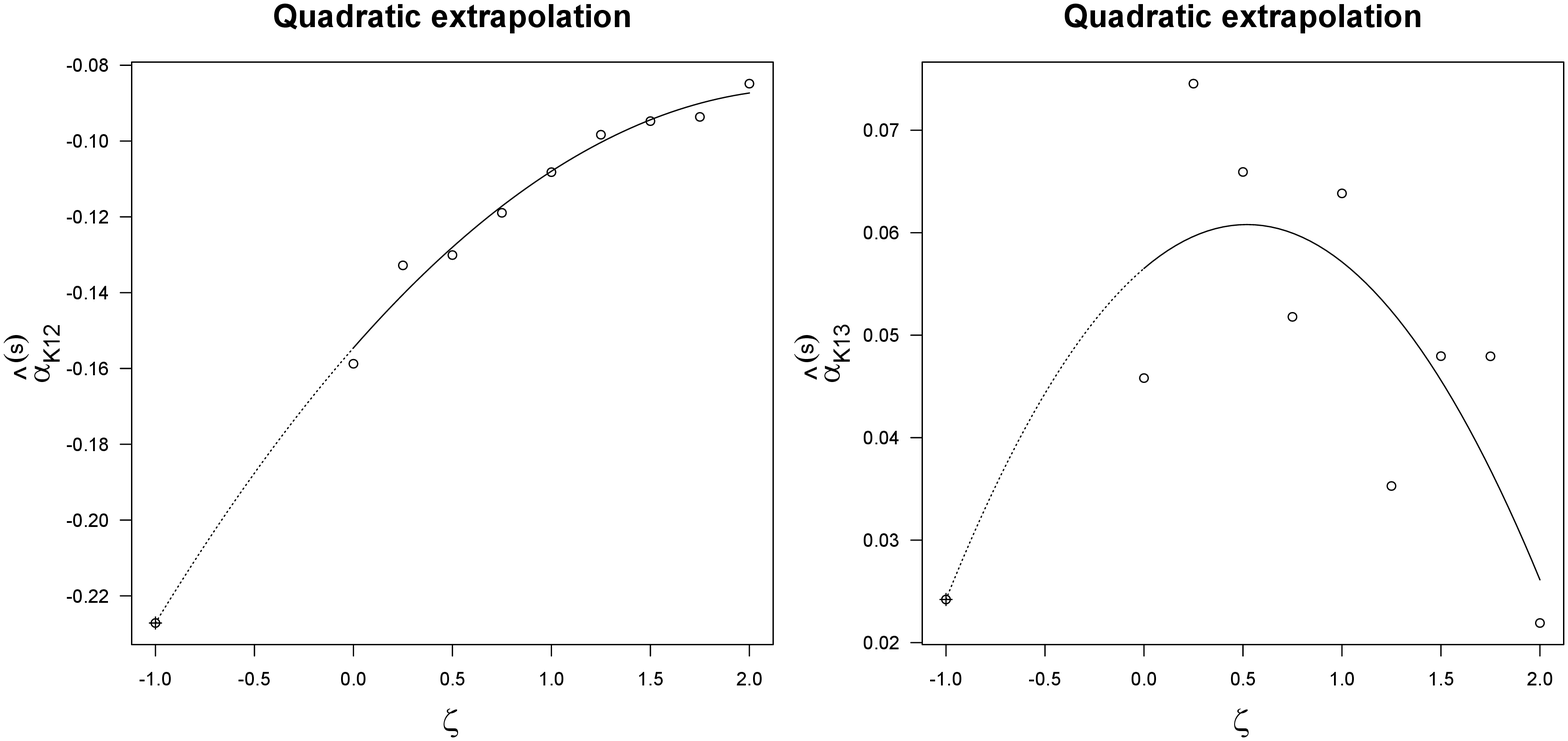}
		\caption{Quadratic extrapolation of estimates $\alpha_{K12}^{(s)}$ and $\alpha_{K13}^{(s)}$ used in SIMEX-3-SA (example from the 11th simulation data set, $s=11$, in scenario with GMM, $C=3$, $M=9$, without correlation between interindividual random effect terms and for health outcome $H_i^{\A}$). Apparently, the extrapolation for $\alpha_{K12}^{(s)}$ seems to be appropriate (left) -- in contrast to the extrapolation for $\alpha_{K13}^{(s)}$ (right).} \label{Fig2_extra_prob}
	\end{center}
\end{figure}

The simulation results are summarised in Tables~\ref{t:HA} - \ref{t:HBcat} for the scenarios with $M=5$ components and $\Corr(u_{ki}, u_{li}) = 0$ for $k\neq l$ for the health outcomes $H^{\A}_i$, $H_i^{\Acat}$, $H_i^{\B}$ and $H_i^{\Bcat}$ as an example. The correction methods performed as follows in these 24 scenarios. The empirical aRI was always lowest for MI-3-SA. Regarding the empirical mean absolute bias, MI-3-SA performed best in 13 scenarios (RC-3-SA: 5, \naive: 8). MI-3-SA performed also best with respect to the empirical maximum absolute bias (MI-3-SA: 15 scenarios, RC-3-SA: 4, SIMEX-3-SA: 4, \naive: 3) and the empirical relative absolute bias (MI-3-SA: 18 scenarios, RC-3-SA: 6, \naive: 0). 

Regarding the empirical relative absolute bias, MI-3-SA performed even better as the gold standard approach based on 7 days in 7 scenarios. This is remarkable, since the gold standard based on 7 days relied on approximately 3 times more repeated measurements. As expected, the gold standard approach based on 28 days performed better than any correction method regarding all measures.

The results in the three scenarios with $M=9$ and $M=5$ with $\Corr(u_{ki}, u_{li}) \neq 0$, $k\neq l$, can be found in Tables SM10 - SM25. In general the results were very similar to the case of $M=5$ with $\Corr(u_{ki}, u_{li}) = 0$ for $k\neq l$. The only differences were that in these scenarios MI-3-SA performed even more often best and the \naiveleer approach less often best. The results for all 96 scenarios are summarised in Figure \ref{Fig1_summ_boxplot}, where the median relative absolute bias is depicted.

To better understand, why the performance of SIMEX-3-SA was so bad, we inspected the SIMEX extrapolation plots visually. A typical example can be found in Figure \ref{Fig2_extra_prob}, where an apparently successful and an unsuccessful extrapolation are contrasted. In the first case effect estimates increase monotonically with higher level of measurement error, the extrapolation seems to be appropriate. In the second case no clear relationship between effect estimate and level of measurement error can be identified, which hinders an appropriate extrapolation. The reason for the missing clear relationship is that the effect estimates additionally depend on the classification function and thus also on the corresponding cluster geometry and the position of individuals in the $M$-dimensional space. That means an increasing measurement error of one single exposure may have a negligible or large effect on the classification of individuals to a specific cluster, which in turn may increase or decrease the effect sizes.

In the simple setting we did not observe such a poor behaviour of the extrapolation function. In contrast, SIMEX-3-SA works even better than the other correction methods (Tables SM1-SM2, see Supplementary Material). Nevertheless, since in this setting a number of simplifications are made, the results cannot be transferred to a realistic scenario (see Appendix \ref{appD} for details).     

For comparison of the bias resulting from one continuous error-prone exposure and the bias resulting from its categorised version, \citet{Gustafson02} proposed the relative bias. This approach was transferred to the setting of $M$ exposures as follows. The resulting estimated coefficients of the \naiveleer and estimated usual intakes as exposures (see subsection \ref{sss:contvscat}) were evaluated using the median of the relative absolute bias introduced in subsection \ref{ss:measures}. That means, first the mean relative absolute bias for the \naiveleer and the error-corrected approach was calculated taking into account the $M$ fitted effect estimates and the corresponding effect estimates of the true exposures for each simulated data set (analogously to $\hat{\alpha}_{Kcc'}$ and $\alpha_{Kcc'}$, $c,c' = 1,...,C, c<c'$). Second, for both approaches the medians across  all scenarios were calculated (i.e. for $M=5$ and $M=9$ exposures, for both assumed correlation structures of the interindividual random effects and of all four outcomes). In most cases, the resulting empirical medians of the mean relative absolute bias using the continuous exposures were higher than the corresponding medians of the MI-3-SA (Table \ref{t:countVScat}). 

\section{Discussion}\label{s:discussion}
We introduced three correction methods for misclassified exposure patterns due to clustering of multiple error-prone exposures. The methods combined cluster algorithms with three well-established correction methods: regression calibration, simulation extrapolation and multiple imputation. In a simulation study, the MI-based approach showed the lowest bias. Furthermore, this approach performed in some scenarios even better than the gold standard approach based on 7 days. 

All methods are based on error models considering the skewness of the intake distributions, covariates, inter- and intraindividual variation. An important further assumption is the unbiasedness of $Y^*_{mit}$ on the original scale which is not always true but commonly used as a working assumption in this field (for an extensive discussion see \citet{Intemann19b}, \citet{Dodd06} and \citet{Carroll06}). To reduce model complexity we assumed the individual random effect terms $u_{mi}$ to be mutually independent. This reduces the computational time for the model fit significantly \citep{Zhang11}, which makes it possible to investigate the proposed methods in a simulation study in the first place. Furthermore, the simplification should have only little influence on the estimates of usual intakes and $\lambda_m$ and $\sigma_{\varepsilon m}$ since the intake data and the same covariates are used in the error models and thus the correlation is partly captured. However, to investigate the influence of mutually correlated $u_{mi}$ on the performance of the correction methods, we generated data with and without correlation in the simulation study. When comparing the median relative absolute bias of MI-3-SA in both cases, the relative difference was 12~\% for $M = 5$ and 5~\% for $M = 9$. Similar numbers were found for RC-3-SA (19 \% for $M = 5$ and 5 \% for $M = 9$). Furthermore, these relative differences were even smaller as for the gold standard approach based on 28 days (38~\% for $M=5$ and 15~\% for $M=9$). This suggests that the simplified assumption does not lead to an unreasonably higher bias.

The simulation results of MI-3-SA in presence of differential and non-differential error agree with the results from Keogh et al. (2012). They showed that their MI-based approach is superior to their RC-based approach. There are two possible reasons, why in our simulation study the MI-based MI-3-SA also led to lower bias than the RC-based RC-3-SA. First, MI-3-SA uses more information since the health outcome is considered in the error model, which is not the case for the RC-3-SA error model. Second, the algorithm itself is superior, since MI-3-SA, unlike RC-3-SA, considers a kind of error propagation from the error in the continuous exposures to the classification of individuals, i.e. higher error variance leads to more uncertainty in the classification variable (step~(3) and (4)). In order to investigate both reasons and to verify, whether the simulation model for $H_i$ prefers MI-3-SA in the simulation study, a sensitivity analysis was conducted. In this additional analysis, MI-3-SA was applied to the same simulation data using error model~(\ref{EM}) without considering $H_i$, which is the RC-3-SA error model. Although the performance was in general slightly worse than that of the original MI-3-SA, the method without considering the health outcome was superior to RC-3-SA, e.g. regarding the relative absolute bias in 77 out of 96 scenarios (see Tables SM10-SM25 for simulation results). Therefore, we conclude that both aforementioned reasons may explain the better performance of MI-3-SA. 

This sensitivity analysis can also be seen as a critical evaluation of the use of the health outcome in the MI error model, which assumes a linear relationship between the health outcome and the transformed exposures. Obviously this choice is arbitrary, especially, when taking into account that in the health model the association between $\K(\Y_i)$ and the health outcome is modelled. However, since error correction is carried out before cluster analysis, we consider the linear relationship as the \naiveleer approach to incorporate additional information and to improve error correction. Furthermore, if really a linear relationship exists, it seems likely that there is also  a relationship between the categorised version of $\Y$ and $H_i$. Nevertheless, if one doubts that considering $H_i$ improves the correction, omitting $H_i$ in the error model is also possible.

In contrast to MI-3-SA and RC-3-SA, the SIMEX-based approach should not be considered since the relationship between the level of additional measurement error $\zeta$ and the resulting effect estimates was too complex to be adequately fitted, or even extrapolated, with a quadratic function. In a sensitivity analysis, we also considered a cubic and a quartic extrapolation function, but the results were even worse (see Tables SM10-SM25 for details).

In conclusion, clustering error-prone exposures led to biased effect estimates, if these clusters were used as a categorical exposure in a health model. The proposed method based on multiple imputation reduced this bias. Therefore, we recommend using this method for such situations.

\begin{appendix}

\section{Differential error in the 3-SA}\label{appA}
We assume the measurement error of the error-prone components is non-differential for the health outcome $H_i$. That means, for $\Y_{it}^{*} = (Y_{1it}^*,..., Y_{Mit}^{*})$ and $\Y_{i} = (Y_{1i},...,Y_{Mi})$ the distributions of $\Y_{it}^{*}|\Y_i, H_i$ and $\Y_{it}^{*}|\Y_i$ are equal. Furthermore, the same should be assumed, if $T_i$ days are considered, i.e. for the random vector $\Y_i^{*} = (Y_{1i1}^*,..., Y_{Mi1}^*, Y_{1i2}^*,..., Y_{Mi2}^*, ..., \newline Y_{1i1T_i}^*,..., Y_{Mi T_i}^*)$ and for the corresponding density functions of $\Y_i^{*}|\Y_i, H_i$ and $\Y_i^{*}|\Y_i$ it is assumed that
\begin{equation*}
	f(\y_i^{*} | \y_i, h_i) = f(\y_i^*|\y_i).
\end{equation*}
To show that the 3-SA approach leads to differential measurement error, we extend the investigation of Buonaccorsi (2010, page 160-163), who showed that categorising a single error-prone exposure leads to differential measurement error. The extension includes two steps. In the first step, we assume $\Y_{it}^{*}$ to include $M$ components measured only once, i.e. $T_i = 1$, and that the $M$-dimensional space can be partitioned into $C$ and $C^*$ disjunct subsets $A_1,...,A_C$ and $\hat{A}_1,...,\hat{A}_{C^*}$ which are used for classification of the individuals, i.e. $K(\y) = c \Leftrightarrow \y \in A_c$ and $\hat{K}(\y) = c^* \Leftrightarrow \y \in A_{c^*}$. It is important to note that it is not necessary that  $C= C^{*}$ or $A_c = \hat{A}_{c^*}$ for $c=c^*$.
Then, the misclassification probability of $\hat{K}$ is given as
\begin{align*}
	P(\hat{K} = c^* | K = c) &=  \frac{P(\y \in A_c, \y^* \in \hat{A}_{c^*} )}{P(\y \in A_c)}\\
	&=   \frac{\int_{\y \in A_c} \int_{\y^* \in \hat{A}_{c^*}} f_{Y, Y^*}(\y, \y^*) d\y^*d\y}{\int_{\y \in A_c} f_Y (\y) d\y}\\
	&= \frac{\int_{\y \in A_c} \int_{\y^* \in \hat{A}_{c^*}} f(\y^* | \y)  f_Y (\y) d\y^* d\y}{\int_{\y \in A_c} f_Y (\y) d\y}\\
	&= \frac{\int_{\y \in A_c} \int_{\y^* \in \hat{A}_{c^*}} f(\y^* | \y) d\y^* f_Y (\y) d\y}{\int_{\y \in A_c} f_Y (\y) d\y}.
\end{align*}
Using the assumption of non-differential error for a given health outcome $H_i= h $ the misclassification probability $P(\hat{K} = c^* | K = c, h)$ is
\begin{equation}
	\frac{\int_{\y \in A_c} \int_{\y^* \in A_{c^*}} f(\y^* | \y, h) d\y^* f_{Y|h} (\y | h) d\y}{\int_{\y \in A_c} f_{Y|h} (\y | h) d\y}
	= \frac{\int_{\y \in A_c} \int_{\y^* \in A_{c^*}} f(\y^* | \y) d\y^* f_{Y|h} (\y | h) d\y}{\int_{\y \in A_c} f_{Y|h} (\y|h) d\y}, \label{eq:non-diff}
\end{equation}
where $f_{Y|h}$ denotes the density of $\Y_i$ given $h$. Here, the same argument as in Buonaccorsi (2010, page 160-163) can be applied: The misclassification probability $P(\hat{K} = c^* | K = c , h)$ depends on $h$, as long as $\Y_{i}$ and $H_i$ are associated, which means in general the misclassification is differential.  In a second step, we extend the case to $T_i \geq 1$. Now we assume that the $M\times T_i$-dimensional space can be partitioned into $C$ and $C^*$ disjunct subsets $\hat{A}_{1 T_i},...,\hat{A}_{C^* T_i}$, which can be used for the classification function $\hat{K}$ as before, i.e. $\hat{K}(\y) = c^*  \Leftrightarrow \y \in \hat{A}_{c^* T_i}$. Then, for a $M \times T_i$ dimensional vector $\y^*$ the misclassification probability $P(\hat{K} = c^* | K = c , h)$ is
\[
\frac{\int_{\y \in A_c} \int_{\y^* \in A_{c^*T_i}} f(\y^* | \y) d\y^* f_{Y|h} (\y | h) d\y}{\int_{\y \in A_c} f_{Y|h} (\y|h) d\y}
\]
as equation~(\ref{eq:non-diff}) and the same reasoning can be applied to this case.

\section{Calculation of the corrective expected value}\label{appB}
As described in Intemann et al. (2019) the corrective expected value $\mu_{mi}(\zeta)$ is defined as the solution of
\begin{align*}
	\E(Y_{mit}^*)  =& (\lambda_m(\E(g_{\lambda_m}(Y_{mit}^*)) + \mu_{mi}(\zeta))+ 1)^{\frac{1}{\lambda_m}}  + \\
	& \frac{1-\lambda_m}{2}  (\lambda_m (\E(g_{\lambda_m}(Y_{mit}^*)) + \mu_{mi}(\zeta)) + 1)^{\frac{1}{\lambda_m} - 2} (1+\zeta) \sigma^2_{\varepsilon m}
\end{align*}
to guarantee the unbiasedness of $	Y^{(l)}_{mit}(\zeta)$. Given the observation $y_{mit}^*$ of $Y_{mit}^*$, $t=1,...,T_i$, and the parameters $\lambda_m$, $\zeta$ and $\sigma_{\varepsilon m}$, the expected value $\mu_{mi}(\zeta)$ is calculated using the corresponding empirical equation
\begin{equation*}
	\bar{y}_{mi\bullet}^*  =  (\lambda(\bar{y^*}^{(g)}_{mi\bullet} + \mu_{mi}(\zeta)) + 1)^{\frac{1}{\lambda_{m}}}  + \frac{1 - \lambda_{m}}{2}(\lambda_{m} (\bar{y^*}^{(g)}_{mi\bullet} + \mu_{mi}(\zeta)) + 1)^{\frac{1}{\lambda_{m}} - 2} \zeta \sigma^2_{\varepsilon_{m}},
\end{equation*}
where $\bar{y}_{mi\bullet}^* = 1/T_i \sum_{1}^{T_i}y_{mit}^*$ and $\bar{y^*}^{(g)}_{mi\bullet} = 1/T_i \sum_{1}^{T_i}g_{\lambda_{m}}(y_{mit}^*)$. The equation can be solved numerically. In doing so, the exact solutions for  $\mu_{mi}(\zeta)$ for $\lambda_{m} =0$ and $\lambda_{m} = 1$ can be used, i.e. $\mu_{mi}(\zeta)$ should be searched in the interval $(\max\{-\zeta \sigma^2_{\varepsilon}/2; -(1/\lambda) + \bar{y^*}^{(g)}_{mi\bullet} )\}, 0)$ if $\lambda_{m} < 1$. For details please see Intemann et al. (2019).

\section{Further analysis of the real 24HDR data (for plausible parameter estimates of the simulation analysis)}\label{appC}
In order to mimic this example in a simulation study, the real data were further analysed. First, the joint distribution of the covariates was estimated assuming a multivariate normal distribution. Second, the correlation matrix between the individual estimates for the random effects $\hat{u}_{1i}, ..., \hat{u}_{9i}$ was calculated. Finally, the linear health models 
\begin{align}\label{HM_HA}
H_i &=  \alpha_0^{\NCI} + \X_i \balpha_X^{\NCI}  + \hat{\Y}_{i}^{\NCI} \balpha_{\hat{Y}} + e_i^{\Hdrei}
\end{align}
and
\begin{align}\label{HM_HB}
H_i & = \alpha_0^{\NCI*} + \X_i \balpha_X^{\NCI*}  +  \hat{\Y}_{i}^{\NCI} \balpha_{\hat{Y}}^* +    \bar{\Y}^*_{i\bullet}  \balpha_{\bar{Y}}^* + e_i^{\Hvier}
\end{align}
for $M=5$ and $M=9$ were fitted to the data, where $\hat{\Y}_{i}^{\NCI} = (\hat{Y}_{1i}^{\NCI}, ..., \hat{Y}_{Mi}^{\NCI})$ denotes the vector of estimators of the usual intake using the NCI-method (as in method RC-3-SA),  $\bar{\Y}^*_{i\bullet} = (\bar{Y}^*_{1i\bullet},..., \bar{Y}^*_{Mi\bullet})$ the individual means, $\X_i$ the covariate vector (all with corresponding coefficients), $e_i^{\Hdrei}$ and $e_i^{\Hvier}$ error terms. The estimated parameters are also summarised in Tables SM8-SM9. The estimated models make it possible to simulate $H_i$ in two ways: first based on the usual intake and additionally based on the reported intake. The latter can be seen as health model implying differential error, since the health outcome also depends on the reports.  

\section{Simulation study in a simple setting}\label{appD}
We conducted an additional simulation study for the proposed correction methods in a very simple setting for three reasons: first, to visualise the 3-SA, second, to demonstrate that the categorisation of erroneous exposures in the 3-SA can be useful to reduce bias compared to using the erroneous continuous exposure and third, to investigate the influence of varying variances of the measurement error, of the individual random effect term and of the error term in the health model. In the simple setting, we assumed only one exposure ($M=1$), two reported days ($T_i=2$), $I=200$ or $I=1000$ individuals and no further covariates. The following models were used to simulate $Y_{1i}, H_i$ and $Y_{1it}^*$:     
\begin{align*}\label{HM_simple}
	Y_{1i} & = u_{1i} \\
	H_i &=  Y_{1i} + e_i^{H} \\
	Y_{1it}^{*} &= Y_{1i} + \varepsilon_{1it}
\end{align*}
with $u_{1i} \sim \mathcal{N}(0, \sigma^2_{u 1})$, $e_i^H \sim \mathcal{N}(0, \sigma^2_{eH})$ and $\varepsilon_{1it} \sim \mathcal{N}(0, \sigma^2_{\varepsilon 1})$ for all combinations of $\sigma^2_{u 1} =1/5, 1, 5$, $\sigma^2_{eH} =1/5, 1, 5$ and $\sigma^2_{\varepsilon 1} =1/5, 1, 5$. That means, we assume $g$ to be the identity function. In each correction algorithm we used a fix classification function $K$ to dichotomise the exposure with the fix cut-off $0$, i.e. $C=2$ without using any cluster algorithm. The true health model of the 3-SA was assumed to be $H_i =  \alpha_{K} \K(Y_{1i}) + e_i$. For the correction methods corresponding health models were used to estimate $\alpha_K$, e.g.  for the \naiveleer 3-SA and RC-3-SA, $\K(\bar{Y}_{i})$ was used instead of the true classification. Both approaches were equivalent since no covariates were considered. As before 1000 simulation data sets were generated and the relative absolute bias of each correction method in each scenario was calculated based on the true 3-SA. Additionally, the results of the correction methods, including the cluster means, the simulated data and the results of the \naiveleer approach only using the erroneous continuous exposures are plotted to visualise the 3-SA in the simple setting (Figure SM1). The comparison of the absolute relative bias can be found in Table SM1 and Table SM2 (see Supplementary Material). They can be summarised as follows: SIMEX-3-SA outperformed the other methods in most cases. Only in a few scenarios with low measurement error variance RC-3-SA and MI-3-SA performed better. In contrast to the realistic scenarios, SIMEX-3-SA seems to work in this simple setting where increasing the level of error variance really leads to an increased misclassification rate. However, if more than two clusters and multiple exposures are considered this simple relationship no longer holds. Therefore, the results from this simple setting cannot be transferred to more complex realistic scenarios.
\end{appendix}

\section*{Acknowledgments}
We gratefully acknowledge the financial support of the European Commission within the Seventh RTD Framework Program Contract No. 266044. We thank the I.Family children and their parents for participating in this extensive examination. We are grateful for the support from school boards, headmasters and communities.

\section*{Funding} 
The study was funded by the Deutsche Forschungsgemeinschaft (DFG, German Research Foundation) -- 391977161. 

\section*{Supplementary Material}
Detailed results of the analysis of real 24HDR data and all results of the simulation study (including the simple setting) can be found in the Supplementary Material.

\section*{Code and data}
Code and data for the simulation study and to mimic the original results of the example can be requested from the author. Due to data protection issues the original data of the example cannot be made available.


\end{document}


\maketitle

\section[Simulation in simple setting]{Simulation in simple setting}

\begin{table}[h]
	\tiny
	\centering
	\caption{Relative bias for $\alpha $\textsubscript{K} in the simple setting of the 3-SA depending on correction method and  $\sigma _{\mathit{eH}}^2,\sigma _{\mathit{\varepsilon 1}}^2$ \textsubscript{ }and  $\sigma _{\mathit{u1}}^2$ (I = 200)}

\begin{tabular}{ll|rrr|rrr|rrr}
	\multicolumn{2}{l|}{M = 1, I = 200,	T\textsubscript{i} = 2, C = 2}
	&
	$\sigma _{\mathit{\varepsilon 1}}^2$ = 1/5&
	&
	&
	$\sigma _{\mathit{\varepsilon 1}}^2$ = 1 &
	&
	&
	$\sigma _{\mathit{\varepsilon 1}}^2$ = 5 &
	&
	\\
		\hline	
	 &&
	naïve/RC &
	MI &
	SIMEX &
	naïve/RC &
	MI &
	SIMEX &
	naïve/RC &
	MI &
	SIMEX\\
	\hline\hline
	$\sigma _{\mathit{eH}}^2$= 1/5 &
	$\sigma _{\mathit{u1}}^2$= 1/5 &
	0.18 &
	0.25 &
	\textbf{0.11} &
	0.46 &
	0.58 &
	\textbf{0.27} &
	0.73 &
	0.8 &
	\textbf{0.62}\\
	&
	$\sigma _{\mathit{u1}}^2$= 1 &
	0.05 &
	0.06 &
	\textbf{0.03} &
	0.18 &
	0.27 &
	\textbf{0.07} &
	0.47 &
	0.59 &
	\textbf{0.28}\\
	&
	$\sigma _{\mathit{u1}}^2$= 5 &
	\textbf{0.01} &
	\textbf{0.01} &
	\textbf{0.01} &
	\textbf{0.05} &
	0.08 &
	0.02 &
	0.18 &
	0.29 &
	\textbf{0.06}\\\hline
	$\sigma _{\mathit{eH}}^2$= 1 &
	$\sigma _{\mathit{u1}}^2$= 1/5 &
	\textbf{0.19} &
	0.23 &
	0.21 &
	0.46 &
	0.56 &
	\textbf{0.33} &
	0.73 &
	0.8 &
	\textbf{0.64}\\
	&
	$\sigma _{\mathit{u1}}^2$= 1 &
	\textbf{0.05} &
	0.06 &
	0.07 &
	0.18 &
	0.24 &
	\textbf{0.1} &
	0.47 &
	0.58 &
	\textbf{0.28}\\
	&
	$\sigma _{\mathit{u1}}^2$=5 &
	\textbf{0.01} &
	\textbf{0.01} &
	0.02 &
	0.04 &
	0.06 &
	\textbf{0.03} &
	0.19 &
	0.27 &
	\textbf{0.07}\\\hline
	$\sigma _{\mathit{eH}}^2$= 5 &
	$\sigma _{\mathit{u1}}^2$=1/5 &
	0.31 &
	\textbf{0.29} &
	0.48 &
	\textbf{0.53} &
	0.56 &
	0.62 &
	\textbf{0.77} &
	0.82 &
	0.81\\
	&
	$\sigma _{\mathit{u1}}^2$= 1 &
	\textbf{0.09} &
	\textbf{0.09} &
	0.14 &
	\textbf{0.18} &
	0.22 &
	0.21 &
	0.47 &
	0.57 &
	\textbf{0.34}\\
	&
	$\sigma _{\mathit{u1}}^2$= 5 &
	\textbf{0.02} &
	\textbf{0.02} &
	0.04 &
	\textbf{0.05} &
	0.05 &
	0.07 &
	0.18 &
	0.25 &
	\textbf{0.11}\\\hline
	
\end{tabular}
\end{table}

\begin{table}[h]
	\tiny
	\centering
	\caption{Relative bias for $\alpha $\textsubscript{K} in the simple setting of the 3-SA depending on correction method and  $\sigma _{\mathit{eH}}^2,\sigma _{\mathit{\varepsilon 1}}^2$ \textsubscript{ }and  $\sigma _{\mathit{u1}}^2$ (I = 1000)}
	
	\begin{tabular}{ll|rrr|rrr|rrr}
		\multicolumn{2}{l|}{M = 1, I = 1000,	T\textsubscript{i} = 2, C = 2}
		&
		$\sigma _{\mathit{\varepsilon 1}}^2$ = 1/5&
		&
		&
		$\sigma _{\mathit{\varepsilon 1}}^2$ = 1 &
		&
		&
		$\sigma _{\mathit{\varepsilon 1}}^2$ = 5 &
		&
		\\
		\hline	
&
&
naïve/RC &
MI &
SIMEX &
naïve/RC &
MI &
SIMEX &
naïve/RC &
MI &
SIMEX\\\hline
$\sigma _{\mathit{eH}}^2$= 1/5 &
$\sigma _{\mathit{u1}}^2$= 1/5 &
0.18 &
0.24 &
\textbf{0.05} &
0.46 &
0.58 &
\textbf{0.28} &
0.72 &
0.8 &
\textbf{0.61}\\
&
$\sigma _{\mathit{u1}}^2$= 1 &
0.05 &
0.06 &
\textbf{0.02} &
0.18 &
0.27 &
\textbf{0.04} &
0.47 &
0.59 &
\textbf{0.28}\\
&
$\sigma _u^2$= 5 &
0.01 &
0.01 &
\textbf{0.00} &
0.05 &
0.08 &
\textbf{0.01} &
0.18 &
0.29 &
\textbf{0.04}\\\hline
$\sigma _{\mathit{eH}}^2$\textsubscript{ }= 1 &
$\sigma _{\mathit{u1}}^2$= 1/5 &
0.18 &
0.23 &
\textbf{0.1} &
0.47 &
0.57 &
\textbf{0.28} &
0.72 &
0.8 &
\textbf{0.6}\\
&
$\sigma _{\mathit{u1}}^2$= 1 &
0.04 &
0.05 &
\textbf{0.03} &
0.18 &
0.24 &
\textbf{0.06} &
0.47 &
0.58 &
\textbf{0.28}\\
&
$\sigma _u^2$= 5 &
\textbf{0.01} &
\textbf{0.01} &
\textbf{0.01} &
0.05 &
0.06 &
\textbf{0.02} &
0.18 &
0.27 &
\textbf{0.04}\\\hline
$\sigma _{\mathit{eH}}^2$\textsubscript{ }= 5 &
$\sigma _{\mathit{u1}}^2$= 1/5 &
\textbf{0.19} &
0.23 &
0.21 &
0.47 &
0.58 &
\textbf{0.35} &
0.72 &
0.79 &
\textbf{0.6}\\
&
$\sigma _{\mathit{u1}}^2$= 1 &
\textbf{0.05} &
\textbf{0.05} &
0.07 &
0.18 &
0.23 &
\textbf{0.1} &
0.47 &
0.57 &
\textbf{0.29}\\
&
$\sigma _{\mathit{u1}}^2$= 5 &
\textbf{0.01} &
\textbf{0.01} &
0.02 &
0.05 &
0.05 &
\textbf{0.03} &
0.18 &
0.24 &
\textbf{0.05}\\\hline
\end{tabular}
\end{table}

\begin{figure}[h]
	\leftskip=-7cm		
		\includegraphics[height=16cm]{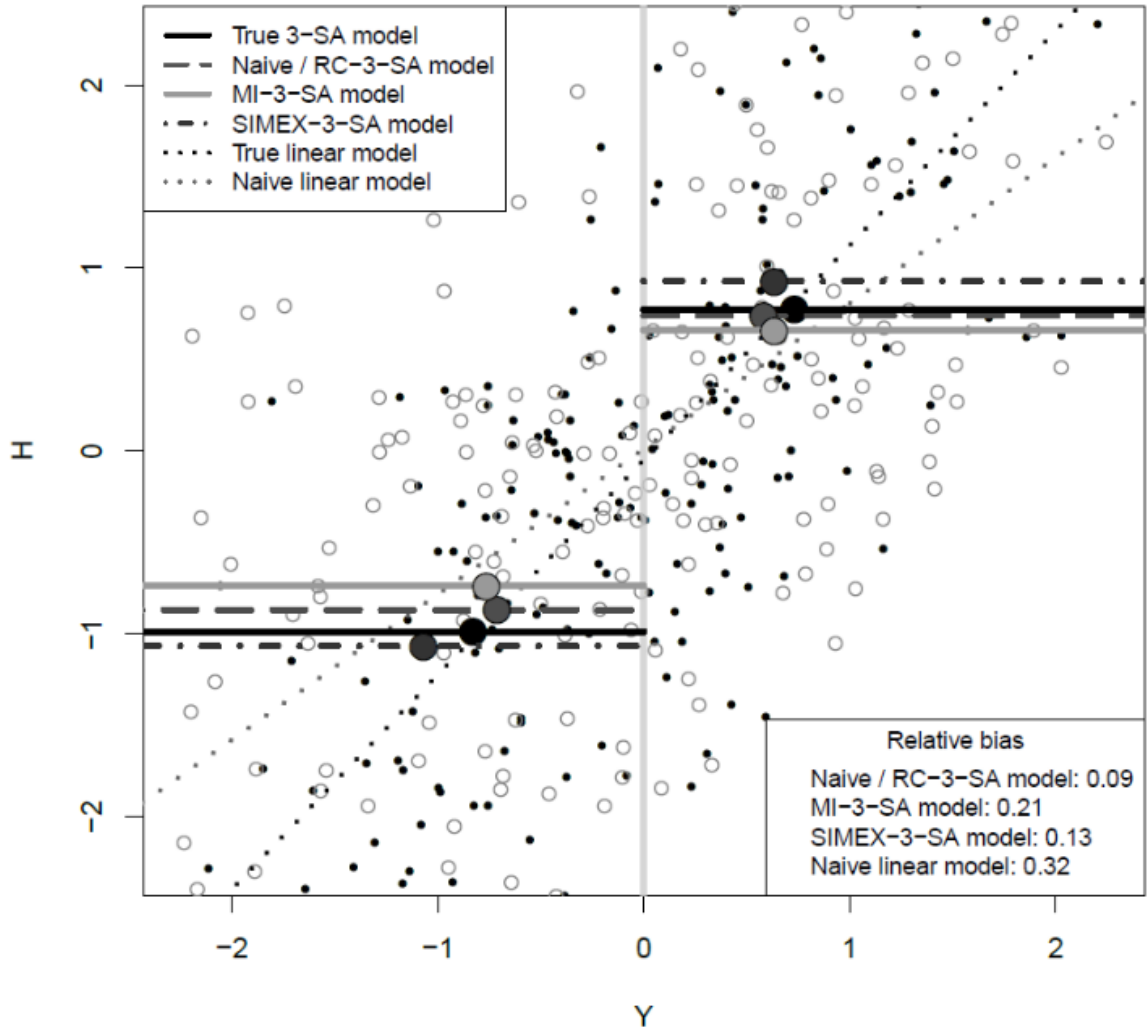}
		\caption{Illustration of the simple setting of 3-SA for $I = 200$, $\sigma^2_{\mathit{\varepsilon 1}} = 1$, $\sigma^2_{\mathit{u 1}} = 1$, 	$\sigma^2_{\mathit{eH}} = 1$ (true (small black dots) and measured data (grey circles)) and the resulting correction models, the true 3-SA model (black line) and the true and naïve health model using the continuous exposure (black and grey dotted line). The vertical grey line is the cut-off used for classification of the exposure. The big black filled dot is the true mean of the exposure clusters drawn on the corresponding regression line, analogue for the correction methods based on multiple imputation (light grey), regression calibration (grey line) and SIMEX (dark grey line). The mean of	the MI method is closest to the true mean. The relative bias of the 3-SA models is much lower than of the naïve method based on the continuous exposure.} \label{Fig_SM1}
\end{figure}

\clearpage
\section{Analysis of real 24HDR data}

\begin{table}[h]
	\tiny
	\caption{
		(a) Fitted error model for RC-3-SA.
		(b) Fitted error model for MI-3-SA considering the continuous health outcome  $H_i$ (corresponding coefficient is the last element of  $\widehat {\bbeta }_m$).
		(c) Fitted error model for MI-3-SA considering the dichotomised version of  $H_i$ (corresponding coefficient is the last element of  $\widehat {\bbeta }_m$)}.
	(a)
\begin{tabular}{|l|l|l|l|l|l|l|l|l|l|}
	\hline
	\centering  $m$ &
	1 &
	2 &
	3 &
	4 &
	5 &
	6 &
	7 &
	8 &
	9\\
	\hline
	\centering  $\widehat {\lambda }_m$ &
	0.36 &
	0.39 &
	1.53 &
	0.30 &
	0.76 &
	0.38 &
	0.39 &
	0.48 &
	0.30\\
	\hline
	\centering  $\widehat {\beta }_{0m}$ &
	25.5 &
	26.8 &
	25.4 &
	{}-0.26 &
	3.00 &
	10.3 &
	8.99 &
	56.4 &
	28.1\\
	\hline
	\centering  $\widehat {\bbeta }_m$ &
  $\left(\begin{matrix}-0.48\\-0.62\\0.35\end{matrix}\right)$&
  $\left(\begin{matrix}-0.11\\-0.51\\0.39\end{matrix}\right)$ &
  $\left(\begin{matrix}0.07\\-0.33\\-0.07\end{matrix}\right)$ &
  $\left(\begin{matrix}0.10\\0.08\\0.15\end{matrix}\right)$ &
 $\left(\begin{matrix}-0.04\\-0.07\\0.10\end{matrix}\right)$ &
 $\left(\begin{matrix}-0.20\\-0.25\\0.39\end{matrix}\right)$ &
  $\left(\begin{matrix}0.02\\-0.17\\0.19\end{matrix}\right)$ &
  $\left(\begin{matrix}-0.12\\-0.35\\-1.15\end{matrix}\right)$ &
 $\left(\begin{matrix}-0.11\\-0.30\\-0.03\end{matrix}\right)$\\
 \hline
	\centering  $\widehat{\sigma }_{\mathit{um}}^2$ &
	8.9 &
	10.5 &
	10.8 &
	0.17 &
	0.11 &
	1.1 &
	2.5 &
	65.5 &
	4.4\\\hline
	\centering  $\widehat {\sigma }_{\mathit{\varepsilon m}}^2$ &
	38.1 &
	51.0 &
	47.5 &
	0.55 &
	0.38 &
	10.3 &
	13.4 &
	194.6 &
	11.3\\\hline
\end{tabular}
\end{table}

\begin{table}[h]
	\tiny
		(b)
	\begin{tabular}{|l|l|l|l|l|l|l|l|l|l|}
		\hline
$m$ &
1 &
2 &
3 &
4 &
5 &
6 &
7 &
8 &
9\\\hline
$\widehat {\lambda }_m$ &
0.36 &
0.39 &
1.53 &
0.30 &
0.76 &
0.38 &
0.39 &
0.48 &
0.30\\\hline
$\widehat {\beta }_{0m}$ &
25.5 &
26.8 &
25.4 &
{}-0.26 &
2.99 &
10.4 &
9.0 &
56.8 &
28.1\\\hline
$\widehat {\bbeta }_m$ &
$\left(\begin{matrix}-0.49\\-0.61\\0.32\\-0.21\end{matrix}\right)$ &
$\left(\begin{matrix}-0.12\\-0.49\\0.31\\-0.41\end{matrix}\right)$ &
$\left(\begin{matrix}0.06\\-0.33\\-0.09\\-0.09\end{matrix}\right)$ &
$\left(\begin{matrix}0.09\\0.08\\0.15\\0.00\end{matrix}\right)$ &
$\left(\begin{matrix}-0.04\\-0.07\\0.10\\0.01\end{matrix}\right)$ &
$\left(\begin{matrix}-0.20\\-0.25\\0.39\\0.01\end{matrix}\right)$ &
$\left(\begin{matrix}0.01\\-0.17\\0.17\\-0.12\end{matrix}\right)$ &
$\left(\begin{matrix}-0.15\\-0.31\\-1.3\\-0.91\end{matrix}\right)$ &
$\left(\begin{matrix}-0.12\\-0.29\\-0.05\\-0.17\end{matrix}\right)$\\\hline
$\widehat {\sigma }_{\mathit{um}}^2$ &
8.9 &
10.3 &
10.8 &
0.17 &
0.11 &
1.1 &
2.5 &
65.4 &
4.4\\\hline
$\widehat {\sigma }_{\mathit{\varepsilon m}}^2$ &
36.1 &
50.8 &
47.5 &
0.55 &
0.38 &
10.3 &
13.4 &
196.4 &
11.3\\\hline
\end{tabular}
\end{table}

\begin{table}[h]
	
	\tiny
		(c)
\begin{tabular}{|l|l|l|l|l|l|l|l|l|l|}
\hline
$m$ &
1 &
2 &
3 &
4 &
5 &
6 &
7 &
8 &
9\\\hline
$\widehat {\lambda }_m$ &
0.36 &
0.39 &
1.53 &
0.30 &
0.76 &
0.38 &
0.39 &
0.48 &
0.28\\\hline
$\widehat {\beta }_{0m}$ &
25.6 &
27.0 &
25.5 &
{}-0.26 &
2.99 &
10.5 &
9.0 &
56.6 &
25.9\\\hline
$\widehat {\bbeta }_m$ &
$\left(\begin{matrix}-0.49\\-0.62\\0.32\\-0.57\end{matrix}\right)$ &
$\left(\begin{matrix}-0.12\\-0.50\\0.31\\-1.25\end{matrix}\right)$ &
$\left(\begin{matrix}0.06\\-0.33\\-0.09\\-0.88\end{matrix}\right)$ &
$\left(\begin{matrix}0.1\\0.08\\0.15\\-0.02\end{matrix}\right)$ &
$\left(\begin{matrix}-0.04\\-0.07\\0.10\\0.00\end{matrix}\right)$ &
$\left(\begin{matrix}-0.21\\-0.25\\0.37\\-0.35\end{matrix}\right)$ &
$\left(\begin{matrix}0.02\\-0.17\\0.18\\-0.03\end{matrix}\right)$ &
$\left(\begin{matrix}-0.13\\-0.35\\-1.2\\-1.09\end{matrix}\right)$ &
$\left(\begin{matrix}-0.10\\-0.26\\-0.02\\-0.02\end{matrix}\right)$\\\hline
$\widehat {\sigma }_{\mathit{um}}^2$ &
8.8 &
10.4 &
10.7 &
0.17 &
0.11 &
1.1 &
2.5 &
65.6 &
3.4\\\hline
$\widehat {\sigma }_{\mathit{\varepsilon m}}^2$ &
36.1 &
51.0 &
47.4 &
0.55 &
0.38 &
10.3 &
13.4 &
195.1 &
9.0\\\hline
\end{tabular}
\end{table}

\begin{table}	
	\caption{Cluster means  $\overline {\y}_{\mathit{Cc}}$ = ( $\overline y_{1c}$,{\dots}, $\overline y_{5c}$), $c = 1,2, 3, M=5$ exposures, obtained from \textit{k}{}-means and Gaussian mixture model (GMM) for correction methods SIMEX-3-SA, RC-3-SA and MI-3-SA,  $I_c$  denotes the number of individuals per cluster}
	\begin{tabular}{|l|l|l|l|l|l|l|l|l|}
\hline
Cluster method &
Correction method &
$I_c$ &
$\overline y_{1c}$ &
$\overline y_{2c}$ &
$\overline y_{3c}$ &
$\overline y_{4c}$ &
$\overline y_{5c}$\\\hline
\textit{k}{}-means &
naïve/SIMEX &
494 &
0.66 &
0.51 &
0.66 &
{}-0.47 &
0.87\\
&
&
748 &
{}-0.3 &
{}-0.2 &
0.01 &
{}-0.18 &
{}-0.25\\
&
&
256 &
{}-0.41 &
{}-0.4 &
{}-1.32 &
1.42 &
{}-0.95\\\hline
&
RC &
704 &
{}-0.26 &
{}-0.01 &
0.12 &
0.17 &
{}-0.17\\
&
&
513 &
0.85 &
0.42 &
0.5 &
{}-0.78 &
0.9\\
&
&
281 &
{}-0.88 &
{}-0.76 &
{}-1.22 &
1 &
{}-1.21\\\hline
&
MI &
512 &
0.85 &
0.41 &
0.5 &
{}-0.79 &
0.9\\
&
(continuous outcome) &
282 &
{}-0.88 &
{}-0.75 &
{}-1.22 &
0.99 &
{}-1.21\\
&
&
704 &
{}-0.26 &
0.01 &
0.12 &
0.18 &
{}-0.17\\\hline
&
M &
282 &
{}-0.88 &
{}-0.75 &
{}-1.22 &
0.99 &
{}-1.21\\
&
(dichotomous &
512 &
0.85 &
0.41 &
0.5 &
{}-0.79 &
0.9\\
&
outcome) &
704 &
{}-0.26 &
0.01 &
0.12 &
0.18 &
{}-0.17\\\hline
GMM &
naïve/SIMEX &
91 &
{}-0.37 &
{}-0.74 &
{}-1.85 &
2.18 &
{}-1.34\\
&
&
997 &
{}-0.24 &
{}-0.13 &
{}-0.11 &
0 &
{}-0.24\\
&
&
410 &
0.58 &
0.44 &
0.63 &
{}-0.46 &
0.79\\\hline
&
RC &
158 &
{}-0.91 &
{}-0.93 &
{}-1.42 &
1.23 &
{}-1.39\\
&
&
949 &
{}-0.23 &
{}-0.06 &
0.01 &
0.11 &
{}-0.17\\
&
&
391 &
0.86 &
0.52 &
0.56 &
{}-0.75 &
0.93\\\hline
&
MI &
159 &
{}-0.91 &
{}-0.92 &
{}-1.41 &
1.23 &
{}-1.39\\
&
(continuous outcome) &
944 &
{}-0.23 &
{}-0.05 &
0.01 &
0.11 &
{}-0.17\\
&
&
395 &
0.86 &
0.49 &
0.56 &
{}-0.75 &
0.93\\\hline
&
MI &
159 &
{}-0.91 &
{}-0.92 &
{}-1.41 &
1.23 &
{}-1.39\\
&
(dichotomous &
944 &
{}-0.23 &
{}-0.05 &
0.01 &
0.11 &
{}-0.17\\
&
outcome) &
395 &
0.86 &
0.49 &
0.56 &
{}-0.75 &
0.93
\\\hline
\end{tabular}
\end{table}

\begin{figure}[h]
	\leftskip=-1cm		
	\includegraphics[width=17cm]{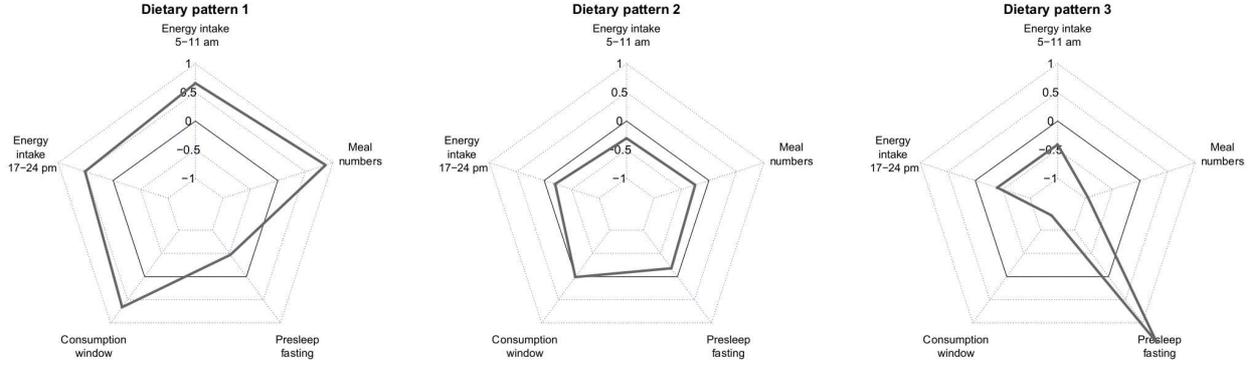}
	
	\caption{Illustration of a radar chart to facilitate presentation of cluster solutions (using the \textit{k}{}-means cluster solution from the naïve/SIMEX approach for M=5 and C=3)}
\label{Fig_SM2}
\end{figure}

\begin{table}[h]
	\caption{Cluster means  $\overline \y_{\mathit{Cc}}$= ( $\overline y_{1c}$,{\dots}, $\overline y_{9c}$), $c = 1, 2, 3$, $M=9$ exposures, obtained from \textit{k}{}-means and Gaussian mixture model (GMM) for correction methods SIMEX-3-SA, RC-3-SA and MI-3-SA,  $I_c$  denotes the number of individuals per cluster}
	\tiny
	\begin{tabular}{llllllllllll}
\hline
Cluster method &
Correction method &
$I_c$ &
$\overline y_{1c}$ &
$\overline y_{2c}$ &
$\overline y_{3c}$ &
$\overline y_{4c}$ &
$\overline y_{5c}$ &
$\overline y_{6c}$ &
$\overline y_{7c}$ &
$\overline y_{8c}$ &
$\overline y_{9c}$\\\hline
k-means &
naïve/SIMEX &
683 &
{}-0.49 &
{}-0.51 &
{}-0.43 &
0.3 &
{}-0.55 &
{}-0.7 &
{}-0.6 &
{}-0.48 &
{}-0.45\\
&
&
532 &
0.71 &
{}-0.08 &
0.51 &
{}-0.37 &
0.61 &
0.34 &
0.21 &
0.71 &
{}-0.14\\
&
&
283 &
{}-0.15 &
1.4 &
0.07 &
{}-0.04 &
0.19 &
1.04 &
1.05 &
{}-0.18 &
1.34\\\hline
&
RC &
600 &
{}-0.65 &
{}-0.63 &
{}-0.44 &
0.49 &
{}-0.69 &
{}-0.83 &
{}-0.68 &
{}-0.66 &
{}-0.58\\
&
&
336 &
{}-0.21 &
1.05 &
0.22 &
0.19 &
0 &
0.87 &
0.89 &
{}-0.27 &
1.01\\
&
&
562 &
0.82 &
0.05 &
0.34 &
{}-0.64 &
0.74 &
0.36 &
0.2 &
0.86 &
0.01\\\hline
&
MI &
553 &
0.82 &
0.03 &
0.35 &
{}-0.65 &
0.75 &
0.35 &
0.18 &
0.86 &
{}-0.01\\
&
(continuous &
595 &
{}-0.66 &
{}-0.64 &
{}-0.45 &
0.48 &
{}-0.7 &
{}-0.83 &
{}-0.69 &
{}-0.66 &
{}-0.59\\
&
outcome) &
350 &
{}-0.18 &
1.05 &
0.22 &
0.21 &
0 &
0.86 &
0.88 &
{}-0.24 &
1.01\\\hline
&
MI &
553 &
0.82 &
0.03 &
0.35 &
{}-0.65 &
0.75 &
0.35 &
0.18 &
0.86 &
{}-0.01\\
&
(dichotomous &
595 &
{}-0.66 &
{}-0.64 &
{}-0.45 &
0.48 &
{}-0.7 &
{}-0.83 &
{}-0.69 &
{}-0.66 &
{}-0.59\\
&
outcome) &
350 &
{}-0.18 &
1.05 &
0.22 &
0.21 &
0 &
0.86 &
0.88 &
{}-0.24 &
1.01\\\hline
GMM &
naïve/SIMEX &
105 &
{}-0.05 &
2.23 &
0.04 &
0.02 &
0.15 &
1.56 &
1.6 &
{}-0.06 &
2.24\\
&
&
744 &
{}-0.45 &
{}-0.46 &
{}-0.38 &
0.26 &
{}-0.49 &
{}-0.62 &
{}-0.54 &
{}-0.44 &
{}-0.41\\
&
&
649 &
0.52 &
0.16 &
0.43 &
{}-0.3 &
0.54 &
0.46 &
0.36 &
0.52 &
0.1\\\hline
&
RC &
344 &
0.9 &
0.96 &
0.46 &
{}-0.51 &
0.79 &
1.14 &
0.93 &
0.86 &
0.88\\
&
&
298 &
{}-0.94 &
{}-0.84 &
{}-0.85 &
0.87 &
{}-1.09 &
{}-1.1 &
{}-0.87 &
{}-0.95 &
{}-0.76\\
&
&
856 &
{}-0.02 &
{}-0.09 &
0.12 &
{}-0.11 &
0.08 &
{}-0.06 &
{}-0.06 &
0 &
{}-0.08\\\hline
&
MI &
238 &
{}-0.14 &
1.31 &
0.26 &
0.16 &
0.12 &
1.08 &
1.06 &
{}-0.19 &
1.26\\
&
(continuous &
614 &
{}-0.67 &
{}-0.58 &
{}-0.45 &
0.5 &
{}-0.7 &
{}-0.78 &
{}-0.63 &
{}-0.67 &
{}-0.52\\
&
outcome) &
646 &
0.69 &
0.06 &
0.33 &
{}-0.54 &
0.62 &
0.34 &
0.2 &
0.71 &
0.02\\\hline
&
MI &
238 &
{}-0.14 &
1.31 &
0.26 &
0.16 &
0.12 &
1.08 &
1.06 &
{}-0.19 &
1.26\\
&
(dichotomous &
614 &
{}-0.67 &
{}-0.58 &
{}-0.45 &
0.5 &
{}-0.7 &
{}-0.78 &
{}-0.63 &
{}-0.67 &
{}-0.52\\
&
outcome) &
646 &
0.69 &
0.06 &
0.33 &
{}-0.54 &
0.62 &
0.34 &
0.2 &
0.71 &
0.02\\\hline
\end{tabular}
\end{table}

\begin{table}[h]
	\caption{Corrected estimates  $\widehat {\balpha }_K$ = ( $\widehat {\alpha }_{\mathit{K12},}\widehat {\alpha }_{\mathit{K13}}$)\textsuperscript{T} for \textit{C} = 3 clusters according to \textit{k}{}-means and Gaussian mixture model (GMM) of the naïve method and the correction methods RC-3-SA, SIMEX-3-SA and MI-3-SA (estimates do not necessarily refer to the same cluster solution for all correction method, see Tables SM4 and SM5 for corresponding cluster means)}
\begin{tabular}{lllll}
	\hline
	Cluster method &
	M &
	Correction method &
	$\widehat {\alpha }_{\mathit{K12}}$ &
	$\widehat {\alpha }_{\mathit{K13}}$\\\hline
	k-means &
	5 &
	naïve &
	0.06 &
	0.06\\
	&
	&
	RC &
	0.08 &
	0.11\\
	&
	&
	SIMEX &
	0.09 &
	0.11\\
	&
	&
	MI &
	{}-0.02 &
	{}-0.03\\\hline{}
	&
	9 &
	naïve &
	{}-0.01 &
	{}-0.09\\
	&
	&
	RC &
	{}-0.16 &
	{}-0.01\\
	&
	&
	SIMEX &
	{}-0.01 &
	{}-0.1\\
	&
	&
	MI &
	{}-0.02 &
	{}-0.11\\\hline
	GMM &
	5 &
	naïve &
	{}-0.03 &
	{}-0.06\\
	&
	&
	RC &
	{}-0.01 &
	0\\
	&
	&
	SIMEX &
	{}-0.05 &
	{}-0.1\\
	&
	&
	MI &
	0.02 &
	0.04\\\hline{ }
	&
	9 &
	naïve &
	0.08 &
	0\\
	&
	&
	RC &
	0.14 &
	0.05\\
	&
	&
	SIMEX &
	0.1 &
	{}-0.05\\
	&
	&
	MI &
	0.09 &
	0.1\\\hline
\end{tabular}
\end{table}

\begin{table}[h]
	\caption{Corrected estimates  $\widehat {\bgamma }_K$ = ( $\widehat {\gamma }_{\mathit{K12},}\widehat {\gamma }_{\mathit{K13}}$)\textsuperscript{T} for C = 3 clusters according to k-means and Gaussian mixture model (GMM) of the naïve method and the correction methods RC-3-SA, SIMEX-3-SA and MI-3-SA (estimates do not necessarily refer to the same cluster solution for all correction method, see Tables 4 and 5 for corresponding cluster means)}
	\begin{tabular}{lllll}
\hline
Cluster method &
M &
Correction method &
$\widehat {\gamma }_{\mathit{K12}}$ &
$\widehat {\gamma }_{\mathit{K13}}$\\\hline
k-means &
5 &
naïve &
0.08 &
{}-0.1\\
&
&
RC &
0.14 &
0.05\\
&
&
SIMEX &
0.11 &
{}-0.07\\
&
&
MI &
0.12 &
0.09\\\hline
&
9 &
naïve &
{}-0.12 &
{}-0.36\\
&
&
RC &
{}-0.47 &
{}-0.24\\
&
&
SIMEX &
{}-0.06 &
{}-0.49\\
&
&
MI &
{}-0.02 &
{}-0.23\\\hline
GMM &
5 &
naïve &
0.17 &
0.1\\
&
&
RC &
0.13 &
0.18\\
&
&
SIMEX &
0.36 &
0.22\\
&
&
MI &
0.16 &
0.18\\\hline
&
9 &
naïve &
0.05 &
{}-0.23\\
&
&
RC &
0.42 &
0.31\\
&
&
SIMEX &
{}-0.28 &
{}-0.73\\
&
&
MI &
0.28 &
0.27\\\hline
\end{tabular}
\end{table}

\begin{table}[h]
	
	\caption{Fitted multivariate normal distribution of the covariates and estimated correlation matrix of the individual random effect terms}
	\begin{tabular}{l|lllllllll}
\hline
Parameter &
Estimate &
&
&
&
&
&
&
&
\\\hline
$\bmu_{\text{cov}}$ &
(12, &
0.5, &
0.5) &
&
&
&
&
&
\\\hline
$\bsigma_{\text{cov}}$ &
 3.8&0&0 & & & & & & \\
		&0&1.2&-0.1 & & &  & & \\
		&0&-0.1&0.2  & & &  & &
\\\hline
Corr($\widehat{u}_{mi}$)&
1 &
0.05 &
0.29 &
{}-0.03 &
0.28 &
0.93 &
0.03 &
0.27 &
0.42\\
&
0.05 &
1 &
0.23 &
{}-0.2 &
0.29 &
0.03 &
0.85 &
0.54 &
0.64\\
&
0.29 &
0.23 &
1 &
{}-0.46 &
0.44 &
0.28 &
0.18 &
0.16 &
0.23\\
&
{}-0.03 &
{}-0.2 &
{}-0.46 &
1 &
{}-0.31 &
{}-0.02 &
{}-0.15 &
{}-0.11 &
{}-0.15\\
&
0.28 &
0.29 &
0.44 &
{}-0.31 &
1 &
0.28 &
0.26 &
0.29 &
0.4\\
&
0.93 &
0.03 &
0.28 &
{}-0.02 &
0.28 &
1 &
0.03 &
0.28 &
0.38\\
&
0.03 &
0.85 &
0.18 &
{}-0.15 &
0.26 &
0.03 &
1 &
0.45 &
0.52\\
&
0.27 &
0.54 &
0.16 &
{}-0.11 &
0.29 &
0.28 &
0.45 &
1 &
0.84\\
&
0.42 &
0.64 &
0.23 &
{}-0.15 &
0.4 &
0.38 &
0.52 &
0.84 &
1\\\hline
\end{tabular}
\end{table}

\begin{table}
\caption{Fitted health models (9) and (10) considering $M=5$ or $M=9$ continuous exposure}
\begin{tabular}{l|rrrr}
	\hline
	Parameter &
	(9) with M=5 &
	(9) with M=9 &
	(10) with M=5 &
	(10) with M=9\\\hline
	$\widehat {\alpha }_0$ &
	0.5728 &
	0.5464 &
	0.6949 &
	0.6581\\\hline
	$\widehat {\balpha }_X^{\mathit{NCI}}$ or  $\widehat {\balpha }_X^{\mathit{NCI}}$ &
	{}-0.0021 &
	{}-0.002 &
	0.0644 &
	0.0599\\
	&
	{}-0.0058 &
	{}-0.0221 &
	{}-0.0512 &
	{}-0.0591\\
	&
	0.0316 &
	0.0307 &
	0.0659 &
	0.0602\\\hline
	$\widehat {\balpha }_{\widehat  Y}$  or   $\widehat {\balpha }_{\widehat  Y}^{\ast }$ &
	3.00E-04 &
	{}-4.00E-04 &
	5.00E-04 &
	0.0012\\
	&
	{}-8.00E-04 &
	{}-0.0011 &
	{}-4.00E-04 &
	8.00E-04\\
	&
	{}-0.0218 &
	{}-0.0157 &
	{}-0.337 &
	{}-0.3252\\
	&
	{}-0.0751 &
	{}-0.0733 &
	{}-0.2695 &
	{}-0.2568\\
	&
	0.0301 &
	0.0037 &
	0.7173 &
	0.7547\\
	&
	&
	0.0048 &
	&
	{}-0.0062\\
	&
	&
	5.00E-04 &
	&
	{}-0.0097\\
	&
	&
	{}-3.00E-04 &
	&
	1.00E-04\\
	&
	&
	3.00E-04 &
	&
	{}-3.00E-04\\\hline
	$\widehat {\balpha }_{\overline Y}^{\ast }$ &
	&
	&
	{}-1.00E-04 &
	{}-4.00E-04\\
	&
	&
	&
	{}-1.00E-04 &
	{}-5.00E-04\\
	&
	&
	&
	0.0813 &
	0.0788\\
	&
	&
	&
	0.06 &
	0.0578\\
	&
	&
	&
	{}-0.2353 &
	{}-0.2551\\
	&
	&
	&
	&
	0.0019\\
	&
	&
	&
	&
	0.0025\\
	&
	&
	&
	&
	{}-2.00E-04\\
	&
	&
	&
	&
	2.00E-04\\\hline
	$\widehat {\sigma }_e$ &
	0.9872 &
	0.988 &
	0.9862 &
	0.9878\\\hline
\end{tabular}
\end{table}

\clearpage
\section{Results of the simulation study}
\subsection{Simulation for  $M=5$  and  $\mathit{Corr}\left(u_{\mathit{ki}},u_{\mathit{li}}\right){\neq}0$  for  $k{\neq}l$}

\begin{table}[h]
\caption{Simulation results of correction methods and gold standard methods based on 7 and 28 days (GS7 and GS28) for scenarios with outcome  $H_i^A$  and cluster method with \textit{C} clusters regarding empirical misclassification rate (MR), adjusted Rand index (aRI), mean absolute bias, maximal absolute bias and median mean relative absolute bias, and number of failed classifications* for 1000 simulation data sets.}
\tiny\begin{tabular}{l|l|l|l|l|l|l|l}
	\hline
	\textit{C} &
	Correction method &
	MR &
	aRI &
	$\overline{{\Delta}}$ &
	$\overline{{\Delta}}_{\mathit{max}}$ &
	$\mathit{med}(\overline{{\Delta}}_{\mathit{rel}})$ &
	Number of failed classifications*\\\hline
	\multicolumn{8}{c}{Cluster method: k-means}\\\hline
	3 &
	naïve &
	37.4 &
	0.2 &
	0.07 &
	0.1 &
	0.92 &
	0\\\hline
	&
	RC &
	36.25 &
	0.22 &
	0.08 &
	0.1 &
	0.86 &
	0\\\hline
	&
	SIMEX-Q** &
	37.39 &
	0.2 &
	0.09 &
	0.13 &
	1.24 &
	1\\\hline
	&
	SIMEX-C** &
	37.39 &
	0.2 &
	0.18 &
	0.25 &
	2.33 &
	1\\\hline
	&
	SIMEX-Q4** &
	37.39 &
	0.2 &
	0.4 &
	0.55 &
	5.45 &
	1\\\hline
	&
	MI &
	36.21 &
	0.22 &
	0.07 &
	0.1 &
	0.73 &
	0\\\hline
	&
	MI-NULL*** &
	36.22 &
	0.22 &
	0.09 &
	0.12 &
	0.91 &
	0\\\hline
	&
	GS7 &
	24.95 &
	0.4 &
	0.05 &
	0.07 &
	0.64 &
	0\\\hline
	&
	GS28 &
	13.3 &
	0.64 &
	0.03 &
	0.05 &
	0.52 &
	0\\\hline
	4 &
	naïve &
	46.5 &
	0.15 &
	0.09 &
	0.16 &
	1.13 &
	0\\\hline
	&
	RC &
	44.26 &
	0.18 &
	0.08 &
	0.13 &
	1 &
	0\\\hline
	&
	SIMEX-Q** &
	46.5 &
	0.15 &
	0.12 &
	0.19 &
	1.45 &
	1\\\hline
	&
	SIMEX-C** &
	46.5 &
	0.15 &
	0.22 &
	0.35 &
	2.65 &
	1\\\hline
	&
	SIMEX-Q4** &
	46.5 &
	0.15 &
	0.49 &
	0.78 &
	5.96 &
	1\\\hline
	&
	MI &
	44.22 &
	0.18 &
	0.08 &
	0.12 &
	0.8 &
	0\\\hline
	&
	MI-NULL*** &
	44.26 &
	0.18 &
	0.1 &
	0.15 &
	0.91 &
	0\\\hline
	&
	GS7 &
	31.94 &
	0.34 &
	0.06 &
	0.1 &
	0.68 &
	0\\\hline
	&
	GS28 &
	17.49 &
	0.59 &
	0.04 &
	0.07 &
	0.53 &
	0\\\hline
	5 &
	naïve &
	52.8 &
	0.12 &
	0.09 &
	0.2 &
	1.32 &
	0\\\hline
	&
	RC &
	50.31 &
	0.15 &
	0.09 &
	0.16 &
	1.14 &
	0\\\hline
	&
	SIMEX-Q** &
	52.81 &
	0.12 &
	0.13 &
	0.23 &
	1.65 &
	1\\\hline
	&
	SIMEX-C** &
	52.81 &
	0.12 &
	0.25 &
	0.44 &
	3.17 &
	1\\\hline
	&
	SIMEX-Q4** &
	52.81 &
	0.12 &
	0.56 &
	0.97 &
	7.02 &
	1\\\hline
	&
	MI &
	50.28 &
	0.15 &
	0.09 &
	0.15 &
	0.86 &
	0\\\hline
	&
	MI-NULL*** &
	50.31 &
	0.15 &
	0.11 &
	0.18 &
	0.93 &
	0\\\hline
	&
	GS7 &
	37.17 &
	0.29 &
	0.07 &
	0.12 &
	0.93 &
	0\\\hline
	&
	GS28 &
	19.91 &
	0.56 &
	0.05 &
	0.08 &
	0.65 &
	0\\\hline
	\multicolumn{8}{c}{  Cluster method: GMM}\\\hline
	3 &
	naïve &
	14.9 &
	0.25 &
	0.15 &
	0.23 &
	1.39 &
	1\\\hline
	&
	RC &
	29.61 &
	0.27 &
	0.08 &
	0.11 &
	1.13 &
	0\\\hline
	&
	SIMEX-Q** &
	14.86 &
	0.25 &
	0.23 &
	0.32 &
	1.94 &
	1\\\hline
	&
	SIMEX-C** &
	14.86 &
	0.25 &
	0.43 &
	0.6 &
	3.73 &
	1\\\hline
	&
	SIMEX-Q4** &
	14.86 &
	0.25 &
	0.95 &
	1.34 &
	8.22 &
	1\\\hline
	&
	MI &
	29.6 &
	0.27 &
	0.07 &
	0.1 &
	0.86 &
	0\\\hline
	&
	MI-NULL*** &
	29.46 &
	0.27 &
	0.08 &
	0.11 &
	0.92 &
	0\\\hline
	&
	GS7 &
	18.01 &
	0.47 &
	0.07 &
	0.09 &
	0.98 &
	0\\\hline
	&
	GS28 &
	10.01 &
	0.69 &
	0.04 &
	0.06 &
	0.78 &
	0\\\hline
	4 &
	naïve &
	19.2 &
	0.23 &
	0.17 &
	0.31 &
	1.41 &
	0\\\hline
	&
	RC &
	37.9 &
	0.21 &
	0.1 &
	0.16 &
	1.11 &
	0\\\hline
	&
	SIMEX-Q** &
	19.18 &
	0.23 &
	0.25 &
	0.39 &
	2.01 &
	1\\\hline
	&
	SIMEX-C** &
	19.18 &
	0.23 &
	0.47 &
	0.75 &
	3.95 &
	1\\\hline
	&
	SIMEX-Q4** &
	19.18 &
	0.23 &
	1.05 &
	1.68 &
	9 &
	1\\\hline
	&
	MI &
	37.88 &
	0.21 &
	0.09 &
	0.14 &
	0.84 &
	0\\\hline
	&
	MI-NULL*** &
	37.92 &
	0.21 &
	0.1 &
	0.17 &
	0.93 &
	0\\\hline
	&
	GS7 &
	22.65 &
	0.43 &
	0.08 &
	0.14 &
	0.9 &
	0\\\hline
	&
	GS28 &
	13 &
	0.66 &
	0.05 &
	0.08 &
	0.62 &
	0\\\hline
	5 &
	naïve &
	29.5 &
	0.22 &
	0.17 &
	0.36 &
	1.5 &
	0\\\hline
	&
	RC &
	42.31 &
	0.19 &
	0.12 &
	0.21 &
	1.17 &
	0\\\hline
	&
	SIMEX-Q** &
	29.45 &
	0.22 &
	0.24 &
	0.43 &
	1.97 &
	1\\\hline
	&
	SIMEX-C** &
	29.45 &
	0.22 &
	0.45 &
	0.8 &
	3.81 &
	1\\\hline
	&
	SIMEX-Q4** &
	29.45 &
	0.22 &
	0.99 &
	1.77 &
	8.64 &
	1\\\hline
	&
	MI &
	42.56 &
	0.19 &
	0.11 &
	0.19 &
	0.9 &
	0\\\hline
	&
	MI-NULL*** &
	42.25 &
	0.19 &
	0.12 &
	0.22 &
	0.93 &
	0\\\hline
	&
	GS7 &
	27.5 &
	0.4 &
	0.1 &
	0.18 &
	0.95 &
	0\\\hline
	&
	GS28 &
	15.46 &
	0.63 &
	0.07 &
	0.12 &
	0.76 &
	0\\\hline
\end{tabular}
\end{table}

*Classifications were considered failed if a cluster was empty.**For SIMEX-3-SA different extrapolation functions were used: quadratic (Q), cubic (C) and quartic (Q4).***MI-NULL denotes the MI method without considering the health outcome in the ME model.

\clearpage
\begin{table}[h]
	\caption{Simulation results of correction methods and gold standard methods based on 7 and 28 days (GS7and GS28) for scenarios with outcome  $H_i^{A-\mathit{cat}}$  and cluster method with \textit{C} clusters regarding empirical misclassification rate (MR), adjusted Rand index (aRI), mean absolute bias, maximal absolute bias and median mean relative absolute bias, and number of failed classifications* for 1000 simulation data sets.}
	\tiny
	\begin{tabular}{|l||l|l|l|l|l|l|l|}
		\hline
		C &
		Correction method &
		MR &
		aRI &
		$\overline{{\Delta}}$ &
		$\overline{{\Delta}}_{\mathit{max}}$ &
		$\mathit{med}(\overline{{\Delta}}_{\mathit{rel}})$ &
		Number of failed classifications*\\\hline
		\multicolumn{8}{c}{Cluster method: k-means}\\\hline
		3 &
		naïve &
		37.4 &
		0.2 &
		0.22 &
		0.33 &
		1.21 &
		0\\\hline
		&
		RC &
		36.25 &
		0.22 &
		0.23 &
		0.31 &
		1.19 &
		0\\\hline
		&
		SIMEX-Q** &
		37.39 &
		0.2 &
		0.33 &
		0.44 &
		1.68 &
		1\\\hline
		&
		SIMEX-C** &
		37.39 &
		0.2 &
		0.62 &
		0.84 &
		3.34 &
		1\\\hline
		&
		SIMEX-Q4** &
		37.39 &
		0.2 &
		1.39 &
		1.89 &
		7.57 &
		1\\\hline
		&
		MI &
		36.25 &
		0.22 &
		0.2 &
		0.27 &
		0.85 &
		0\\\hline
		&
		MI-NULL*** &
		36.24 &
		0.22 &
		0.22 &
		0.3 &
		0.91 &
		0\\\hline
		&
		GS7 &
		24.95 &
		0.4 &
		0.17 &
		0.23 &
		0.89 &
		0\\\hline
		&
		GS28 &
		13.3 &
		0.64 &
		0.11 &
		0.15 &
		0.68 &
		0\\\hline
		4 &
		naïve &
		46.5 &
		0.15 &
		0.26 &
		0.48 &
		1.43 &
		0\\\hline
		&
		RC &
		44.26 &
		0.18 &
		0.26 &
		0.41 &
		1.26 &
		0\\\hline
		&
		SIMEX-Q** &
		46.5 &
		0.15 &
		0.39 &
		0.61 &
		1.91 &
		1\\\hline
		&
		SIMEX-C** &
		46.5 &
		0.15 &
		0.77 &
		1.2 &
		3.93 &
		1\\\hline
		&
		SIMEX-Q4** &
		46.5 &
		0.15 &
		1.71 &
		2.71 &
		8.69 &
		1\\\hline
		&
		MI &
		44.28 &
		0.18 &
		0.22 &
		0.35 &
		0.97 &
		0\\\hline
		&
		MI-NULL*** &
		44.26 &
		0.18 &
		0.24 &
		0.39 &
		0.93 &
		0\\\hline
		&
		GS7 &
		31.94 &
		0.34 &
		0.21 &
		0.33 &
		0.95 &
		0\\\hline
		&
		GS28 &
		17.49 &
		0.59 &
		0.15 &
		0.23 &
		0.77 &
		0\\\hline
		5 &
		naïve &
		52.8 &
		0.12 &
		0.3 &
		0.62 &
		1.79 &
		0\\\hline
		&
		RC &
		50.31 &
		0.15 &
		0.3 &
		0.51 &
		1.46 &
		0\\\hline
		&
		SIMEX-Q** &
		52.81 &
		0.12 &
		0.45 &
		0.78 &
		2.29 &
		1\\\hline
		&
		SIMEX-C** &
		52.81 &
		0.12 &
		0.87 &
		1.5 &
		4.67 &
		1\\\hline
		&
		SIMEX-Q4** &
		52.81 &
		0.12 &
		1.96 &
		3.37 &
		10.23 &
		1\\\hline
		&
		MI &
		50.37 &
		0.15 &
		0.25 &
		0.44 &
		0.98 &
		0\\\hline
		&
		MI-NULL*** &
		50.31 &
		0.15 &
		0.28 &
		0.48 &
		0.94 &
		0\\\hline
		&
		GS7 &
		37.17 &
		0.29 &
		0.24 &
		0.41 &
		1.24 &
		0\\\hline
		&
		GS28 &
		19.91 &
		0.56 &
		0.17 &
		0.3 &
		0.91 &
		0\\\hline
		\multicolumn{8}{c}{Cluster method: GMM}\\\hline
		3 &
		naïve &
		14.9 &
		0.25 &
		1.19 &
		1.78 &
		1.66 &
		1\\\hline
		&
		RC &
		29.61 &
		0.27 &
		0.27 &
		0.38 &
		1.25 &
		0\\\hline
		&
		SIMEX-Q** &
		14.86 &
		0.25 &
		1.86 &
		2.78 &
		2.51 &
		1\\\hline
		&
		SIMEX-C** &
		14.86 &
		0.25 &
		3.78 &
		5.64 &
		5.08 &
		1\\\hline
		&
		SIMEX-Q4** &
		14.86 &
		0.25 &
		8.78 &
		13.03 &
		12.81 &
		1\\\hline
		&
		MI &
		29.45 &
		0.27 &
		0.22 &
		0.32 &
		0.97 &
		0\\\hline
		&
		MI-NULL*** &
		29.46 &
		0.27 &
		0.23 &
		0.32 &
		0.92 &
		0\\\hline
		&
		GS7 &
		18.01 &
		0.47 &
		0.25 &
		0.35 &
		1.08 &
		0\\\hline
		&
		GS28 &
		10.01 &
		0.69 &
		0.14 &
		0.2 &
		0.78 &
		0\\\hline
		4 &
		naïve &
		19.2 &
		0.23 &
		1.34 &
		2.59 &
		1.88 &
		0\\\hline
		&
		RC &
		37.9 &
		0.21 &
		0.35 &
		0.58 &
		1.44 &
		0\\\hline
		&
		SIMEX-Q** &
		19.18 &
		0.23 &
		2.05 &
		3.74 &
		2.55 &
		1\\\hline
		&
		SIMEX-C** &
		19.18 &
		0.23 &
		4 &
		7.32 &
		5.56 &
		1\\\hline
		&
		SIMEX-Q4** &
		19.18 &
		0.23 &
		9.37 &
		17.45 &
		13.8 &
		1\\\hline
		&
		MI &
		37.97 &
		0.21 &
		0.29 &
		0.51 &
		1.03 &
		0\\\hline
		&
		MI-NULL*** &
		37.92 &
		0.21 &
		0.32 &
		0.53 &
		0.96 &
		0\\\hline
		&
		GS7 &
		22.65 &
		0.43 &
		0.35 &
		0.63 &
		1.15 &
		0\\\hline
		&
		GS28 &
		13 &
		0.66 &
		0.22 &
		0.38 &
		0.85 &
		0\\\hline
		5 &
		naïve &
		29.5 &
		0.22 &
		1.39 &
		3.21 &
		1.99 &
		0\\\hline
		&
		RC &
		42.31 &
		0.19 &
		0.52 &
		1.06 &
		1.51 &
		0\\\hline
		&
		SIMEX-Q** &
		29.45 &
		0.22 &
		2 &
		4.03 &
		2.76 &
		1\\\hline
		&
		SIMEX-C** &
		29.45 &
		0.22 &
		3.58 &
		7.02 &
		5.28 &
		1\\\hline
		&
		SIMEX-Q4** &
		29.45 &
		0.22 &
		7.8 &
		15.39 &
		13.28 &
		1\\\hline
		&
		MI &
		42.45 &
		0.19 &
		0.46 &
		0.92 &
		1.13 &
		0\\\hline
		&
		MI-NULL*** &
		42.25 &
		0.19 &
		0.47 &
		0.91 &
		1 &
		0\\\hline
		&
		GS7 &
		27.5 &
		0.4 &
		0.46 &
		1.02 &
		1.3 &
		0\\\hline
		&
		GS28 &
		15.46 &
		0.63 &
		0.33 &
		0.63 &
		1.03 &
		0\\\hline
	\end{tabular}
\end{table}
*Classifications were considered failed if a cluster was empty.**For SIMEX-3-SA different extrapolation functions were used: quadratic (Q), cubic (C) and quartic (Q4).***MI-NULL denotes the MI method without considering the health outcome in the ME model.

\clearpage\begin{table}[h]
	\caption{Simulation results of correction methods and gold standard methods based on 7 and 28 days (GS7 and GS28) for scenarios with outcome  $H_i^B$  and cluster method with \textit{C} clusters regarding empirical misclassification rate (MR), adjusted Rand index (aRI), mean absolute bias, maximal absolute bias and median mean relative absolute bias, and number of failed classifications* for 1000 simulation data sets.}
	\tiny
	\begin{tabular}{l|l|l|l|l|l|l|l}
		\hline
		C &
		Correction method &
		MR &
		aRI &
		$\overline{{\Delta}}$ &
		$\overline{{\Delta}}_{\mathit{max}}$ &
		$\mathit{med}(\overline{{\Delta}}_{\mathit{rel}})$ &
		Number of failed classifications*\\\hline
		\multicolumn{8}{c}{Cluster method: k-means}\\\hline
		3 &
		Naïve &
		37.4 &
		0.2 &
		0.11 &
		0.17 &
		0.97 &
		0\\\hline
		&
		RC &
		36.25 &
		0.22 &
		0.22 &
		0.3 &
		1.05 &
		0\\\hline
		&
		SIMEX-Q** &
		37.39 &
		0.2 &
		0.14 &
		0.18 &
		1.13 &
		1\\\hline
		&
		SIMEX-C** &
		37.39 &
		0.2 &
		0.22 &
		0.29 &
		1.85 &
		1\\\hline
		&
		SIMEX-Q4** &
		37.39 &
		0.2 &
		0.44 &
		0.6 &
		3.81 &
		1\\\hline
		&
		MI &
		36.21 &
		0.22 &
		0.21 &
		0.29 &
		0.99 &
		0\\\hline
		&
		MI-NULL*** &
		36.23 &
		0.22 &
		0.21 &
		0.3 &
		1 &
		0\\\hline
		&
		GS7 &
		24.95 &
		0.4 &
		0.05 &
		0.07 &
		0.45 &
		0\\\hline
		&
		GS28 &
		13.3 &
		0.64 &
		0.03 &
		0.05 &
		0.3 &
		0\\\hline
		4 &
		Naïve &
		46.5 &
		0.15 &
		0.14 &
		0.25 &
		1.25 &
		0\\\hline
		&
		RC &
		44.26 &
		0.18 &
		0.23 &
		0.35 &
		1.06 &
		0\\\hline
		&
		SIMEX-Q** &
		46.5 &
		0.15 &
		0.16 &
		0.26 &
		1.47 &
		1\\\hline
		&
		SIMEX-C** &
		46.5 &
		0.15 &
		0.26 &
		0.42 &
		2.58 &
		1\\\hline
		&
		SIMEX-Q4** &
		46.5 &
		0.15 &
		0.53 &
		0.85 &
		5.54 &
		1\\\hline
		&
		MI &
		44.23 &
		0.18 &
		0.23 &
		0.35 &
		0.98 &
		0\\\hline
		&
		MI-NULL*** &
		44.27 &
		0.18 &
		0.23 &
		0.35 &
		0.98 &
		0\\\hline
		&
		GS7 &
		31.94 &
		0.34 &
		0.06 &
		0.1 &
		0.6 &
		0\\\hline
		&
		GS28 &
		17.49 &
		0.59 &
		0.04 &
		0.07 &
		0.43 &
		0\\\hline
		5 &
		Naïve &
		52.8 &
		0.12 &
		0.18 &
		0.37 &
		1.52 &
		0\\\hline
		&
		RC &
		50.31 &
		0.15 &
		0.26 &
		0.42 &
		1.24 &
		0\\\hline
		&
		SIMEX-Q** &
		52.81 &
		0.12 &
		0.21 &
		0.37 &
		1.74 &
		1\\\hline
		&
		SIMEX-C** &
		52.81 &
		0.12 &
		0.32 &
		0.56 &
		2.9 &
		1\\\hline
		&
		SIMEX-Q4** &
		52.81 &
		0.12 &
		0.63 &
		1.09 &
		6.23 &
		1\\\hline
		&
		MI &
		50.28 &
		0.15 &
		0.24 &
		0.39 &
		1 &
		0\\\hline
		&
		MI-NULL*** &
		50.32 &
		0.15 &
		0.24 &
		0.39 &
		1 &
		0\\\hline
		&
		GS7 &
		37.17 &
		0.29 &
		0.08 &
		0.13 &
		0.64 &
		0\\\hline
		&
		GS28 &
		19.91 &
		0.56 &
		0.05 &
		0.08 &
		0.53 &
		0\\\hline
		\multicolumn{8}{c}{Cluster method: GMM}\\\hline
		3 &
		naïve &
		14.9 &
		0.25 &
		0.28 &
		0.42 &
		1.19 &
		1\\\hline
		&
		RC &
		29.61 &
		0.27 &
		0.2 &
		0.27 &
		0.89 &
		0\\\hline
		&
		SIMEX-Q** &
		14.86 &
		0.25 &
		0.33 &
		0.47 &
		1.41 &
		1\\\hline
		&
		SIMEX-C** &
		14.86 &
		0.25 &
		0.52 &
		0.72 &
		2.12 &
		1\\\hline
		&
		SIMEX-Q4** &
		14.86 &
		0.25 &
		1.06 &
		1.48 &
		4.45 &
		1\\\hline
		&
		MI &
		29.6 &
		0.27 &
		0.22 &
		0.29 &
		0.96 &
		0\\\hline
		&
		MI-NULL*** &
		29.46 &
		0.27 &
		0.21 &
		0.29 &
		0.95 &
		0\\\hline
		&
		GS7 &
		18.01 &
		0.47 &
		0.07 &
		0.09 &
		0.38 &
		0\\\hline
		&
		GS28 &
		10.01 &
		0.69 &
		0.04 &
		0.06 &
		0.24 &
		0\\\hline
		4 &
		naïve &
		19.2 &
		0.23 &
		0.29 &
		0.54 &
		1.4 &
		0\\\hline
		&
		RC &
		37.9 &
		0.21 &
		0.23 &
		0.36 &
		0.98 &
		0\\\hline
		&
		SIMEX-Q** &
		19.18 &
		0.23 &
		0.36 &
		0.58 &
		1.69 &
		1\\\hline
		&
		SIMEX-C** &
		19.18 &
		0.23 &
		0.56 &
		0.91 &
		2.91 &
		1\\\hline
		&
		SIMEX-Q4** &
		19.18 &
		0.23 &
		1.15 &
		1.85 &
		6.06 &
		1\\\hline
		&
		MI &
		37.88 &
		0.21 &
		0.24 &
		0.37 &
		0.97 &
		0\\\hline
		&
		MI-NULL*** &
		37.92 &
		0.21 &
		0.24 &
		0.37 &
		0.98 &
		0\\\hline
		&
		GS7 &
		22.65 &
		0.43 &
		0.09 &
		0.14 &
		0.58 &
		0\\\hline
		&
		GS28 &
		13 &
		0.66 &
		0.06 &
		0.09 &
		0.42 &
		0\\\hline
		5 &
		naïve &
		29.5 &
		0.22 &
		0.28 &
		0.62 &
		1.64 &
		0\\\hline
		&
		RC &
		42.31 &
		0.19 &
		0.26 &
		0.45 &
		1.17 &
		0\\\hline
		&
		SIMEX-Q** &
		29.45 &
		0.22 &
		0.35 &
		0.64 &
		1.88 &
		1\\\hline
		&
		SIMEX-C** &
		29.45 &
		0.22 &
		0.56 &
		1.01 &
		3.18 &
		1\\\hline
		&
		SIMEX-Q4** &
		29.45 &
		0.22 &
		1.15 &
		2.05 &
		6.94 &
		1\\\hline
		&
		MI &
		42.56 &
		0.19 &
		0.26 &
		0.44 &
		0.99 &
		0\\\hline
		&
		MI-NULL*** &
		42.25 &
		0.19 &
		0.26 &
		0.44 &
		0.98 &
		0\\\hline
		&
		GS7 &
		27.5 &
		0.4 &
		0.1 &
		0.19 &
		0.76 &
		0\\\hline
		&
		GS28 &
		15.46 &
		0.63 &
		0.07 &
		0.13 &
		0.54 &
		0\\\hline
	\end{tabular}
\end{table}

*Classifications were considered failed if a cluster was empty.**For SIMEX-3-SA different extrapolation functions were used: quadratic (Q), cubic (C) and quartic (Q4).***MI-NULL denotes the MI method without considering the health outcome in the ME model.

\clearpage\begin{table}[h]
	\caption{Simulation results of correction methods and gold standard methods based on 7 and 28 days (GS7 and GS28) for scenarios with outcome  $H_i^{B-\mathit{cat}}$  and cluster method with \textit{C} clusters regarding empirical misclassification rate (MR), adjusted Rand index (aRI), mean absolute bias, maximal absolute bias and median mean relative absolute bias, and number of failed classifications* for 1000 simulation data sets.}
	\tiny
	\begin{tabular}{l|l|l|l|l|l|l|l|l|l|}
		\hline
		\textit{C} &
		Correction method &
		MR &
		aRI &
		$\overline{{\Delta}}$ &
		$\overline{{\Delta}}_{\mathit{max}}$ &
		$\mathit{med}(\overline{{\Delta}}_{\mathit{rel}})$ &
		Number of failed classifications*\\\hline
		\multicolumn{8}{c}{Cluster method: k-means}\\\hline
		3 &
		naïve &
		37.4 &
		0.2 &
		0.27 &
		0.4 &
		1.17 &
		0\\\hline
		&
		RC &
		36.25 &
		0.22 &
		0.44 &
		0.61 &
		1.12 &
		0\\\hline
		&
		SIMEX-Q** &
		37.39 &
		0.2 &
		0.36 &
		0.5 &
		1.51 &
		1\\\hline
		&
		SIMEX-C** &
		37.39 &
		0.2 &
		0.64 &
		0.89 &
		2.67 &
		1\\\hline
		&
		SIMEX-Q4** &
		37.39 &
		0.2 &
		1.42 &
		1.95 &
		5.94 &
		1\\\hline
		&
		MI &
		36.31 &
		0.22 &
		0.41 &
		0.57 &
		1.03 &
		0\\\hline
		&
		MI-NULL*** &
		36.24 &
		0.22 &
		0.42 &
		0.58 &
		0.98 &
		0\\\hline
		&
		GS7 &
		24.95 &
		0.4 &
		0.16 &
		0.22 &
		0.73 &
		0\\\hline
		&
		GS28 &
		13.3 &
		0.64 &
		0.11 &
		0.15 &
		0.5 &
		0\\\hline
		4 &
		naïve &
		46.5 &
		0.15 &
		0.32 &
		0.59 &
		1.57 &
		0\\\hline
		&
		RC &
		44.26 &
		0.18 &
		0.47 &
		0.74 &
		1.2 &
		0\\\hline
		&
		SIMEX-Q** &
		46.5 &
		0.15 &
		0.44 &
		0.7 &
		2.05 &
		1\\\hline
		&
		SIMEX-C** &
		46.5 &
		0.15 &
		0.78 &
		1.23 &
		3.7 &
		1\\\hline
		&
		SIMEX-Q4** &
		46.5 &
		0.15 &
		1.69 &
		2.67 &
		8.43 &
		1\\\hline
		&
		MI &
		44.38 &
		0.18 &
		0.44 &
		0.69 &
		1.07 &
		0\\\hline
		&
		MI-NULL*** &
		44.28 &
		0.18 &
		0.45 &
		0.7 &
		0.99 &
		0\\\hline
		&
		GS7 &
		31.94 &
		0.34 &
		0.2 &
		0.31 &
		0.92 &
		0\\\hline
		&
		GS28 &
		17.49 &
		0.59 &
		0.14 &
		0.23 &
		0.69 &
		0\\\hline
		5 &
		naïve &
		52.8 &
		0.12 &
		0.41 &
		0.86 &
		1.78 &
		0\\\hline
		&
		RC &
		50.31 &
		0.15 &
		0.53 &
		0.89 &
		1.4 &
		0\\\hline
		&
		SIMEX-Q** &
		52.81 &
		0.12 &
		0.55 &
		0.95 &
		2.21 &
		1\\\hline
		&
		SIMEX-C** &
		52.81 &
		0.12 &
		0.94 &
		1.6 &
		4.11 &
		1\\\hline
		&
		SIMEX-Q4** &
		52.81 &
		0.12 &
		1.99 &
		3.39 &
		8.68 &
		1\\\hline
		&
		MI &
		50.49 &
		0.15 &
		0.48 &
		0.82 &
		1.12 &
		0\\\hline
		&
		MI-NULL*** &
		50.31 &
		0.15 &
		0.48 &
		0.81 &
		1 &
		0\\\hline
		&
		GS7 &
		37.17 &
		0.29 &
		0.23 &
		0.4 &
		0.96 &
		0\\\hline
		&
		GS28 &
		19.91 &
		0.56 &
		0.16 &
		0.28 &
		0.78 &
		0\\\hline
		\multicolumn{8}{c}{  Cluster method: GMM}\\\hline
		3 &
		naïve &
		14.9 &
		0.25 &
		1.29 &
		1.94 &
		1.35 &
		1\\\hline
		&
		RC &
		29.61 &
		0.27 &
		0.44 &
		0.61 &
		1.04 &
		0\\\hline
		&
		SIMEX-Q** &
		14.86 &
		0.25 &
		1.75 &
		2.59 &
		1.88 &
		1\\\hline
		&
		SIMEX-C** &
		14.86 &
		0.25 &
		3.14 &
		4.61 &
		3.62 &
		1\\\hline
		&
		SIMEX-Q4** &
		14.86 &
		0.25 &
		6.73 &
		9.93 &
		7.85 &
		1\\\hline
		&
		MI &
		29.5 &
		0.27 &
		0.41 &
		0.56 &
		0.91 &
		0\\\hline
		&
		MI-NULL*** &
		29.46 &
		0.27 &
		0.43 &
		0.59 &
		0.96 &
		0\\\hline
		&
		GS7 &
		18.01 &
		0.47 &
		0.27 &
		0.38 &
		0.68 &
		0\\\hline
		&
		GS28 &
		10.01 &
		0.69 &
		0.14 &
		0.2 &
		0.46 &
		0\\\hline
		4 &
		naïve &
		19.2 &
		0.23 &
		1.35 &
		2.6 &
		1.65 &
		0\\\hline
		&
		RC &
		37.9 &
		0.21 &
		0.53 &
		0.87 &
		1.12 &
		0\\\hline
		&
		SIMEX-Q** &
		19.18 &
		0.23 &
		1.87 &
		3.25 &
		2.19 &
		1\\\hline
		&
		SIMEX-C** &
		19.18 &
		0.23 &
		3.21 &
		5.7 &
		4.06 &
		1\\\hline
		&
		SIMEX-Q4** &
		19.18 &
		0.23 &
		6.98 &
		12.4 &
		10.57 &
		1\\\hline
		&
		MI &
		38.04 &
		0.21 &
		0.52 &
		0.85 &
		1.01 &
		0\\\hline
		&
		MI-NULL*** &
		37.92 &
		0.21 &
		0.53 &
		0.84 &
		0.99 &
		0\\\hline
		&
		GS7 &
		22.65 &
		0.43 &
		0.34 &
		0.64 &
		0.95 &
		0\\\hline
		&
		GS28 &
		13 &
		0.66 &
		0.19 &
		0.3 &
		0.69 &
		0\\\hline
		5 &
		naïve &
		29.5 &
		0.22 &
		1.34 &
		3.1 &
		1.87 &
		0\\\hline
		&
		RC &
		42.31 &
		0.19 &
		0.69 &
		1.39 &
		1.28 &
		0\\\hline
		&
		SIMEX-Q** &
		29.45 &
		0.22 &
		1.83 &
		3.65 &
		2.46 &
		1\\\hline
		&
		SIMEX-C** &
		29.45 &
		0.22 &
		3.1 &
		6.07 &
		4.93 &
		1\\\hline
		&
		SIMEX-Q4** &
		29.45 &
		0.22 &
		6.69 &
		13.07 &
		11.59 &
		1\\\hline
		&
		MI &
		42.7 &
		0.19 &
		0.68 &
		1.26 &
		1.13 &
		0\\\hline
		&
		MI-NULL*** &
		42.25 &
		0.19 &
		0.66 &
		1.25 &
		1.03 &
		0\\\hline
		&
		GS7 &
		27.5 &
		0.4 &
		0.46 &
		0.93 &
		1.08 &
		0\\\hline
		&
		GS28 &
		15.46 &
		0.63 &
		0.35 &
		0.74 &
		0.85 &
		0\\\hline
	\end{tabular}
\end{table}
*Classifications were considered failed if a cluster was empty.**For SIMEX-3-SA different extrapolation functions were used: quadratic (Q), cubic (C) and quartic (Q4).***MI-NULL denotes the MI method without considering the health outcome in the ME model.

\clearpage
\subsection{Simulation for  $M=5$  and  $\mathit{Corr}\left(u_{\mathit{ki}},u_{\mathit{li}}\right)=0$  for  $k{\neq}l$}

\begin{table}[h]
	 \caption{Simulation results of correction methods and gold standard methods based on 7 and 28 days (GS7 and GS28) for scenarios with outcome  $H_i^A$  and cluster method with \textit{C} clusters regarding empirical misclassification rate (MR), adjusted Rand index (aRI), mean absolute bias, maximal absolute bias and median mean relative absolute bias, and number of failed classifications* for 1000 simulation data sets.}
	\tiny \begin{tabular}{l|l|l|l|l|l|l|l}
		\hline
		\textit{C} &
		Correction method &
		MR &
		aRI &
		$\overline{{\Delta}}$ &
		$\overline{{\Delta}}_{\mathit{max}}$ &
		$\mathit{med}(\overline{{\Delta}}_{\mathit{rel}})$ &
		Number of failed classifications*\\\hline
		\multicolumn{8}{c}{  Cluster method: k-means}\\\hline
		3 &
		naïve &
		39.8 &
		0.16 &
		0.07 &
		0.11 &
		0.85 &
		0\\\hline
		&
		RC &
		36.82 &
		0.21 &
		0.08 &
		0.11 &
		0.77 &
		0\\\hline
		&
		SIMEX-Q** &
		39.72 &
		0.16 &
		0.1 &
		0.13 &
		1.11 &
		1\\\hline
		&
		SIMEX-C** &
		39.72 &
		0.16 &
		0.18 &
		0.25 &
		2.15 &
		1\\\hline
		&
		SIMEX-Q4** &
		39.72 &
		0.16 &
		0.41 &
		0.56 &
		4.88 &
		1\\\hline
		&
		MI &
		36.71 &
		0.21 &
		0.08 &
		0.11 &
		0.66 &
		0\\\hline
		&
		MI-NULL*** &
		36.81 &
		0.21 &
		0.1 &
		0.14 &
		0.88 &
		0\\\hline
		&
		GS7 &
		27.13 &
		0.35 &
		0.06 &
		0.08 &
		0.57 &
		0\\\hline
		&
		GS28 &
		14.78 &
		0.61 &
		0.04 &
		0.05 &
		0.4 &
		0\\\hline
		4 &
		naïve &
		47.6 &
		0.13 &
		0.09 &
		0.16 &
		1.04 &
		0\\\hline
		&
		RC &
		45.01 &
		0.17 &
		0.09 &
		0.14 &
		0.88 &
		0\\\hline
		&
		SIMEX-Q** &
		47.59 &
		0.13 &
		0.12 &
		0.18 &
		1.27 &
		1\\\hline
		&
		SIMEX-C** &
		47.59 &
		0.13 &
		0.22 &
		0.35 &
		2.44 &
		1\\\hline
		&
		SIMEX-Q4** &
		47.59 &
		0.13 &
		0.48 &
		0.77 &
		5.55 &
		1\\\hline
		&
		MI &
		44.96 &
		0.17 &
		0.09 &
		0.13 &
		0.73 &
		0\\\hline
		&
		MI-NULL*** &
		45.04 &
		0.17 &
		0.11 &
		0.18 &
		0.89 &
		0\\\hline
		&
		GS7 &
		33.32 &
		0.31 &
		0.07 &
		0.1 &
		0.69 &
		0\\\hline
		&
		GS28 &
		18.53 &
		0.57 &
		0.05 &
		0.07 &
		0.46 &
		0\\\hline
		5 &
		naïve &
		53 &
		0.11 &
		0.1 &
		0.2 &
		1.2 &
		0\\\hline
		&
		RC &
		50.6 &
		0.14 &
		0.1 &
		0.17 &
		1.01 &
		0\\\hline
		&
		SIMEX-Q** &
		53.01 &
		0.11 &
		0.13 &
		0.23 &
		1.51 &
		1\\\hline
		&
		SIMEX-C** &
		53.01 &
		0.11 &
		0.25 &
		0.44 &
		2.96 &
		1\\\hline
		&
		SIMEX-Q4** &
		53.01 &
		0.11 &
		0.56 &
		0.97 &
		6.43 &
		1\\\hline
		&
		MI &
		50.5 &
		0.14 &
		0.09 &
		0.16 &
		0.78 &
		0\\\hline
		&
		MI-NULL*** &
		50.61 &
		0.14 &
		0.12 &
		0.2 &
		0.9 &
		0\\\hline
		&
		GS7 &
		37.6 &
		0.28 &
		0.07 &
		0.13 &
		0.76 &
		0\\\hline
		&
		GS28 &
		21.15 &
		0.54 &
		0.05 &
		0.09 &
		0.62 &
		0\\\hline
		\multicolumn{8}{c}{  Cluster method: GMM}\\\hline
		3 &
		naïve &
		14.8 &
		0.2 &
		0.2 &
		0.3 &
		1.16 &
		5\\\hline
		&
		RC &
		31.47 &
		0.25 &
		0.09 &
		0.13 &
		0.87 &
		0\\\hline
		&
		SIMEX-Q** &
		14.83 &
		0.2 &
		0.27 &
		0.39 &
		1.64 &
		5\\\hline
		&
		SIMEX-C** &
		14.83 &
		0.2 &
		0.5 &
		0.7 &
		3.14 &
		5\\\hline
		&
		SIMEX-Q4** &
		14.83 &
		0.2 &
		1.1 &
		1.54 &
		6.98 &
		5\\\hline
		&
		MI &
		31.6 &
		0.25 &
		0.08 &
		0.11 &
		0.68 &
		0\\\hline
		&
		MI-NULL*** &
		31.43 &
		0.25 &
		0.11 &
		0.15 &
		0.88 &
		0\\\hline
		&
		GS7 &
		12.69 &
		0.44 &
		0.12 &
		0.16 &
		0.75 &
		0\\\hline
		&
		GS28 &
		7.98 &
		0.68 &
		0.07 &
		0.1 &
		0.54 &
		0\\\hline
		4 &
		naïve &
		16.8 &
		0.19 &
		0.22 &
		0.4 &
		1.26 &
		1\\\hline
		&
		RC &
		37.59 &
		0.21 &
		0.1 &
		0.17 &
		0.97 &
		0\\\hline
		&
		SIMEX-Q** &
		16.81 &
		0.19 &
		0.3 &
		0.48 &
		1.63 &
		1\\\hline
		&
		SIMEX-C** &
		16.81 &
		0.19 &
		0.55 &
		0.88 &
		3.2 &
		1\\\hline
		&
		SIMEX-Q4** &
		16.81 &
		0.19 &
		1.18 &
		1.89 &
		7.19 &
		1\\\hline
		&
		MI &
		37.53 &
		0.21 &
		0.1 &
		0.16 &
		0.73 &
		0\\\hline
		&
		MI-NULL*** &
		37.54 &
		0.21 &
		0.12 &
		0.2 &
		0.89 &
		0\\\hline
		&
		GS7 &
		17.16 &
		0.41 &
		0.12 &
		0.19 &
		0.81 &
		0\\\hline
		&
		GS28 &
		10.85 &
		0.65 &
		0.08 &
		0.13 &
		0.63 &
		0\\\hline
		5 &
		naïve &
		22.8 &
		0.2 &
		0.2 &
		0.42 &
		1.53 &
		0\\\hline
		&
		RC &
		42.29 &
		0.19 &
		0.12 &
		0.22 &
		1.03 &
		0\\\hline
		&
		SIMEX-Q** &
		22.82 &
		0.2 &
		0.29 &
		0.5 &
		2.02 &
		1\\\hline
		&
		SIMEX-C** &
		22.82 &
		0.2 &
		0.54 &
		0.94 &
		3.9 &
		1\\\hline
		&
		SIMEX-Q4** &
		22.82 &
		0.2 &
		1.19 &
		2.05 &
		8.38 &
		1\\\hline
		&
		MI &
		42.6 &
		0.19 &
		0.11 &
		0.2 &
		0.79 &
		0\\\hline
		&
		MI-NULL*** &
		42.44 &
		0.19 &
		0.15 &
		0.25 &
		0.91 &
		0\\\hline
		&
		GS7 &
		26.28 &
		0.38 &
		0.11 &
		0.2 &
		0.94 &
		0\\\hline
		&
		GS28 &
		15.94 &
		0.61 &
		0.07 &
		0.13 &
		0.62 &
		0\\\hline
	\end{tabular}
\end{table}

*Classifications were considered failed if a cluster was empty.**For SIMEX-3-SA different extrapolation functions were used: quadratic (Q), cubic (C) and quartic (Q4).***MI-NULL denotes the MI method without considering the health outcome in the ME model.

\clearpage
\begin{table}[h]
	 \caption{Simulation results of correction methods and gold standard methods based on 7 and 28 days (GS7 and GS28) for scenarios with outcome  $H_i^{A-\mathit{cat}}$  and cluster method with \textit{C} clusters regarding empirical misclassification rate (MR), adjusted Rand index (aRI), mean absolute bias, maximal absolute bias and median mean relative absolute bias, and number of failed classifications* for 1000 simulation data sets.}

	\tiny\begin{tabular}{l|l|l|l|l|l|l|l}
		\hline
		\textit{C} &
		Correction method &
		MR &
		aRI &
		$\overline{{\Delta}}$ &
		$\overline{{\Delta}}_{\mathit{max}}$ &
		$\mathit{med}(\overline{{\Delta}}_{\mathit{rel}})$ &
		Number of failed classifications*\\\hline
		\multicolumn{8}{c}{  Cluster method: k-means}\\\hline
		3 &
		naïve &
		39.8 &
		0.16 &
		0.23 &
		0.34 &
		1.12 &
		0\\\hline
		&
		RC &
		36.82 &
		0.21 &
		0.24 &
		0.33 &
		1.09 &
		0\\\hline
		&
		SIMEX-Q** &
		39.72 &
		0.16 &
		0.34 &
		0.46 &
		1.51 &
		1\\\hline
		&
		SIMEX-C** &
		39.72 &
		0.16 &
		0.63 &
		0.87 &
		2.9 &
		1\\\hline
		&
		SIMEX-Q4** &
		39.72 &
		0.16 &
		1.41 &
		1.92 &
		6.76 &
		1\\\hline
		&
		MI &
		36.8 &
		0.21 &
		0.21 &
		0.28 &
		0.83 &
		0\\\hline
		&
		MI-NULL*** &
		36.8 &
		0.21 &
		0.24 &
		0.33 &
		0.91 &
		0\\\hline
		&
		GS7 &
		27.13 &
		0.35 &
		0.18 &
		0.25 &
		0.81 &
		0\\\hline
		&
		GS28 &
		14.78 &
		0.61 &
		0.13 &
		0.18 &
		0.62 &
		0\\\hline
		4 &
		naïve &
		47.6 &
		0.13 &
		0.27 &
		0.48 &
		1.37 &
		0\\\hline
		&
		RC &
		45.01 &
		0.17 &
		0.27 &
		0.42 &
		1.21 &
		0\\\hline
		&
		SIMEX-Q** &
		47.59 &
		0.13 &
		0.4 &
		0.63 &
		1.84 &
		1\\\hline
		&
		SIMEX-C** &
		47.59 &
		0.13 &
		0.77 &
		1.21 &
		3.6 &
		1\\\hline
		&
		SIMEX-Q4** &
		47.59 &
		0.13 &
		1.73 &
		2.74 &
		8.02 &
		1\\\hline
		&
		MI &
		45.05 &
		0.16 &
		0.23 &
		0.37 &
		0.89 &
		0\\\hline
		&
		MI-NULL*** &
		45.01 &
		0.17 &
		0.27 &
		0.43 &
		0.92 &
		0\\\hline
		&
		GS7 &
		33.32 &
		0.31 &
		0.22 &
		0.34 &
		1.02 &
		0\\\hline
		&
		GS28 &
		18.53 &
		0.57 &
		0.16 &
		0.25 &
		0.7 &
		0\\\hline
		5 &
		naïve &
		53 &
		0.11 &
		0.3 &
		0.62 &
		1.64 &
		0\\\hline
		&
		RC &
		50.6 &
		0.14 &
		0.3 &
		0.52 &
		1.32 &
		0\\\hline
		&
		SIMEX-Q** &
		53.01 &
		0.11 &
		0.46 &
		0.79 &
		2.09 &
		1\\\hline
		&
		SIMEX-C** &
		53.01 &
		0.11 &
		0.89 &
		1.54 &
		4.25 &
		1\\\hline
		&
		SIMEX-Q4** &
		53.01 &
		0.11 &
		1.98 &
		3.42 &
		9.51 &
		1\\\hline
		&
		MI &
		50.62 &
		0.14 &
		0.26 &
		0.46 &
		0.93 &
		0\\\hline
		&
		MI-NULL*** &
		50.6 &
		0.14 &
		0.3 &
		0.52 &
		0.93 &
		0\\\hline
		&
		GS7 &
		37.6 &
		0.28 &
		0.24 &
		0.42 &
		1.11 &
		0\\\hline
		&
		GS28 &
		21.15 &
		0.54 &
		0.18 &
		0.31 &
		0.87 &
		0\\\hline
		\multicolumn{8}{c}{  Cluster method: GMM}\\\hline
		3 &
		naïve &
		14.8 &
		0.2 &
		2.12 &
		3.19 &
		1.37 &
		5\\\hline
		&
		RC &
		31.47 &
		0.25 &
		0.35 &
		0.47 &
		1.12 &
		0\\\hline
		&
		SIMEX-Q** &
		14.83 &
		0.2 &
		2.91 &
		4.42 &
		2.13 &
		5\\\hline
		&
		SIMEX-C** &
		14.83 &
		0.2 &
		5.55 &
		8.44 &
		4.53 &
		5\\\hline
		&
		SIMEX-Q4** &
		14.83 &
		0.2 &
		12.08 &
		18.45 &
		11.23 &
		5\\\hline
		&
		MI &
		31.62 &
		0.25 &
		0.25 &
		0.35 &
		0.85 &
		0\\\hline
		&
		MI-NULL*** &
		31.43 &
		0.25 &
		0.33 &
		0.44 &
		0.9 &
		0\\\hline
		&
		GS7 &
		12.69 &
		0.44 &
		0.89 &
		1.31 &
		1.02 &
		0\\\hline
		&
		GS28 &
		7.98 &
		0.68 &
		0.46 &
		0.7 &
		0.83 &
		0\\\hline
		4 &
		naïve &
		16.8 &
		0.19 &
		1.91 &
		3.68 &
		1.72 &
		1\\\hline
		&
		RC &
		37.59 &
		0.21 &
		0.39 &
		0.63 &
		1.24 &
		0\\\hline
		&
		SIMEX-Q** &
		16.81 &
		0.19 &
		2.74 &
		4.96 &
		2.31 &
		1\\\hline
		&
		SIMEX-C** &
		16.81 &
		0.19 &
		5.23 &
		9.43 &
		5.07 &
		1\\\hline
		&
		SIMEX-Q4** &
		16.81 &
		0.19 &
		11.7 &
		21.19 &
		12.92 &
		1\\\hline
		&
		MI &
		37.69 &
		0.21 &
		0.35 &
		0.57 &
		0.9 &
		0\\\hline
		&
		MI-NULL*** &
		37.54 &
		0.21 &
		0.39 &
		0.65 &
		0.93 &
		0\\\hline
		&
		GS7 &
		17.16 &
		0.41 &
		0.85 &
		1.44 &
		1.14 &
		0\\\hline
		&
		GS28 &
		10.85 &
		0.65 &
		0.52 &
		0.88 &
		0.93 &
		0\\\hline
		5 &
		naïve &
		22.8 &
		0.2 &
		1.47 &
		3.38 &
		1.93 &
		0\\\hline
		&
		RC &
		42.29 &
		0.19 &
		0.49 &
		0.95 &
		1.32 &
		0\\\hline
		&
		SIMEX-Q** &
		22.82 &
		0.2 &
		2.14 &
		4.49 &
		2.49 &
		1\\\hline
		&
		SIMEX-C** &
		22.82 &
		0.2 &
		3.98 &
		8.32 &
		5.27 &
		1\\\hline
		&
		SIMEX-Q4** &
		22.82 &
		0.2 &
		8.99 &
		18.72 &
		13.04 &
		1\\\hline
		&
		MI &
		42.61 &
		0.19 &
		0.41 &
		0.8 &
		0.95 &
		0\\\hline
		&
		MI-NULL*** &
		42.44 &
		0.19 &
		0.49 &
		0.91 &
		0.94 &
		0\\\hline
		&
		GS7 &
		26.28 &
		0.38 &
		0.56 &
		1.03 &
		1.2 &
		0\\\hline
		&
		GS28 &
		15.94 &
		0.61 &
		0.32 &
		0.63 &
		0.92 &
		0\\\hline
	\end{tabular}
\end{table}
*Classifications were considered failed if a cluster was empty.**For SIMEX-3-SA different extrapolation functions were used: quadratic (Q), cubic (C) and quartic (Q4).***MI-NULL denotes the MI method without considering the health outcome in the ME model.

\clearpage\begin{table}[h]
	 \caption{Simulation results of correction methods and gold standard methods based on 7 and 28 days (GS7 and GS28) for scenarios with outcome  $H_i^B$  and cluster method with \textit{C} clusters regarding empirical misclassification rate (MR), adjusted Rand index (aRI), mean absolute bias, maximal absolute bias and median mean relative absolute bias, and number of failed classifications* for 1000 simulation data sets.}

	\tiny\begin{tabular}{l|l|l|l|l|l|l|l}
		\hline
		\textit{C} &
		Correction method &
		MR &
		aRI &
		$\overline{{\Delta}}$ &
		$\overline{{\Delta}}_{\mathit{max}}$ &
		$\mathit{med}(\overline{{\Delta}}_{\mathit{rel}})$ &
		Number of failed classifications*\\\hline
		\multicolumn{8}{c}{  Cluster method: k-means}\\\hline
		3 &
		naïve &
		39.8 &
		0.16 &
		0.23 &
		0.35 &
		0.91 &
		0\\\hline
		&
		RC &
		36.82 &
		0.21 &
		0.29 &
		0.41 &
		0.83 &
		0\\\hline
		&
		SIMEX-Q** &
		39.72 &
		0.16 &
		0.23 &
		0.32 &
		0.92 &
		1\\\hline
		&
		SIMEX-C** &
		39.72 &
		0.16 &
		0.29 &
		0.4 &
		1.22 &
		1\\\hline
		&
		SIMEX-Q4** &
		39.72 &
		0.16 &
		0.5 &
		0.68 &
		2.14 &
		1\\\hline
		&
		MI &
		36.72 &
		0.21 &
		0.35 &
		0.48 &
		0.94 &
		0\\\hline
		&
		MI-NULL*** &
		36.8 &
		0.21 &
		0.35 &
		0.48 &
		0.94 &
		0\\\hline
		&
		GS7 &
		27.13 &
		0.35 &
		0.08 &
		0.11 &
		0.28 &
		0\\\hline
		&
		GS28 &
		14.78 &
		0.61 &
		0.04 &
		0.06 &
		0.16 &
		0\\\hline
		4 &
		naïve &
		47.6 &
		0.13 &
		0.23 &
		0.41 &
		0.99 &
		0\\\hline
		&
		RC &
		45.01 &
		0.17 &
		0.29 &
		0.45 &
		0.87 &
		0\\\hline
		&
		SIMEX-Q** &
		47.59 &
		0.13 &
		0.23 &
		0.36 &
		1.03 &
		1\\\hline
		&
		SIMEX-C** &
		47.59 &
		0.13 &
		0.31 &
		0.48 &
		1.49 &
		1\\\hline
		&
		SIMEX-Q4** &
		47.59 &
		0.13 &
		0.57 &
		0.89 &
		2.95 &
		1\\\hline
		&
		MI &
		44.95 &
		0.17 &
		0.33 &
		0.51 &
		0.94 &
		0\\\hline
		&
		MI-NULL*** &
		45.06 &
		0.16 &
		0.33 &
		0.52 &
		0.94 &
		0\\\hline
		&
		GS7 &
		33.32 &
		0.31 &
		0.09 &
		0.14 &
		0.36 &
		0\\\hline
		&
		GS28 &
		18.53 &
		0.57 &
		0.05 &
		0.08 &
		0.21 &
		0\\\hline
		5 &
		naïve &
		53 &
		0.11 &
		0.25 &
		0.51 &
		1.05 &
		0\\\hline
		&
		RC &
		50.6 &
		0.14 &
		0.3 &
		0.5 &
		0.91 &
		0\\\hline
		&
		SIMEX-Q** &
		53.01 &
		0.11 &
		0.26 &
		0.45 &
		1.12 &
		1\\\hline
		&
		SIMEX-C** &
		53.01 &
		0.11 &
		0.35 &
		0.6 &
		1.72 &
		1\\\hline
		&
		SIMEX-Q4** &
		53.01 &
		0.11 &
		0.66 &
		1.13 &
		3.48 &
		1\\\hline
		&
		MI &
		50.53 &
		0.14 &
		0.34 &
		0.56 &
		0.95 &
		0\\\hline
		&
		MI-NULL*** &
		50.61 &
		0.14 &
		0.34 &
		0.56 &
		0.95 &
		0\\\hline
		&
		GS7 &
		37.6 &
		0.28 &
		0.09 &
		0.16 &
		0.39 &
		0\\\hline
		&
		GS28 &
		21.15 &
		0.54 &
		0.06 &
		0.1 &
		0.27 &
		0\\\hline
		\multicolumn{8}{c}{  Cluster method: GMM}\\\hline
		3 &
		naïve &
		14.8 &
		0.2 &
		0.39 &
		0.59 &
		1.07 &
		5\\\hline
		&
		RC &
		31.47 &
		0.25 &
		0.31 &
		0.43 &
		0.83 &
		0\\\hline
		&
		SIMEX-Q** &
		14.83 &
		0.2 &
		0.44 &
		0.61 &
		1.28 &
		5\\\hline
		&
		SIMEX-C** &
		14.83 &
		0.2 &
		0.65 &
		0.9 &
		1.91 &
		5\\\hline
		&
		SIMEX-Q4** &
		14.83 &
		0.2 &
		1.26 &
		1.77 &
		3.86 &
		5\\\hline
		&
		MI &
		31.6 &
		0.25 &
		0.36 &
		0.49 &
		0.94 &
		0\\\hline
		&
		MI-NULL*** &
		31.43 &
		0.25 &
		0.36 &
		0.5 &
		0.94 &
		0\\\hline
		&
		GS7 &
		12.69 &
		0.44 &
		0.14 &
		0.2 &
		0.38 &
		0\\\hline
		&
		GS28 &
		7.98 &
		0.68 &
		0.08 &
		0.11 &
		0.21 &
		0\\\hline
		4 &
		naïve &
		16.8 &
		0.19 &
		0.42 &
		0.77 &
		1.16 &
		1\\\hline
		&
		RC &
		37.59 &
		0.21 &
		0.34 &
		0.52 &
		0.84 &
		0\\\hline
		&
		SIMEX-Q** &
		16.81 &
		0.19 &
		0.47 &
		0.75 &
		1.42 &
		1\\\hline
		&
		SIMEX-C** &
		16.81 &
		0.19 &
		0.69 &
		1.1 &
		2.28 &
		1\\\hline
		&
		SIMEX-Q4** &
		16.81 &
		0.19 &
		1.37 &
		2.17 &
		4.57 &
		1\\\hline
		&
		MI &
		37.53 &
		0.21 &
		0.4 &
		0.61 &
		0.94 &
		0\\\hline
		&
		MI-NULL*** &
		37.54 &
		0.21 &
		0.4 &
		0.61 &
		0.94 &
		0\\\hline
		&
		GS7 &
		17.16 &
		0.41 &
		0.15 &
		0.24 &
		0.47 &
		0\\\hline
		&
		GS28 &
		10.85 &
		0.65 &
		0.09 &
		0.15 &
		0.28 &
		0\\\hline
		5 &
		naïve &
		22.8 &
		0.2 &
		0.44 &
		0.94 &
		1.19 &
		0\\\hline
		&
		RC &
		42.29 &
		0.19 &
		0.34 &
		0.57 &
		0.89 &
		0\\\hline
		&
		SIMEX-Q** &
		22.82 &
		0.2 &
		0.48 &
		0.85 &
		1.39 &
		1\\\hline
		&
		SIMEX-C** &
		22.82 &
		0.2 &
		0.68 &
		1.22 &
		2.19 &
		1\\\hline
		&
		SIMEX-Q4** &
		22.82 &
		0.2 &
		1.31 &
		2.32 &
		4.37 &
		1\\\hline
		&
		MI &
		42.6 &
		0.19 &
		0.4 &
		0.66 &
		0.94 &
		0\\\hline
		&
		MI-NULL*** &
		42.44 &
		0.19 &
		0.4 &
		0.67 &
		0.94 &
		0\\\hline
		&
		GS7 &
		26.28 &
		0.38 &
		0.13 &
		0.23 &
		0.47 &
		0\\\hline
		&
		GS28 &
		15.94 &
		0.61 &
		0.08 &
		0.14 &
		0.31 &
		0\\\hline
	\end{tabular}
\end{table}
*Classifications were considered failed if a cluster was empty.**For SIMEX-3-SA different extrapolation functions were used: quadratic (Q), cubic (C) and quartic (Q4).***MI-NULL denotes the MI method without considering the health outcome in the ME model.

\clearpage\begin{table}[h]
	 \caption{Simulation results of correction methods and gold standard methods based on 7 and 28 days (GS7 and GS28) for scenarios with outcome  $H_i^{B-\mathit{cat}}$  and cluster method with \textit{C} clusters regarding empirical misclassification rate (MR), adjusted Rand index (aRI), mean absolute bias, maximal absolute bias and median mean relative absolute bias, and number of failed classifications* for 1000 simulation data sets.}
	 
	\tiny\begin{tabular}{l|l|l|l|l|l|l|l}
		\hline
		\textit{C} &
		Correction method &
		MR &
		aRI &
		$\overline{{\Delta}}$ &
		$\overline{{\Delta}}_{\mathit{max}}$ &
		$\mathit{med}(\overline{{\Delta}}_{\mathit{rel}})$ &
		Number of failed classifications*\\\hline
		\multicolumn{8}{c}{  Cluster method: k-means}\\\hline
		3 &
		naïve &
		39.8 &
		0.16 &
		0.47 &
		0.71 &
		1.03 &
		0\\\hline
		&
		RC &
		36.82 &
		0.21 &
		0.59 &
		0.81 &
		0.88 &
		0\\\hline
		&
		SIMEX-Q** &
		39.72 &
		0.16 &
		0.52 &
		0.72 &
		1.15 &
		1\\\hline
		&
		SIMEX-C** &
		39.72 &
		0.16 &
		0.77 &
		1.06 &
		1.78 &
		1\\\hline
		&
		SIMEX-Q4** &
		39.72 &
		0.16 &
		1.5 &
		2.05 &
		3.56 &
		1\\\hline
		&
		MI &
		36.68 &
		0.21 &
		0.6 &
		0.83 &
		0.83 &
		0\\\hline
		&
		MI-NULL*** &
		36.82 &
		0.21 &
		0.67 &
		0.93 &
		0.94 &
		0\\\hline
		&
		GS7 &
		27.13 &
		0.35 &
		0.21 &
		0.28 &
		0.41 &
		0\\\hline
		&
		GS28 &
		14.78 &
		0.61 &
		0.14 &
		0.19 &
		0.26 &
		0\\\hline
		4 &
		naïve &
		47.6 &
		0.13 &
		0.47 &
		0.87 &
		1.21 &
		0\\\hline
		&
		RC &
		45.01 &
		0.17 &
		0.6 &
		0.94 &
		0.93 &
		0\\\hline
		&
		SIMEX-Q** &
		47.59 &
		0.13 &
		0.55 &
		0.87 &
		1.37 &
		1\\\hline
		&
		SIMEX-C** &
		47.59 &
		0.13 &
		0.87 &
		1.36 &
		2.31 &
		1\\\hline
		&
		SIMEX-Q4** &
		47.59 &
		0.13 &
		1.77 &
		2.76 &
		4.9 &
		1\\\hline
		&
		MI &
		44.87 &
		0.17 &
		0.6 &
		0.94 &
		0.87 &
		0\\\hline
		&
		MI-NULL*** &
		45.04 &
		0.16 &
		0.66 &
		1.03 &
		0.95 &
		0\\\hline
		&
		GS7 &
		33.32 &
		0.31 &
		0.24 &
		0.38 &
		0.56 &
		0\\\hline
		&
		GS28 &
		18.53 &
		0.57 &
		0.16 &
		0.26 &
		0.34 &
		0\\\hline
		5 &
		naïve &
		53 &
		0.11 &
		0.53 &
		1.08 &
		1.33 &
		0\\\hline
		&
		RC &
		50.6 &
		0.14 &
		0.62 &
		1.05 &
		1.02 &
		0\\\hline
		&
		SIMEX-Q** &
		53.01 &
		0.11 &
		0.62 &
		1.07 &
		1.61 &
		1\\\hline
		&
		SIMEX-C** &
		53.01 &
		0.11 &
		0.98 &
		1.7 &
		2.72 &
		1\\\hline
		&
		SIMEX-Q4** &
		53.01 &
		0.11 &
		2.03 &
		3.51 &
		5.8 &
		1\\\hline
		&
		MI &
		50.49 &
		0.14 &
		0.61 &
		1.02 &
		0.88 &
		0\\\hline
		&
		MI-NULL*** &
		50.59 &
		0.14 &
		0.68 &
		1.13 &
		0.95 &
		0\\\hline
		&
		GS7 &
		37.6 &
		0.28 &
		0.27 &
		0.47 &
		0.61 &
		0\\\hline
		&
		GS28 &
		21.15 &
		0.54 &
		0.19 &
		0.33 &
		0.47 &
		0\\\hline
		\multicolumn{8}{c}{  Cluster method: GMM}\\\hline
		3 &
		naïve &
		14.8 &
		0.2 &
		2.77 &
		4.16 &
		1.15 &
		5\\\hline
		&
		RC &
		31.47 &
		0.25 &
		0.74 &
		1.02 &
		0.9 &
		0\\\hline
		&
		SIMEX-Q** &
		14.83 &
		0.2 &
		3.5 &
		5.29 &
		1.55 &
		5\\\hline
		&
		SIMEX-C** &
		14.83 &
		0.2 &
		6.13 &
		9.28 &
		2.9 &
		5\\\hline
		&
		SIMEX-Q4** &
		14.83 &
		0.2 &
		12.7 &
		19.11 &
		6.5 &
		5\\\hline
		&
		MI &
		31.75 &
		0.25 &
		0.7 &
		1 &
		0.85 &
		0\\\hline
		&
		MI-NULL*** &
		31.43 &
		0.25 &
		0.75 &
		1.04 &
		0.94 &
		0\\\hline
		&
		GS7 &
		12.69 &
		0.44 &
		1.37 &
		2.07 &
		0.65 &
		0\\\hline
		&
		GS28 &
		7.98 &
		0.68 &
		0.85 &
		1.33 &
		0.41 &
		0\\\hline
		4 &
		naïve &
		16.8 &
		0.19 &
		2.61 &
		5.01 &
		1.5 &
		1\\\hline
		&
		RC &
		37.59 &
		0.21 &
		0.87 &
		1.42 &
		0.91 &
		0\\\hline
		&
		SIMEX-Q** &
		16.81 &
		0.19 &
		3.39 &
		6.21 &
		1.88 &
		1\\\hline
		&
		SIMEX-C** &
		16.81 &
		0.19 &
		6.05 &
		11.05 &
		3.78 &
		1\\\hline
		&
		SIMEX-Q4** &
		16.81 &
		0.19 &
		13.02 &
		23.59 &
		9.72 &
		1\\\hline
		&
		MI &
		37.46 &
		0.22 &
		0.82 &
		1.37 &
		0.86 &
		0\\\hline
		&
		MI-NULL*** &
		37.54 &
		0.21 &
		0.92 &
		1.48 &
		0.95 &
		0\\\hline
		&
		GS7 &
		17.16 &
		0.41 &
		1.14 &
		2.09 &
		0.82 &
		0\\\hline
		&
		GS28 &
		10.85 &
		0.65 &
		0.8 &
		1.53 &
		0.56 &
		0\\\hline
		5 &
		naïve &
		22.8 &
		0.2 &
		2.08 &
		4.77 &
		1.51 &
		0\\\hline
		&
		RC &
		42.29 &
		0.19 &
		0.91 &
		1.75 &
		0.97 &
		0\\\hline
		&
		SIMEX-Q** &
		22.82 &
		0.2 &
		2.55 &
		5.61 &
		1.84 &
		1\\\hline
		&
		SIMEX-C** &
		22.82 &
		0.2 &
		4.25 &
		9.29 &
		3.59 &
		1\\\hline
		&
		SIMEX-Q4** &
		22.82 &
		0.2 &
		9.11 &
		19.89 &
		8.68 &
		1\\\hline
		&
		MI &
		42.46 &
		0.19 &
		0.89 &
		1.73 &
		0.89 &
		0\\\hline
		&
		MI-NULL*** &
		42.44 &
		0.19 &
		1.04 &
		1.92 &
		0.96 &
		0\\\hline
		&
		GS7 &
		26.28 &
		0.38 &
		0.64 &
		1.31 &
		0.81 &
		0\\\hline
		&
		GS28 &
		15.94 &
		0.61 &
		0.49 &
		1 &
		0.58 &
		0\\\hline
	\end{tabular}
\end{table}
*Classifications were considered failed if a cluster was empty.**For SIMEX-3-SA different extrapolation functions were used: quadratic (Q), cubic (C) and quartic (Q4).***MI-NULL denotes the MI method without considering the health outcome in the ME model.

\clearpage
\subsection{Simulation for  $M=9$  and  $\mathit{Corr}\left(u_{\mathit{ki}},u_{\mathit{li}}\right){\neq}0$  for  $k{\neq}l$}

\begin{table}[h]
	 \caption{Simulation results of correction methods and gold standard methods based on 7 and 28 days (GS7 and GS28) for scenarios with outcome  $H_i^A$  and cluster method with \textit{C} clusters regarding empirical misclassification rate (MR), adjusted Rand index (aRI), mean absolute bias, maximal absolute bias and median mean relative absolute bias, and number of failed classifications* for 1000 simulation data sets.}	 
	 
	\tiny\begin{tabular}{l|l|l|l|l|l|l|l}
		\hline
		\textit{C} &
		Correction method &
		MR &
		aRI &
		$\overline{{\Delta}}$ &
		$\overline{{\Delta}}_{\mathit{max}}$ &
		$\mathit{med}(\overline{{\Delta}}_{\mathit{rel}})$ &
		Number of failed classifications*\\\hline
		\multicolumn{8}{c}{Cluster method: k-means}\\\hline
		3 &
		naïve &
		31.8 &
		0.29 &
		0.06 &
		0.1 &
		0.68 &
		0\\\hline
		&
		RC &
		32.9 &
		0.27 &
		0.07 &
		0.09 &
		0.67 &
		0\\\hline
		&
		SIMEX-Q** &
		31.79 &
		0.29 &
		0.09 &
		0.12 &
		0.83 &
		2\\\hline
		&
		SIMEX-C** &
		31.79 &
		0.29 &
		0.17 &
		0.23 &
		1.62 &
		2\\\hline
		&
		SIMEX-Q4** &
		31.79 &
		0.29 &
		0.37 &
		0.5 &
		3.71 &
		2\\\hline
		&
		MI &
		32.85 &
		0.27 &
		0.06 &
		0.08 &
		0.54 &
		0\\\hline
		&
		MI-NULL*** &
		32.9 &
		0.27 &
		0.09 &
		0.13 &
		0.8 &
		0\\\hline
		&
		GS7 &
		19.44 &
		0.52 &
		0.05 &
		0.06 &
		0.42 &
		0\\\hline
		&
		GS28 &
		10.19 &
		0.73 &
		0.03 &
		0.04 &
		0.31 &
		0\\\hline
		4 &
		naïve &
		41.1 &
		0.21 &
		0.07 &
		0.13 &
		0.96 &
		0\\\hline
		&
		RC &
		40.14 &
		0.22 &
		0.08 &
		0.12 &
		0.89 &
		0\\\hline
		&
		SIMEX-Q** &
		41.12 &
		0.21 &
		0.1 &
		0.16 &
		1.31 &
		2\\\hline
		&
		SIMEX-C** &
		41.12 &
		0.21 &
		0.2 &
		0.32 &
		2.56 &
		2\\\hline
		&
		SIMEX-Q4** &
		41.12 &
		0.21 &
		0.46 &
		0.72 &
		5.65 &
		2\\\hline
		&
		MI &
		40.18 &
		0.22 &
		0.07 &
		0.11 &
		0.68 &
		0\\\hline
		&
		MI-NULL*** &
		40.16 &
		0.22 &
		0.1 &
		0.16 &
		0.84 &
		0\\\hline
		&
		GS7 &
		25.37 &
		0.44 &
		0.05 &
		0.08 &
		0.6 &
		0\\\hline
		&
		GS28 &
		12.82 &
		0.69 &
		0.04 &
		0.06 &
		0.39 &
		0\\\hline
		5 &
		naïve &
		47.4 &
		0.18 &
		0.08 &
		0.17 &
		1.22 &
		0\\\hline
		&
		RC &
		45.67 &
		0.2 &
		0.09 &
		0.15 &
		1.01 &
		0\\\hline
		&
		SIMEX-Q** &
		47.36 &
		0.18 &
		0.12 &
		0.21 &
		1.57 &
		2\\\hline
		&
		SIMEX-C** &
		47.36 &
		0.18 &
		0.23 &
		0.4 &
		3.05 &
		2\\\hline
		&
		SIMEX-Q4** &
		47.36 &
		0.18 &
		0.52 &
		0.89 &
		6.48 &
		2\\\hline
		&
		MI &
		45.74 &
		0.2 &
		0.08 &
		0.13 &
		0.81 &
		0\\\hline
		&
		MI-NULL*** &
		45.68 &
		0.2 &
		0.1 &
		0.17 &
		0.87 &
		0\\\hline
		&
		GS7 &
		29.1 &
		0.41 &
		0.06 &
		0.1 &
		0.76 &
		0\\\hline
		&
		GS28 &
		15.18 &
		0.66 &
		0.04 &
		0.07 &
		0.53 &
		0\\\hline
		\multicolumn{8}{c}{  Cluster method: GMM}\\\hline
		3 &
		naïve &
		22.7 &
		0.35 &
		0.1 &
		0.15 &
		0.95 &
		0\\\hline
		&
		RC &
		30.43 &
		0.28 &
		0.07 &
		0.1 &
		0.8 &
		0\\\hline
		&
		SIMEX-Q** &
		22.66 &
		0.35 &
		0.15 &
		0.21 &
		1.25 &
		2\\\hline
		&
		SIMEX-C** &
		22.66 &
		0.35 &
		0.29 &
		0.4 &
		2.47 &
		2\\\hline
		&
		SIMEX-Q4** &
		22.66 &
		0.35 &
		0.64 &
		0.9 &
		5.62 &
		2\\\hline
		&
		MI &
		30.72 &
		0.28 &
		0.06 &
		0.08 &
		0.6 &
		0\\\hline
		&
		MI-NULL*** &
		30.45 &
		0.28 &
		0.09 &
		0.12 &
		0.8 &
		0\\\hline
		&
		GS7 &
		17.67 &
		0.55 &
		0.05 &
		0.07 &
		0.42 &
		0\\\hline
		&
		GS28 &
		9.53 &
		0.75 &
		0.03 &
		0.04 &
		0.33 &
		0\\\hline
		4 &
		naïve &
		27.2 &
		0.32 &
		0.13 &
		0.24 &
		1.29 &
		0\\\hline
		&
		RC &
		38.42 &
		0.23 &
		0.08 &
		0.13 &
		0.9 &
		0\\\hline
		&
		SIMEX-Q** &
		27.21 &
		0.32 &
		0.19 &
		0.31 &
		1.68 &
		2\\\hline
		&
		SIMEX-C** &
		27.21 &
		0.32 &
		0.36 &
		0.6 &
		3.33 &
		2\\\hline
		&
		SIMEX-Q4** &
		27.21 &
		0.32 &
		0.81 &
		1.34 &
		7.41 &
		2\\\hline
		&
		MI &
		38.51 &
		0.23 &
		0.07 &
		0.11 &
		0.71 &
		0\\\hline
		&
		MI-NULL*** &
		38.45 &
		0.23 &
		0.1 &
		0.16 &
		0.83 &
		0\\\hline
		&
		GS7 &
		23.58 &
		0.46 &
		0.06 &
		0.09 &
		0.63 &
		0\\\hline
		&
		GS28 &
		12.52 &
		0.69 &
		0.04 &
		0.06 &
		0.42 &
		0\\\hline
		5 &
		naïve &
		31.7 &
		0.29 &
		0.16 &
		0.35 &
		1.5 &
		3\\\hline
		&
		RC &
		43.03 &
		0.21 &
		0.09 &
		0.16 &
		0.98 &
		0\\\hline
		&
		SIMEX-Q** &
		31.72 &
		0.29 &
		0.23 &
		0.42 &
		1.85 &
		2\\\hline
		&
		SIMEX-C** &
		31.72 &
		0.29 &
		0.43 &
		0.78 &
		3.69 &
		2\\\hline
		&
		SIMEX-Q4** &
		31.72 &
		0.29 &
		0.93 &
		1.71 &
		7.99 &
		2\\\hline
		&
		MI &
		42.96 &
		0.21 &
		0.08 &
		0.14 &
		0.8 &
		0\\\hline
		&
		MI-NULL*** &
		43 &
		0.21 &
		0.11 &
		0.19 &
		0.85 &
		0\\\hline
		&
		GS7 &
		27.11 &
		0.43 &
		0.07 &
		0.12 &
		0.78 &
		0\\\hline
		&
		GS28 &
		14.74 &
		0.66 &
		0.04 &
		0.08 &
		0.49 &
		0\\\hline
	\end{tabular}
\end{table}
*Classifications were considered failed if a cluster was empty.**For SIMEX-3-SA different extrapolation functions were used: quadratic (Q), cubic (C) and quartic (Q4).***MI-NULL denotes the MI method without considering the health outcome in the ME model.

\clearpage\begin{table}[h]
	 \caption{Simulation results of correction methods and gold standard methods based on 7 and 28 days (GS7 and GS28) for scenarios with outcome  $H_i^{A-\mathit{cat}}$  and cluster method with \textit{C} clusters regarding empirical misclassification rate (MR), adjusted Rand index (aRI), mean absolute bias, maximal absolute bias and median mean relative absolute bias, and number of failed classifications* for 1000 simulation data sets.}
	 
	\tiny\begin{tabular}{l|l|l|l|l|l|l|l}
		\hline
		\textit{C} &
		Correction method &
		MR &
		aRI &
		$\overline{{\Delta}}$ &
		$\overline{{\Delta}}_{\mathit{max}}$ &
		$\mathit{med}(\overline{{\Delta}}_{\mathit{rel}})$ &
		Number of failed classifications*\\\hline
		\multicolumn{8}{c}{  Cluster method: k-means}\\\hline
		3 &
		naïve &
		31.8 &
		0.29 &
		0.2 &
		0.3 &
		1.02 &
		0\\\hline
		&
		RC &
		32.9 &
		0.27 &
		0.21 &
		0.29 &
		1 &
		0\\\hline
		&
		SIMEX-Q** &
		31.79 &
		0.29 &
		0.3 &
		0.41 &
		1.32 &
		2\\\hline
		&
		SIMEX-C** &
		31.79 &
		0.29 &
		0.58 &
		0.8 &
		2.5 &
		2\\\hline
		&
		SIMEX-Q4** &
		31.79 &
		0.29 &
		1.29 &
		1.77 &
		5.47 &
		2\\\hline
		&
		MI &
		32.98 &
		0.27 &
		0.18 &
		0.24 &
		0.75 &
		0\\\hline
		&
		MI-NULL*** &
		32.92 &
		0.27 &
		0.22 &
		0.31 &
		0.85 &
		0\\\hline
		&
		GS7 &
		19.44 &
		0.52 &
		0.16 &
		0.21 &
		0.67 &
		0\\\hline
		&
		GS28 &
		10.19 &
		0.73 &
		0.11 &
		0.15 &
		0.49 &
		0\\\hline
		4 &
		naïve &
		41.1 &
		0.21 &
		0.23 &
		0.42 &
		1.43 &
		0\\\hline
		&
		RC &
		40.14 &
		0.22 &
		0.24 &
		0.39 &
		1.18 &
		0\\\hline
		&
		SIMEX-Q** &
		41.12 &
		0.21 &
		0.35 &
		0.55 &
		1.7 &
		2\\\hline
		&
		SIMEX-C** &
		41.12 &
		0.21 &
		0.68 &
		1.07 &
		3.43 &
		2\\\hline
		&
		SIMEX-Q4** &
		41.12 &
		0.21 &
		1.55 &
		2.42 &
		7.84 &
		2\\\hline
		&
		MI &
		40.21 &
		0.22 &
		0.2 &
		0.32 &
		0.88 &
		0\\\hline
		&
		MI-NULL*** &
		40.15 &
		0.22 &
		0.24 &
		0.38 &
		0.88 &
		0\\\hline
		&
		GS7 &
		25.37 &
		0.44 &
		0.18 &
		0.28 &
		0.92 &
		0\\\hline
		&
		GS28 &
		12.82 &
		0.69 &
		0.12 &
		0.2 &
		0.6 &
		0\\\hline
		5 &
		naïve &
		47.4 &
		0.18 &
		0.28 &
		0.57 &
		1.72 &
		0\\\hline
		&
		RC &
		45.67 &
		0.2 &
		0.28 &
		0.47 &
		1.31 &
		0\\\hline
		&
		SIMEX-Q** &
		47.36 &
		0.18 &
		0.43 &
		0.74 &
		2.22 &
		2\\\hline
		&
		SIMEX-C** &
		47.36 &
		0.18 &
		0.83 &
		1.43 &
		4.38 &
		2\\\hline
		&
		SIMEX-Q4** &
		47.36 &
		0.18 &
		1.86 &
		3.19 &
		9.74 &
		2\\\hline
		&
		MI &
		45.75 &
		0.2 &
		0.23 &
		0.4 &
		1.02 &
		0\\\hline
		&
		MI-NULL*** &
		45.67 &
		0.2 &
		0.26 &
		0.46 &
		0.92 &
		0\\\hline
		&
		GS7 &
		29.1 &
		0.41 &
		0.2 &
		0.35 &
		1.06 &
		0\\\hline
		&
		GS28 &
		15.18 &
		0.66 &
		0.15 &
		0.25 &
		0.73 &
		0\\\hline
		\multicolumn{8}{c}{  Cluster method: GMM}\\\hline
		3 &
		naïve &
		22.7 &
		0.35 &
		0.6 &
		0.9 &
		1.3 &
		0\\\hline
		&
		RC &
		30.43 &
		0.28 &
		0.23 &
		0.32 &
		1.09 &
		0\\\hline
		&
		SIMEX-Q** &
		22.66 &
		0.35 &
		0.89 &
		1.33 &
		1.83 &
		2\\\hline
		&
		SIMEX-C** &
		22.66 &
		0.35 &
		1.74 &
		2.55 &
		4.05 &
		2\\\hline
		&
		SIMEX-Q4** &
		22.66 &
		0.35 &
		4.1 &
		5.87 &
		9.51 &
		2\\\hline
		&
		MI &
		30.6 &
		0.28 &
		0.19 &
		0.27 &
		0.83 &
		0\\\hline
		&
		MI-NULL*** &
		30.45 &
		0.28 &
		0.22 &
		0.3 &
		0.85 &
		0\\\hline
		&
		GS7 &
		17.67 &
		0.55 &
		0.16 &
		0.22 &
		0.6 &
		0\\\hline
		&
		GS28 &
		9.53 &
		0.75 &
		0.12 &
		0.16 &
		0.49 &
		0\\\hline
		4 &
		naïve &
		27.2 &
		0.32 &
		0.93 &
		1.81 &
		1.69 &
		0\\\hline
		&
		RC &
		38.42 &
		0.23 &
		0.26 &
		0.42 &
		1.3 &
		0\\\hline
		&
		SIMEX-Q** &
		27.21 &
		0.32 &
		1.22 &
		2.43 &
		2.3 &
		2\\\hline
		&
		SIMEX-C** &
		27.21 &
		0.32 &
		2.37 &
		4.59 &
		4.84 &
		2\\\hline
		&
		SIMEX-Q4** &
		27.21 &
		0.32 &
		5.45 &
		10.21 &
		11.55 &
		2\\\hline
		&
		MI &
		38.56 &
		0.23 &
		0.24 &
		0.39 &
		0.97 &
		0\\\hline
		&
		MI-NULL*** &
		38.45 &
		0.23 &
		0.25 &
		0.41 &
		0.88 &
		0\\\hline
		&
		GS7 &
		23.58 &
		0.46 &
		0.19 &
		0.31 &
		0.91 &
		0\\\hline
		&
		GS28 &
		12.52 &
		0.69 &
		0.14 &
		0.22 &
		0.62 &
		0\\\hline
		5 &
		naïve &
		31.7 &
		0.29 &
		1.55 &
		3.65 &
		1.95 &
		3\\\hline
		&
		RC &
		43.03 &
		0.21 &
		0.32 &
		0.57 &
		1.36 &
		0\\\hline
		&
		SIMEX-Q** &
		31.72 &
		0.29 &
		2.12 &
		4.76 &
		2.51 &
		2\\\hline
		&
		SIMEX-C** &
		31.72 &
		0.29 &
		3.97 &
		8.66 &
		5.64 &
		2\\\hline
		&
		SIMEX-Q4** &
		31.72 &
		0.29 &
		8.58 &
		18.46 &
		13.83 &
		2\\\hline
		&
		MI &
		43.1 &
		0.21 &
		0.44 &
		0.96 &
		1.41 &
		0\\\hline
		&
		MI-NULL*** &
		43 &
		0.21 &
		0.44 &
		0.85 &
		1.06 &
		0\\\hline
		&
		GS7 &
		27.11 &
		0.43 &
		0.26 &
		0.47 &
		1.04 &
		0\\\hline
		&
		GS28 &
		14.74 &
		0.66 &
		0.16 &
		0.29 &
		0.76 &
		0\\\hline
	\end{tabular}
\end{table}
*Classifications were considered failed if a cluster was empty.**For SIMEX-3-SA different extrapolation functions were used: quadratic (Q), cubic (C) and quartic (Q4).***MI-NULL denotes the MI method without considering the health outcome in the ME model.

\clearpage\begin{table}[h]
	 \caption{Simulation results of correction methods and gold standard methods based on 7 and 28 days (GS7 and GS28) for scenarios with outcome  $H_i^B$  and cluster method with \textit{C} clusters regarding empirical misclassification rate (MR), adjusted Rand index (aRI), mean absolute bias, maximal absolute bias and median mean relative absolute bias, and number of failed classifications* for 1000 simulation data sets.}
	\tiny\begin{tabular}{l|l|l|l|l|l|l|l}
		\hline
		\textit{C} &
		Correction method &
		MR &
		aRI &
		$\overline{{\Delta}}$ &
		$\overline{{\Delta}}_{\mathit{max}}$ &
		$\mathit{med}(\overline{{\Delta}}_{\mathit{rel}})$ &
		Number of failed classifications*\\\hline
		\multicolumn{8}{c}{  Cluster method: k-means}\\\hline
		3 &
		naïve &
		31.8 &
		0.29 &
		0.09 &
		0.14 &
		1.37 &
		0\\\hline
		&
		RC &
		32.9 &
		0.27 &
		0.13 &
		0.18 &
		1.16 &
		0\\\hline
		&
		SIMEX-Q** &
		31.79 &
		0.29 &
		0.12 &
		0.16 &
		1.62 &
		2\\\hline
		&
		SIMEX-C** &
		31.79 &
		0.29 &
		0.2 &
		0.27 &
		2.51 &
		2\\\hline
		&
		SIMEX-Q4** &
		31.79 &
		0.29 &
		0.42 &
		0.57 &
		5.32 &
		2\\\hline
		&
		MI &
		32.85 &
		0.27 &
		0.13 &
		0.18 &
		0.97 &
		0\\\hline
		&
		MI-NULL*** &
		32.9 &
		0.27 &
		0.13 &
		0.17 &
		0.98 &
		0\\\hline
		&
		GS7 &
		19.44 &
		0.52 &
		0.05 &
		0.06 &
		0.57 &
		0\\\hline
		&
		GS28 &
		10.19 &
		0.73 &
		0.03 &
		0.05 &
		0.44 &
		0\\\hline
		4 &
		naïve &
		41.1 &
		0.21 &
		0.11 &
		0.21 &
		1.42 &
		0\\\hline
		&
		RC &
		40.14 &
		0.22 &
		0.15 &
		0.23 &
		1.15 &
		0\\\hline
		&
		SIMEX-Q** &
		41.12 &
		0.21 &
		0.15 &
		0.23 &
		1.73 &
		2\\\hline
		&
		SIMEX-C** &
		41.12 &
		0.21 &
		0.24 &
		0.38 &
		2.81 &
		2\\\hline
		&
		SIMEX-Q4** &
		41.12 &
		0.21 &
		0.5 &
		0.78 &
		5.65 &
		2\\\hline
		&
		MI &
		40.16 &
		0.22 &
		0.14 &
		0.22 &
		0.98 &
		0\\\hline
		&
		MI-NULL*** &
		40.14 &
		0.22 &
		0.14 &
		0.22 &
		1 &
		0\\\hline
		&
		GS7 &
		25.37 &
		0.44 &
		0.05 &
		0.09 &
		0.71 &
		0\\\hline
		&
		GS28 &
		12.82 &
		0.69 &
		0.04 &
		0.06 &
		0.56 &
		0\\\hline
		5 &
		naïve &
		47.4 &
		0.18 &
		0.16 &
		0.34 &
		2.04 &
		0\\\hline
		&
		RC &
		45.67 &
		0.2 &
		0.17 &
		0.28 &
		1.46 &
		0\\\hline
		&
		SIMEX-Q** &
		47.36 &
		0.18 &
		0.21 &
		0.36 &
		2.12 &
		2\\\hline
		&
		SIMEX-C** &
		47.36 &
		0.18 &
		0.31 &
		0.54 &
		3.42 &
		2\\\hline
		&
		SIMEX-Q4** &
		47.36 &
		0.18 &
		0.6 &
		1.03 &
		6.79 &
		2\\\hline
		&
		MI &
		45.74 &
		0.2 &
		0.15 &
		0.25 &
		1.02 &
		0\\\hline
		&
		MI-NULL*** &
		45.66 &
		0.2 &
		0.15 &
		0.25 &
		1.03 &
		0\\\hline
		&
		GS7 &
		29.1 &
		0.41 &
		0.06 &
		0.11 &
		0.8 &
		0\\\hline
		&
		GS28 &
		15.18 &
		0.66 &
		0.04 &
		0.08 &
		0.62 &
		0\\\hline
		\multicolumn{8}{c}{  Cluster method: GMM}\\\hline
		3 &
		naïve &
		22.7 &
		0.35 &
		0.19 &
		0.28 &
		1.76 &
		0\\\hline
		&
		RC &
		30.43 &
		0.28 &
		0.13 &
		0.18 &
		1.28 &
		0\\\hline
		&
		SIMEX-Q** &
		22.66 &
		0.35 &
		0.24 &
		0.34 &
		2.23 &
		2\\\hline
		&
		SIMEX-C** &
		22.66 &
		0.35 &
		0.37 &
		0.52 &
		3.49 &
		2\\\hline
		&
		SIMEX-Q4** &
		22.66 &
		0.35 &
		0.74 &
		1.03 &
		6.95 &
		2\\\hline
		&
		MI &
		30.72 &
		0.28 &
		0.12 &
		0.16 &
		1 &
		0\\\hline
		&
		MI-NULL*** &
		30.45 &
		0.28 &
		0.12 &
		0.16 &
		1.01 &
		0\\\hline
		&
		GS7 &
		17.67 &
		0.55 &
		0.05 &
		0.07 &
		0.7 &
		0\\\hline
		&
		GS28 &
		9.53 &
		0.75 &
		0.04 &
		0.05 &
		0.45 &
		0\\\hline
		4 &
		naïve &
		27.2 &
		0.32 &
		0.25 &
		0.47 &
		1.83 &
		0\\\hline
		&
		RC &
		38.42 &
		0.23 &
		0.15 &
		0.23 &
		1.21 &
		0\\\hline
		&
		SIMEX-Q** &
		27.21 &
		0.32 &
		0.3 &
		0.52 &
		2.29 &
		2\\\hline
		&
		SIMEX-C** &
		27.21 &
		0.32 &
		0.47 &
		0.79 &
		3.75 &
		2\\\hline
		&
		SIMEX-Q4** &
		27.21 &
		0.32 &
		0.95 &
		1.56 &
		7.68 &
		2\\\hline
		&
		MI &
		38.51 &
		0.23 &
		0.14 &
		0.22 &
		0.98 &
		0\\\hline
		&
		MI-NULL*** &
		38.45 &
		0.23 &
		0.15 &
		0.22 &
		1 &
		0\\\hline
		&
		GS7 &
		23.58 &
		0.46 &
		0.06 &
		0.1 &
		0.75 &
		0\\\hline
		&
		GS28 &
		12.52 &
		0.69 &
		0.04 &
		0.07 &
		0.55 &
		0\\\hline
		5 &
		naïve &
		31.7 &
		0.29 &
		0.3 &
		0.65 &
		2.08 &
		3\\\hline
		&
		RC &
		43.03 &
		0.21 &
		0.17 &
		0.3 &
		1.34 &
		0\\\hline
		&
		SIMEX-Q** &
		31.72 &
		0.29 &
		0.37 &
		0.7 &
		2.47 &
		2\\\hline
		&
		SIMEX-C** &
		31.72 &
		0.29 &
		0.55 &
		1.02 &
		3.91 &
		2\\\hline
		&
		SIMEX-Q4** &
		31.72 &
		0.29 &
		1.07 &
		1.96 &
		8.02 &
		2\\\hline
		&
		MI &
		42.96 &
		0.21 &
		0.16 &
		0.28 &
		1 &
		0\\\hline
		&
		MI-NULL*** &
		43 &
		0.21 &
		0.16 &
		0.28 &
		1.01 &
		0\\\hline
		&
		GS7 &
		27.11 &
		0.43 &
		0.07 &
		0.13 &
		0.87 &
		0\\\hline
		&
		GS28 &
		14.74 &
		0.66 &
		0.05 &
		0.08 &
		0.59 &
		0\\\hline
	\end{tabular}
\end{table}
*Classifications were considered failed if a cluster was empty.**For SIMEX-3-SA different extrapolation functions were used: quadratic (Q), cubic (C) and quartic (Q4).***MI-NULL denotes the MI method without considering the health outcome in the ME model.

\clearpage\begin{table}[h]
	 
\caption{Simulation results of correction methods and gold standard methods based on 7 and 28 days (GS7 and GS28) for scenarios with outcome  $H_i^{B-\mathit{cat}}$  and cluster method with \textit{C} clusters regarding empirical misclassification rate (MR), adjusted Rand index (aRI), mean absolute bias, maximal absolute bias and median mean relative absolute bias, and number of failed classifications* for 1000 simulation data sets.}
	 
	\tiny\begin{tabular}{l|l|l|l|l|l|l|l}
		\hline
		\textit{C} &
		Correction method &
		MR &
		aRI &
		$\overline{{\Delta}}$ &
		$\overline{{\Delta}}_{\mathit{max}}$ &
		$\mathit{med}(\overline{{\Delta}}_{\mathit{rel}})$ &
		Number of failed classifications*\\\hline
		\multicolumn{8}{c}{  Cluster method: k-means}\\\hline
		3 &
		naïve &
		31.8 &
		0.29 &
		0.22 &
		0.33 &
		1.37 &
		0\\\hline
		&
		RC &
		32.9 &
		0.27 &
		0.29 &
		0.4 &
		1.26 &
		0\\\hline
		&
		SIMEX-Q** &
		31.79 &
		0.29 &
		0.32 &
		0.44 &
		1.92 &
		2\\\hline
		&
		SIMEX-C** &
		31.79 &
		0.29 &
		0.59 &
		0.81 &
		3.63 &
		2\\\hline
		&
		SIMEX-Q4** &
		31.79 &
		0.29 &
		1.3 &
		1.76 &
		8.03 &
		2\\\hline
		&
		MI &
		32.94 &
		0.27 &
		0.26 &
		0.36 &
		1 &
		0\\\hline
		&
		MI-NULL*** &
		32.9 &
		0.27 &
		0.27 &
		0.37 &
		0.97 &
		0\\\hline
		&
		GS7 &
		19.44 &
		0.52 &
		0.15 &
		0.2 &
		0.78 &
		0\\\hline
		&
		GS28 &
		10.19 &
		0.73 &
		0.1 &
		0.14 &
		0.55 &
		0\\\hline
		4 &
		naïve &
		41.1 &
		0.21 &
		0.28 &
		0.51 &
		1.66 &
		0\\\hline
		&
		RC &
		40.14 &
		0.22 &
		0.33 &
		0.52 &
		1.29 &
		0\\\hline
		&
		SIMEX-Q** &
		41.12 &
		0.21 &
		0.41 &
		0.64 &
		2 &
		2\\\hline
		&
		SIMEX-C** &
		41.12 &
		0.21 &
		0.72 &
		1.14 &
		3.74 &
		2\\\hline
		&
		SIMEX-Q4** &
		41.12 &
		0.21 &
		1.55 &
		2.44 &
		7.95 &
		2\\\hline
		&
		MI &
		40.2 &
		0.22 &
		0.31 &
		0.48 &
		1.11 &
		0\\\hline
		&
		MI-NULL*** &
		40.15 &
		0.22 &
		0.31 &
		0.48 &
		1 &
		0\\\hline
		&
		GS7 &
		25.37 &
		0.44 &
		0.17 &
		0.27 &
		0.97 &
		0\\\hline
		&
		GS28 &
		12.82 &
		0.69 &
		0.12 &
		0.19 &
		0.68 &
		0\\\hline
		5 &
		naïve &
		47.4 &
		0.18 &
		0.38 &
		0.8 &
		2.15 &
		0\\\hline
		&
		RC &
		45.67 &
		0.2 &
		0.38 &
		0.65 &
		1.62 &
		0\\\hline
		&
		SIMEX-Q** &
		47.36 &
		0.18 &
		0.53 &
		0.92 &
		2.71 &
		2\\\hline
		&
		SIMEX-C** &
		47.36 &
		0.18 &
		0.91 &
		1.55 &
		4.84 &
		2\\\hline
		&
		SIMEX-Q4** &
		47.36 &
		0.18 &
		1.86 &
		3.22 &
		10.24 &
		2\\\hline
		&
		MI &
		45.81 &
		0.2 &
		0.34 &
		0.59 &
		1.35 &
		0\\\hline
		&
		MI-NULL*** &
		45.7 &
		0.2 &
		0.33 &
		0.55 &
		1.03 &
		0\\\hline
		&
		GS7 &
		29.1 &
		0.41 &
		0.2 &
		0.35 &
		1.12 &
		0\\\hline
		&
		GS28 &
		15.18 &
		0.66 &
		0.14 &
		0.25 &
		0.82 &
		0\\\hline
		\multicolumn{8}{c}{  Cluster method: GMM}\\\hline
		3 &
		naïve &
		22.7 &
		0.35 &
		0.78 &
		1.18 &
		1.69 &
		0\\\hline
		&
		RC &
		30.43 &
		0.28 &
		0.3 &
		0.41 &
		1.34 &
		0\\\hline
		&
		SIMEX-Q** &
		22.66 &
		0.35 &
		1.04 &
		1.51 &
		2.31 &
		2\\\hline
		&
		SIMEX-C** &
		22.66 &
		0.35 &
		1.7 &
		2.45 &
		4.51 &
		2\\\hline
		&
		SIMEX-Q4** &
		22.66 &
		0.35 &
		3.57 &
		5.15 &
		9.79 &
		2\\\hline
		&
		MI &
		30.41 &
		0.28 &
		0.27 &
		0.38 &
		1.11 &
		0\\\hline
		&
		MI-NULL*** &
		30.45 &
		0.28 &
		0.27 &
		0.37 &
		1 &
		0\\\hline
		&
		GS7 &
		17.67 &
		0.55 &
		0.16 &
		0.22 &
		0.83 &
		0\\\hline
		&
		GS28 &
		9.53 &
		0.75 &
		0.11 &
		0.15 &
		0.63 &
		0\\\hline
		4 &
		naïve &
		27.2 &
		0.32 &
		1.21 &
		2.35 &
		1.93 &
		0\\\hline
		&
		RC &
		38.42 &
		0.23 &
		0.35 &
		0.55 &
		1.38 &
		0\\\hline
		&
		SIMEX-Q** &
		27.21 &
		0.32 &
		1.49 &
		2.91 &
		2.47 &
		2\\\hline
		&
		SIMEX-C** &
		27.21 &
		0.32 &
		2.52 &
		4.84 &
		4.88 &
		2\\\hline
		&
		SIMEX-Q4** &
		27.21 &
		0.32 &
		5.1 &
		9.71 &
		11.08 &
		2\\\hline
		&
		MI &
		38.45 &
		0.23 &
		0.34 &
		0.54 &
		1.19 &
		0\\\hline
		&
		MI-NULL*** &
		38.45 &
		0.23 &
		0.34 &
		0.53 &
		1.05 &
		0\\\hline
		&
		GS7 &
		23.58 &
		0.46 &
		0.19 &
		0.3 &
		0.97 &
		0\\\hline
		&
		GS28 &
		12.52 &
		0.69 &
		0.13 &
		0.21 &
		0.7 &
		0\\\hline
		5 &
		naïve &
		31.7 &
		0.29 &
		1.96 &
		4.62 &
		2.26 &
		3\\\hline
		&
		RC &
		43.03 &
		0.21 &
		0.44 &
		0.79 &
		1.54 &
		0\\\hline
		&
		SIMEX-Q** &
		31.72 &
		0.29 &
		2.43 &
		5.66 &
		2.68 &
		2\\\hline
		&
		SIMEX-C** &
		31.72 &
		0.29 &
		4.04 &
		9.08 &
		5.08 &
		2\\\hline
		&
		SIMEX-Q4** &
		31.72 &
		0.29 &
		8.15 &
		17.87 &
		12.18 &
		2\\\hline
		&
		MI &
		43.08 &
		0.21 &
		0.54 &
		1.06 &
		1.55 &
		0\\\hline
		&
		MI-NULL*** &
		43 &
		0.21 &
		0.52 &
		1 &
		1.18 &
		0\\\hline
		&
		GS7 &
		27.11 &
		0.43 &
		0.25 &
		0.47 &
		1.15 &
		0\\\hline
		&
		GS28 &
		14.74 &
		0.66 &
		0.15 &
		0.27 &
		0.76 &
		0\\\hline
	\end{tabular}
\end{table}
*Classifications were considered failed if a cluster was empty.**For SIMEX-3-SA different extrapolation functions were used: quadratic (Q), cubic (C) and quartic (Q4).***MI-NULL denotes the MI method without considering the health outcome in the ME model.

\clearpage
\subsection{Simulation for  $M=9$  and  $\mathit{Corr}\left(u_{\mathit{ki}},u_{\mathit{li}}\right)=0$  for  $k{\neq}l$}

\begin{table}[h]
	 \caption{Simulation results of correction methods and gold standard methods based on 7 and 28 days (GS7 and GS28) for scenarios with outcome  $H_i^A$  and cluster method with \textit{C} clusters regarding empirical misclassification rate (MR), adjusted Rand index (aRI), mean absolute bias, maximal absolute bias and median mean relative absolute bias, and number of failed classifications* for 1000 simulation data sets.}

	\tiny\begin{tabular}{l|l|l|l|l|l|l|l}
		\hline
		\textit{C} &
		Correction method &
		MR &
		aRI &
		$\overline{{\Delta}}$ &
		$\overline{{\Delta}}_{\mathit{max}}$ &
		$\mathit{med}(\overline{{\Delta}}_{\mathit{rel}})$ &
		Number of failed classifications*\\\hline
		\multicolumn{8}{c}{  Cluster method: k-means}\\\hline
		3 &
		naïve &
		39.5 &
		0.17 &
		0.07 &
		0.11 &
		1.02 &
		0\\\hline
		&
		RC &
		35.06 &
		0.23 &
		0.08 &
		0.11 &
		0.9 &
		0\\\hline
		&
		SIMEX-Q** &
		39.39 &
		0.17 &
		0.1 &
		0.13 &
		1.27 &
		1\\\hline
		&
		SIMEX-C** &
		39.39 &
		0.17 &
		0.18 &
		0.24 &
		2.43 &
		1\\\hline
		&
		SIMEX-Q4** &
		39.39 &
		0.17 &
		0.39 &
		0.53 &
		5.62 &
		1\\\hline
		&
		MI &
		35.02 &
		0.23 &
		0.07 &
		0.1 &
		0.71 &
		0\\\hline
		&
		MI-NULL*** &
		35.03 &
		0.23 &
		0.09 &
		0.12 &
		0.9 &
		0\\\hline
		&
		GS7 &
		25.96 &
		0.38 &
		0.05 &
		0.07 &
		0.72 &
		0\\\hline
		&
		GS28 &
		13.76 &
		0.63 &
		0.04 &
		0.05 &
		0.58 &
		0\\\hline
		4 &
		naïve &
		47 &
		0.14 &
		0.09 &
		0.16 &
		1.16 &
		0\\\hline
		&
		RC &
		42.66 &
		0.19 &
		0.09 &
		0.14 &
		1.06 &
		0\\\hline
		&
		SIMEX-Q** &
		46.99 &
		0.14 &
		0.12 &
		0.18 &
		1.39 &
		1\\\hline
		&
		SIMEX-C** &
		46.99 &
		0.14 &
		0.22 &
		0.35 &
		2.73 &
		1\\\hline
		&
		SIMEX-Q4** &
		46.99 &
		0.14 &
		0.47 &
		0.76 &
		6.02 &
		1\\\hline
		&
		MI &
		42.57 &
		0.19 &
		0.08 &
		0.13 &
		0.79 &
		0\\\hline
		&
		MI-NULL*** &
		42.65 &
		0.19 &
		0.09 &
		0.14 &
		0.92 &
		0\\\hline
		&
		GS7 &
		31.87 &
		0.33 &
		0.06 &
		0.1 &
		0.8 &
		0\\\hline
		&
		GS28 &
		16.98 &
		0.6 &
		0.04 &
		0.07 &
		0.57 &
		0\\\hline
		5 &
		naïve &
		52.3 &
		0.12 &
		0.1 &
		0.2 &
		1.27 &
		0\\\hline
		&
		RC &
		47.86 &
		0.16 &
		0.09 &
		0.16 &
		1.19 &
		0\\\hline
		&
		SIMEX-Q** &
		52.32 &
		0.12 &
		0.13 &
		0.23 &
		1.54 &
		1\\\hline
		&
		SIMEX-C** &
		52.32 &
		0.12 &
		0.25 &
		0.43 &
		2.96 &
		1\\\hline
		&
		SIMEX-Q4** &
		52.32 &
		0.12 &
		0.55 &
		0.95 &
		6.74 &
		1\\\hline
		&
		MI &
		47.88 &
		0.16 &
		0.08 &
		0.14 &
		0.82 &
		0\\\hline
		&
		MI-NULL*** &
		47.88 &
		0.16 &
		0.1 &
		0.17 &
		0.93 &
		0\\\hline
		&
		GS7 &
		35.93 &
		0.31 &
		0.07 &
		0.12 &
		0.85 &
		0\\\hline
		&
		GS28 &
		19.39 &
		0.58 &
		0.05 &
		0.08 &
		0.63 &
		0\\\hline
		\multicolumn{8}{c}{  Cluster method: GMM}\\\hline
		3 &
		naïve &
		15 &
		0.2 &
		0.21 &
		0.31 &
		1.12 &
		1\\\hline
		&
		RC &
		30.39 &
		0.27 &
		0.09 &
		0.12 &
		1 &
		0\\\hline
		&
		SIMEX-Q** &
		14.94 &
		0.2 &
		0.27 &
		0.39 &
		1.57 &
		1\\\hline
		&
		SIMEX-C** &
		14.94 &
		0.2 &
		0.49 &
		0.69 &
		2.82 &
		1\\\hline
		&
		SIMEX-Q4** &
		14.94 &
		0.2 &
		1.05 &
		1.47 &
		6.27 &
		1\\\hline
		&
		MI &
		30.35 &
		0.27 &
		0.07 &
		0.1 &
		0.74 &
		0\\\hline
		&
		MI-NULL*** &
		30.25 &
		0.27 &
		0.09 &
		0.13 &
		0.91 &
		0\\\hline
		&
		GS7 &
		15.59 &
		0.46 &
		0.1 &
		0.14 &
		0.84 &
		0\\\hline
		&
		GS28 &
		10.06 &
		0.68 &
		0.06 &
		0.09 &
		0.68 &
		0\\\hline
		4 &
		naïve &
		15.7 &
		0.19 &
		0.23 &
		0.44 &
		1.41 &
		0\\\hline
		&
		RC &
		38.25 &
		0.22 &
		0.1 &
		0.16 &
		1.15 &
		0\\\hline
		&
		SIMEX-Q** &
		15.65 &
		0.19 &
		0.32 &
		0.52 &
		1.78 &
		1\\\hline
		&
		SIMEX-C** &
		15.65 &
		0.19 &
		0.58 &
		0.93 &
		3.33 &
		1\\\hline
		&
		SIMEX-Q4** &
		15.65 &
		0.19 &
		1.29 &
		2.07 &
		7.36 &
		1\\\hline
		&
		MI &
		38.31 &
		0.22 &
		0.09 &
		0.14 &
		0.82 &
		0\\\hline
		&
		MI-NULL*** &
		38.34 &
		0.22 &
		0.1 &
		0.16 &
		0.92 &
		0\\\hline
		&
		GS7 &
		17.53 &
		0.44 &
		0.12 &
		0.2 &
		1.03 &
		0\\\hline
		&
		GS28 &
		11.62 &
		0.66 &
		0.07 &
		0.12 &
		0.7 &
		0\\\hline
		5 &
		naïve &
		16.6 &
		0.18 &
		0.24 &
		0.52 &
		1.53 &
		2\\\hline
		&
		RC &
		42.92 &
		0.2 &
		0.11 &
		0.2 &
		1.25 &
		0\\\hline
		&
		SIMEX-Q** &
		16.6 &
		0.18 &
		0.34 &
		0.6 &
		2.04 &
		1\\\hline
		&
		SIMEX-C** &
		16.6 &
		0.18 &
		0.63 &
		1.09 &
		4.02 &
		1\\\hline
		&
		SIMEX-Q4** &
		16.6 &
		0.18 &
		1.36 &
		2.38 &
		8.76 &
		1\\\hline
		&
		MI &
		43.08 &
		0.2 &
		0.1 &
		0.18 &
		0.85 &
		0\\\hline
		&
		MI-NULL*** &
		43.01 &
		0.2 &
		0.11 &
		0.2 &
		0.94 &
		0\\\hline
		&
		GS7 &
		19.79 &
		0.43 &
		0.13 &
		0.23 &
		1.05 &
		0\\\hline
		&
		GS28 &
		13.14 &
		0.65 &
		0.08 &
		0.15 &
		0.82 &
		0\\\hline
	\end{tabular}
\end{table}
*Classifications were considered failed if a cluster was empty.**For SIMEX-3-SA different extrapolation functions were used: quadratic (Q), cubic (C) and quartic (Q4).***MI-NULL denotes the MI method without considering the health outcome in the ME model.

\clearpage\begin{table}[h]
	 
	 \caption{Simulation results of correction methods and gold standard methods based on 7 and 28 days (GS7 and GS28) for scenarios with outcome  $H_i^{A-\mathit{cat}}$  and cluster method with \textit{C} clusters regarding empirical misclassification rate (MR), adjusted Rand index (aRI), mean absolute bias, maximal absolute bias and median mean relative absolute bias, and number of failed classifications* for 1000 simulation data sets.}
	 
	\tiny\begin{tabular}{l|l|l|l|l|l|l|l}
		\hline
		\textit{C} &
		Correction method &
		MR &
		aRI &
		$\overline{{\Delta}}$ &
		$\overline{{\Delta}}_{\mathit{max}}$ &
		$\mathit{med}(\overline{{\Delta}}_{\mathit{rel}})$ &
		Number of failed classifications*\\\hline
		\multicolumn{8}{c}{  Cluster method: k-means}\\\hline
		3 &
		naïve &
		39.5 &
		0.17 &
		0.23 &
		0.34 &
		1.27 &
		0\\\hline
		&
		RC &
		35.06 &
		0.23 &
		0.24 &
		0.33 &
		1.26 &
		0\\\hline
		&
		SIMEX-Q** &
		39.39 &
		0.17 &
		0.32 &
		0.44 &
		1.67 &
		1\\\hline
		&
		SIMEX-C** &
		39.39 &
		0.17 &
		0.6 &
		0.82 &
		3.1 &
		1\\\hline
		&
		SIMEX-Q4** &
		39.39 &
		0.17 &
		1.32 &
		1.78 &
		7.03 &
		1\\\hline
		&
		MI &
		35.06 &
		0.23 &
		0.2 &
		0.27 &
		0.85 &
		0\\\hline
		&
		MI-NULL*** &
		35.04 &
		0.23 &
		0.22 &
		0.3 &
		0.91 &
		0\\\hline
		&
		GS7 &
		25.96 &
		0.38 &
		0.17 &
		0.24 &
		0.89 &
		0\\\hline
		&
		GS28 &
		13.76 &
		0.63 &
		0.13 &
		0.18 &
		0.73 &
		0\\\hline
		4 &
		naïve &
		47 &
		0.14 &
		0.27 &
		0.49 &
		1.49 &
		0\\\hline
		&
		RC &
		42.66 &
		0.19 &
		0.27 &
		0.43 &
		1.39 &
		0\\\hline
		&
		SIMEX-Q** &
		46.99 &
		0.14 &
		0.4 &
		0.63 &
		2 &
		1\\\hline
		&
		SIMEX-C** &
		46.99 &
		0.14 &
		0.73 &
		1.16 &
		3.97 &
		1\\\hline
		&
		SIMEX-Q4** &
		46.99 &
		0.14 &
		1.59 &
		2.52 &
		8.29 &
		1\\\hline
		&
		MI &
		42.64 &
		0.19 &
		0.22 &
		0.35 &
		0.92 &
		0\\\hline
		&
		MI-NULL*** &
		42.63 &
		0.19 &
		0.25 &
		0.39 &
		0.94 &
		0\\\hline
		&
		GS7 &
		31.87 &
		0.33 &
		0.21 &
		0.33 &
		1.07 &
		0\\\hline
		&
		GS28 &
		16.98 &
		0.6 &
		0.15 &
		0.24 &
		0.82 &
		0\\\hline
		5 &
		naïve &
		52.3 &
		0.12 &
		0.31 &
		0.64 &
		1.68 &
		0\\\hline
		&
		RC &
		47.86 &
		0.16 &
		0.3 &
		0.51 &
		1.5 &
		0\\\hline
		&
		SIMEX-Q** &
		52.32 &
		0.12 &
		0.45 &
		0.79 &
		2.08 &
		1\\\hline
		&
		SIMEX-C** &
		52.32 &
		0.12 &
		0.87 &
		1.51 &
		4.03 &
		1\\\hline
		&
		SIMEX-Q4** &
		52.32 &
		0.12 &
		1.91 &
		3.32 &
		9.03 &
		1\\\hline
		&
		MI &
		47.92 &
		0.16 &
		0.25 &
		0.43 &
		0.94 &
		0\\\hline
		&
		MI-NULL*** &
		47.9 &
		0.16 &
		0.27 &
		0.46 &
		0.96 &
		0\\\hline
		&
		GS7 &
		35.93 &
		0.31 &
		0.23 &
		0.4 &
		1.15 &
		0\\\hline
		&
		GS28 &
		19.39 &
		0.58 &
		0.17 &
		0.3 &
		0.88 &
		0\\\hline
		\multicolumn{8}{c}{  Cluster method: GMM}\\\hline
		3 &
		naïve &
		15 &
		0.2 &
		2.43 &
		3.65 &
		1.32 &
		1\\\hline
		&
		RC &
		30.39 &
		0.27 &
		0.28 &
		0.39 &
		1.24 &
		0\\\hline
		&
		SIMEX-Q** &
		14.94 &
		0.2 &
		3.26 &
		4.9 &
		1.99 &
		1\\\hline
		&
		SIMEX-C** &
		14.94 &
		0.2 &
		6.22 &
		9.23 &
		4.57 &
		1\\\hline
		&
		SIMEX-Q4** &
		14.94 &
		0.2 &
		13.64 &
		20.17 &
		10.74 &
		1\\\hline
		&
		MI &
		30.3 &
		0.27 &
		0.23 &
		0.32 &
		0.94 &
		0\\\hline
		&
		MI-NULL*** &
		30.25 &
		0.27 &
		0.25 &
		0.36 &
		0.94 &
		0\\\hline
		&
		GS7 &
		15.59 &
		0.46 &
		0.69 &
		1.1 &
		1.06 &
		0\\\hline
		&
		GS28 &
		10.06 &
		0.68 &
		0.36 &
		0.59 &
		0.85 &
		0\\\hline
		4 &
		naïve &
		15.7 &
		0.19 &
		2.81 &
		5.39 &
		1.73 &
		0\\\hline
		&
		RC &
		38.25 &
		0.22 &
		0.33 &
		0.54 &
		1.44 &
		0\\\hline
		&
		SIMEX-Q** &
		15.65 &
		0.19 &
		3.64 &
		7.02 &
		2.42 &
		1\\\hline
		&
		SIMEX-C** &
		15.65 &
		0.19 &
		6.89 &
		12.95 &
		5.76 &
		1\\\hline
		&
		SIMEX-Q4** &
		15.65 &
		0.19 &
		15.41 &
		28.55 &
		15.56 &
		1\\\hline
		&
		MI &
		38.47 &
		0.22 &
		0.26 &
		0.42 &
		1.02 &
		0\\\hline
		&
		MI-NULL*** &
		38.34 &
		0.22 &
		0.31 &
		0.51 &
		0.95 &
		0\\\hline
		&
		GS7 &
		17.53 &
		0.44 &
		0.91 &
		1.81 &
		1.18 &
		0\\\hline
		&
		GS28 &
		11.62 &
		0.66 &
		0.61 &
		1.19 &
		0.96 &
		0\\\hline
		5 &
		naïve &
		16.6 &
		0.18 &
		2.87 &
		6.58 &
		1.94 &
		2\\\hline
		&
		RC &
		42.92 &
		0.2 &
		0.43 &
		0.85 &
		1.41 &
		0\\\hline
		&
		SIMEX-Q** &
		16.6 &
		0.18 &
		4.13 &
		8.63 &
		2.62 &
		1\\\hline
		&
		SIMEX-C** &
		16.6 &
		0.18 &
		7.74 &
		16.26 &
		6.55 &
		1\\\hline
		&
		SIMEX-Q4** &
		16.6 &
		0.18 &
		17.33 &
		36.71 &
		17.54 &
		1\\\hline
		&
		MI &
		43.19 &
		0.2 &
		0.31 &
		0.6 &
		1.07 &
		0\\\hline
		&
		MI-NULL*** &
		43.01 &
		0.2 &
		0.37 &
		0.73 &
		1 &
		0\\\hline
		&
		GS7 &
		19.79 &
		0.43 &
		0.9 &
		2.13 &
		1.36 &
		0\\\hline
		&
		GS28 &
		13.14 &
		0.65 &
		0.61 &
		1.48 &
		1.03 &
		0\\\hline
	\end{tabular}
\end{table}
*Classifications were considered failed if a cluster was empty.**For SIMEX-3-SA different extrapolation functions were used: quadratic (Q), cubic (C) and quartic (Q4).***MI-NULL denotes the MI method without considering the health outcome in the ME model.

\clearpage\begin{table}[h]
	 \caption{Simulation results of correction methods and gold standard methods based on 7 and 28 days (GS7 and GS28) for scenarios with outcome  $H_i^B$  and cluster method with \textit{C} clusters regarding empirical misclassification rate (MR), adjusted Rand index (aRI), mean absolute bias, maximal absolute bias and median mean relative absolute bias, and number of failed classifications* for 1000 simulation data sets.}

	\tiny\begin{tabular}{l|l|l|l|l|l|l|l}
		\hline
		\textit{C} &
		Correction method &
		MR &
		aRI &
		$\overline{{\Delta}}$ &
		$\overline{{\Delta}}_{\mathit{max}}$ &
		$\mathit{med}(\overline{{\Delta}}_{\mathit{rel}})$ &
		Number of failed classifications*\\\hline
		\multicolumn{8}{c}{Cluster method: k-means}\\\hline
		3 &
		naïve &
		39.5 &
		0.17 &
		0.22 &
		0.34 &
		1.12 &
		0\\\hline
		&
		RC &
		35.06 &
		0.23 &
		0.25 &
		0.34 &
		0.96 &
		0\\\hline
		&
		SIMEX-Q** &
		39.39 &
		0.17 &
		0.24 &
		0.33 &
		1.21 &
		1\\\hline
		&
		SIMEX-C** &
		39.39 &
		0.17 &
		0.29 &
		0.4 &
		1.51 &
		1\\\hline
		&
		SIMEX-Q4** &
		39.39 &
		0.17 &
		0.48 &
		0.65 &
		2.71 &
		1\\\hline
		&
		MI &
		35.02 &
		0.23 &
		0.26 &
		0.36 &
		0.97 &
		0\\\hline
		&
		MI-NULL*** &
		35.07 &
		0.23 &
		0.26 &
		0.36 &
		0.96 &
		0\\\hline
		&
		GS7 &
		25.96 &
		0.38 &
		0.07 &
		0.1 &
		0.33 &
		0\\\hline
		&
		GS28 &
		13.76 &
		0.63 &
		0.04 &
		0.06 &
		0.21 &
		0\\\hline
		4 &
		naïve &
		47 &
		0.14 &
		0.24 &
		0.44 &
		1.34 &
		0\\\hline
		&
		RC &
		42.66 &
		0.19 &
		0.27 &
		0.41 &
		1.01 &
		0\\\hline
		&
		SIMEX-Q** &
		46.99 &
		0.14 &
		0.27 &
		0.42 &
		1.51 &
		1\\\hline
		&
		SIMEX-C** &
		46.99 &
		0.14 &
		0.35 &
		0.54 &
		2.06 &
		1\\\hline
		&
		SIMEX-Q4** &
		46.99 &
		0.14 &
		0.59 &
		0.93 &
		3.71 &
		1\\\hline
		&
		MI &
		42.58 &
		0.19 &
		0.27 &
		0.42 &
		0.97 &
		0\\\hline
		&
		MI-NULL*** &
		42.67 &
		0.19 &
		0.28 &
		0.42 &
		0.97 &
		0\\\hline
		&
		GS7 &
		31.87 &
		0.33 &
		0.08 &
		0.12 &
		0.41 &
		0\\\hline
		&
		GS28 &
		16.98 &
		0.6 &
		0.05 &
		0.08 &
		0.27 &
		0\\\hline
		5 &
		naïve &
		52.3 &
		0.12 &
		0.26 &
		0.52 &
		1.52 &
		0\\\hline
		&
		RC &
		47.86 &
		0.16 &
		0.26 &
		0.44 &
		1.07 &
		0\\\hline
		&
		SIMEX-Q** &
		52.32 &
		0.12 &
		0.3 &
		0.51 &
		1.7 &
		1\\\hline
		&
		SIMEX-C** &
		52.32 &
		0.12 &
		0.39 &
		0.67 &
		2.48 &
		1\\\hline
		&
		SIMEX-Q4** &
		52.32 &
		0.12 &
		0.68 &
		1.16 &
		4.6 &
		1\\\hline
		&
		MI &
		47.93 &
		0.16 &
		0.27 &
		0.45 &
		0.97 &
		0\\\hline
		&
		MI-NULL*** &
		47.88 &
		0.16 &
		0.27 &
		0.45 &
		0.96 &
		0\\\hline
		&
		GS7 &
		35.93 &
		0.31 &
		0.09 &
		0.15 &
		0.48 &
		0\\\hline
		&
		GS28 &
		19.39 &
		0.58 &
		0.06 &
		0.1 &
		0.33 &
		0\\\hline
		\multicolumn{8}{c}{  Cluster method: GMM}\\\hline
		3 &
		naïve &
		15 &
		0.2 &
		0.43 &
		0.65 &
		1.35 &
		1\\\hline
		&
		RC &
		30.39 &
		0.27 &
		0.27 &
		0.37 &
		0.98 &
		0\\\hline
		&
		SIMEX-Q** &
		14.94 &
		0.2 &
		0.51 &
		0.71 &
		1.59 &
		1\\\hline
		&
		SIMEX-C** &
		14.94 &
		0.2 &
		0.7 &
		0.98 &
		2.16 &
		1\\\hline
		&
		SIMEX-Q4** &
		14.94 &
		0.2 &
		1.32 &
		1.85 &
		4.09 &
		1\\\hline
		&
		MI &
		30.35 &
		0.27 &
		0.28 &
		0.39 &
		0.97 &
		0\\\hline
		&
		MI-NULL*** &
		30.25 &
		0.27 &
		0.28 &
		0.39 &
		0.97 &
		0\\\hline
		&
		GS7 &
		15.59 &
		0.46 &
		0.12 &
		0.17 &
		0.37 &
		0\\\hline
		&
		GS28 &
		10.06 &
		0.68 &
		0.07 &
		0.1 &
		0.22 &
		0\\\hline
		4 &
		naïve &
		15.7 &
		0.19 &
		0.49 &
		0.9 &
		1.69 &
		0\\\hline
		&
		RC &
		38.25 &
		0.22 &
		0.29 &
		0.45 &
		0.98 &
		0\\\hline
		&
		SIMEX-Q** &
		15.65 &
		0.19 &
		0.59 &
		0.94 &
		1.98 &
		1\\\hline
		&
		SIMEX-C** &
		15.65 &
		0.19 &
		0.81 &
		1.32 &
		2.85 &
		1\\\hline
		&
		SIMEX-Q4** &
		15.65 &
		0.19 &
		1.5 &
		2.42 &
		5.5 &
		1\\\hline
		&
		MI &
		38.31 &
		0.22 &
		0.3 &
		0.46 &
		0.97 &
		0\\\hline
		&
		MI-NULL*** &
		38.34 &
		0.22 &
		0.31 &
		0.47 &
		0.97 &
		0\\\hline
		&
		GS7 &
		17.53 &
		0.44 &
		0.14 &
		0.23 &
		0.5 &
		0\\\hline
		&
		GS28 &
		11.62 &
		0.66 &
		0.09 &
		0.14 &
		0.31 &
		0\\\hline
		5 &
		naïve &
		16.6 &
		0.18 &
		0.5 &
		1.05 &
		1.9 &
		2\\\hline
		&
		RC &
		42.92 &
		0.2 &
		0.29 &
		0.49 &
		1.05 &
		0\\\hline
		&
		SIMEX-Q** &
		16.6 &
		0.18 &
		0.62 &
		1.07 &
		2.21 &
		1\\\hline
		&
		SIMEX-C** &
		16.6 &
		0.18 &
		0.86 &
		1.5 &
		3.21 &
		1\\\hline
		&
		SIMEX-Q4** &
		16.6 &
		0.18 &
		1.58 &
		2.8 &
		6.51 &
		1\\\hline
		&
		MI &
		43.08 &
		0.2 &
		0.3 &
		0.5 &
		0.97 &
		0\\\hline
		&
		MI-NULL*** &
		43.01 &
		0.2 &
		0.3 &
		0.51 &
		0.97 &
		0\\\hline
		&
		GS7 &
		19.79 &
		0.43 &
		0.15 &
		0.26 &
		0.6 &
		0\\\hline
		&
		GS28 &
		13.14 &
		0.65 &
		0.09 &
		0.17 &
		0.4 &
		0\\\hline
	\end{tabular}
\end{table}
*Classifications were considered failed if a cluster was empty.**For SIMEX-3-SA different extrapolation functions were used: quadratic (Q), cubic (C) and quartic (Q4).***MI-NULL denotes the MI method without considering the health outcome in the ME model.

\clearpage\begin{table}[h]
	\caption{Simulation results of correction methods and gold standard methods based on 7 and 28 days (GS7 and GS28) for scenarios with outcome  $H_i^{B-\mathit{cat}}$  and cluster method with \textit{C} clusters regarding empirical misclassification rate (MR), adjusted Rand index (aRI), mean absolute bias, maximal absolute bias and median mean relative absolute bias, and number of failed classifications* for 1000 simulation data sets.}
	\tiny\begin{tabular}{l|l|l|l|l|l|l|l}
		\hline
		\textit{C} &
		Correction method &
		MR &
		aRI &
		$\overline{{\Delta}}$ &
		$\overline{{\Delta}}_{\mathit{max}}$ &
		$\mathit{med}(\overline{{\Delta}}_{\mathit{rel}})$ &
		Number of failed classifications*\\\hline
		\multicolumn{8}{c}{  Cluster method: k-means}\\\hline
		3 &
		naïve &
		39.5 &
		0.17 &
		0.43 &
		0.65 &
		1.2 &
		0\\\hline
		&
		RC &
		35.06 &
		0.23 &
		0.49 &
		0.68 &
		1.05 &
		0\\\hline
		&
		SIMEX-Q** &
		39.39 &
		0.17 &
		0.5 &
		0.69 &
		1.36 &
		1\\\hline
		&
		SIMEX-C** &
		39.39 &
		0.17 &
		0.72 &
		0.99 &
		2.05 &
		1\\\hline
		&
		SIMEX-Q4** &
		39.39 &
		0.17 &
		1.38 &
		1.91 &
		4.09 &
		1\\\hline
		&
		MI &
		35.07 &
		0.23 &
		0.46 &
		0.64 &
		0.91 &
		0\\\hline
		&
		MI-NULL*** &
		35.04 &
		0.23 &
		0.5 &
		0.68 &
		0.97 &
		0\\\hline
		&
		GS7 &
		25.96 &
		0.38 &
		0.19 &
		0.27 &
		0.51 &
		0\\\hline
		&
		GS28 &
		13.76 &
		0.63 &
		0.13 &
		0.18 &
		0.35 &
		0\\\hline
		4 &
		naïve &
		47.00 &
		0.14 &
		0.48 &
		0.87 &
		1.49 &
		0\\\hline
		&
		RC &
		42.66 &
		0.19 &
		0.52 &
		0.81 &
		1.07 &
		0\\\hline
		&
		SIMEX-Q** &
		46.99 &
		0.14 &
		0.58 &
		0.91 &
		1.66 &
		1\\\hline
		&
		SIMEX-C** &
		46.99 &
		0.14 &
		0.86 &
		1.36 &
		2.55 &
		1\\\hline
		&
		SIMEX-Q4** &
		46.99 &
		0.14 &
		1.69 &
		2.68 &
		5.12 &
		1\\\hline
		&
		MI &
		42.66 &
		0.19 &
		0.5 &
		0.77 &
		0.94 &
		0\\\hline
		&
		MI-NULL*** &
		42.66 &
		0.19 &
		0.53 &
		0.82 &
		0.97 &
		0\\\hline
		&
		GS7 &
		31.87 &
		0.33 &
		0.22 &
		0.35 &
		0.61 &
		0\\\hline
		&
		GS28 &
		16.98 &
		0.6 &
		0.15 &
		0.24 &
		0.44 &
		0\\\hline
		5 &
		naïve &
		52.3 &
		0.12 &
		0.52 &
		1.05 &
		1.73 &
		0\\\hline
		&
		RC &
		47.86 &
		0.16 &
		0.54 &
		0.91 &
		1.19 &
		0\\\hline
		&
		SIMEX-Q** &
		52.32 &
		0.12 &
		0.64 &
		1.1 &
		1.88 &
		1\\\hline
		&
		SIMEX-C** &
		52.32 &
		0.12 &
		0.98 &
		1.7 &
		3.21 &
		1\\\hline
		&
		SIMEX-Q4** &
		52.32 &
		0.12 &
		1.93 &
		3.35 &
		6.57 &
		1\\\hline
		&
		MI &
		47.89 &
		0.16 &
		0.5 &
		0.85 &
		0.97 &
		0\\\hline
		&
		MI-NULL*** &
		47.91 &
		0.16 &
		0.53 &
		0.89 &
		0.96 &
		0\\\hline
		&
		GS7 &
		35.93 &
		0.31 &
		0.25 &
		0.44 &
		0.78 &
		0\\\hline
		&
		GS28 &
		19.39 &
		0.58 &
		0.18 &
		0.31 &
		0.56 &
		0\\\hline
		\multicolumn{8}{c}{  Cluster method: GMM}\\\hline
		3 &
		naïve &
		15 &
		0.2 &
		3.31 &
		4.96 &
		1.28 &
		1\\\hline
		&
		RC &
		30.39 &
		0.27 &
		0.55 &
		0.75 &
		1.01 &
		0\\\hline
		&
		SIMEX-Q** &
		14.94 &
		0.2 &
		4.01 &
		6.11 &
		1.54 &
		1\\\hline
		&
		SIMEX-C** &
		14.94 &
		0.2 &
		6.92 &
		10.44 &
		3.02 &
		1\\\hline
		&
		SIMEX-Q4** &
		14.94 &
		0.2 &
		14.4 &
		21.68 &
		7.04 &
		1\\\hline
		&
		MI &
		30.08 &
		0.27 &
		0.53 &
		0.72 &
		0.91 &
		0\\\hline
		&
		MI-NULL*** &
		30.25 &
		0.27 &
		0.55 &
		0.76 &
		0.97 &
		0\\\hline
		&
		GS7 &
		15.59 &
		0.46 &
		0.89 &
		1.44 &
		0.62 &
		0\\\hline
		&
		GS28 &
		10.06 &
		0.68 &
		0.51 &
		0.85 &
		0.41 &
		0\\\hline
		4 &
		naïve &
		15.7 &
		0.19 &
		3.71 &
		7.08 &
		1.66 &
		0\\\hline
		&
		RC &
		38.25 &
		0.22 &
		0.6 &
		0.96 &
		1.08 &
		0\\\hline
		&
		SIMEX-Q** &
		15.65 &
		0.19 &
		4.76 &
		8.72 &
		2.14 &
		1\\\hline
		&
		SIMEX-C** &
		15.65 &
		0.19 &
		8.34 &
		15.38 &
		4.75 &
		1\\\hline
		&
		SIMEX-Q4** &
		15.65 &
		0.19 &
		17.63 &
		32.68 &
		11.5 &
		1\\\hline
		&
		MI &
		38.21 &
		0.22 &
		0.59 &
		0.92 &
		0.94 &
		0\\\hline
		&
		MI-NULL*** &
		38.34 &
		0.22 &
		0.65 &
		1.05 &
		0.97 &
		0\\\hline
		&
		GS7 &
		17.53 &
		0.44 &
		1.3 &
		2.66 &
		0.85 &
		0\\\hline
		&
		GS28 &
		11.62 &
		0.66 &
		0.81 &
		1.65 &
		0.58 &
		0\\\hline
		5 &
		naïve &
		16.6 &
		0.18 &
		3.69 &
		8.34 &
		2.08 &
		2\\\hline
		&
		RC &
		42.92 &
		0.2 &
		0.74 &
		1.44 &
		1.2 &
		0\\\hline
		&
		SIMEX-Q** &
		16.6 &
		0.18 &
		4.96 &
		10.17 &
		2.43 &
		1\\\hline
		&
		SIMEX-C** &
		16.6 &
		0.18 &
		8.84 &
		18.03 &
		5.73 &
		1\\\hline
		&
		SIMEX-Q4** &
		16.6 &
		0.18 &
		18.98 &
		38.47 &
		14.16 &
		1\\\hline
		&
		MI &
		43.06 &
		0.2 &
		0.66 &
		1.24 &
		0.99 &
		0\\\hline
		&
		MI-NULL*** &
		43.01 &
		0.2 &
		0.72 &
		1.38 &
		0.99 &
		0\\\hline
		&
		GS7 &
		19.79 &
		0.43 &
		1.29 &
		3 &
		0.98 &
		0\\\hline
		&
		GS28 &
		13.14 &
		0.65 &
		0.69 &
		1.67 &
		0.73 &
		0\\\hline
	\end{tabular}
\end{table}
*Classifications were considered failed if a cluster was empty.**For SIMEX-3-SA different extrapolation functions were used: quadratic (Q), cubic (C) and quartic (Q4).***MI-NULL denotes the MI method without considering the health outcome in the ME model.